\documentclass{aa}
\usepackage[varg]{txfonts}
\usepackage{natbib}
\usepackage{graphicx}
\usepackage{mathabx}
\usepackage{amsmath}
\usepackage[normalem]{ulem}

\usepackage{color}
\usepackage{multirow}
\usepackage{longtable}
\usepackage{pdflscape}
\usepackage{subcaption}
\usepackage{array}

\DeclareCaptionLabelSeparator{persp}{.$\enskip$}

\bibpunct{(}{)}{;}{a}{}{,}

\begin{document}

\title{Deep, multi-band photometry of low-mass stars to reveal young clusters: a blind study of the NGC~2264 region}

\author{L. Venuti\inst{1,2} \and F. Damiani\inst{1} \and L. Prisinzano\inst{1}}

\institute{INAF -- Osservatorio Astronomico di Palermo G.\,S. Vaiana, Piazza del Parlamento 1, 90134 Palermo, Italy \and Eberhard Karls Universit\"at, Institut f\"ur Astronomie und Astrophysik, Sand 1, 72076 T\"ubingen, Germany \\ e-mail: laura.venuti@astro.uni-tuebingen.de}

\date{Received 18 April 2018 / Accepted 4 November 2018}

\abstract {Thanks to their extensive and homogeneous sky coverage, deep, large-scale, multi-wavelength surveys are uniquely suited to statistically identify and map young star clusters in our Galaxy. Such studies are crucial to address themes like the initial mass function, or the modes and dynamics of star cluster formation and evolution. }{We aim to test a purely photometric approach to statistically identify a young clustered population embedded in a large population of field stars, with no prior knowledge on the nature of stars in the field. We conducted our blind test study on the NGC~2264 region, which hosts a well-known, richly populated young cluster ($\sim$3~Myr-old) and several active star-forming sites.}{We selected a large (4~deg$^2$) area around the NGC~2264 cluster, and assembled an extensive $r,i,J$ catalog of the field from pre-existing large-scale surveys, notably Pan-STARRS1 and UKIDSS. We then mapped the stellar color locus on the ($i-J$, $r-i$) diagram to select M-type stars, which {offer the following observational advantages with respect to more massive stars: i) they comprise a significant fraction of the Galactic stellar population; ii) their pre-main sequence phase lasts significantly longer than for higher-mass stars; iii) they exhibit the strongest luminosity evolution from the pre-main sequence to the main sequence; iv) their observed $r,i,J$ colors provide a direct and empirical estimate of $A_V$}. A comparative analysis of the photometric and spatial properties of M-type stars as a function of $A_V$ enabled us to probe the structure and stellar content of our field.}{Using only $r,i,J$ photometry, we could identify two distinct populations in our field: a diffuse field population and a clustered population in the center of the field. The presence of a concentration of occulting material, spatially associated with the clustered population, allowed us to derive an estimate of its distance (800--900~pc) and age ($\sim${0.5--5}~Myr); these values are overall consistent with the literature parameters for the NGC~2264 star-forming region. The extracted clustered population exhibits a hierarchical structure, with two main clumps and peaks in number density of objects around the most extincted locations within the field. An excellent agreement is found between the observed substructures for the clustered population and a map of the NGC~2264 subregions reported in the literature. Our selection of clustered members is coherent with the literature census of the NGC~2264 cluster for about 95\% of the objects located in the inner regions of the field, where the estimated contamination rate by field stars in our sample is only 2\%. In addition, the availability of a uniform dataset for a large area around the NGC~2264 region enabled us to discover a population of about a hundred stars with indications of statistical membership to the cluster, therefore extending the low-mass population census of NGC~2264 to distances of 10--15~pc from the cluster cores.}{By making use solely of deep, multi-band ($r,i,J$) photometry, without assuming any further knowledge on the stellar population of our field, we were able to statistically identify and reconstruct the structure of a very young cluster that has been a prime target for star formation studies over several decades. The method tested here can be readily applied to surveys such as Pan-STARRS and the future LSST to undertake a first complete census of low-mass, young stellar populations down to distances of several kpc across the Galactic plane.}


\titlerunning{Deep, multi-band photometry of low-mass stars: a blind study of NGC 2264}

\maketitle

\section{Introduction} \label{sec:introduction}

Over the past two decades, significant advancement in tomographic studies of our Galaxy was prompted by the advent of large-area, multi-band sky surveys (see \citealp{ivezic2012} for a review). To name a few, the Sloan Digital Sky Survey \citep[SDSS;][]{york2000} mapped an area of $\sim$7500~deg$^2$ in the North Galactic Cap, using five broad-band filters designed for the survey (more specifically, $u$,$g$,$r$,$i$,$z$, which together cover the spectral range from 3000 to {$\sim$10000} \AA; \citealp{F96SDSS}), to a depth of $\sim$22.2~mag in the $r$-band ($\lambda_\mathrm{eff}$\,=\,6261~\AA). The INT Photometric H$\alpha$ Survey \citep[IPHAS;][]{iphas} mapped 1800~deg$^2$ in the Northern Galactic Plane, down to $r$$\sim$21~mag, using the SDSS $r$-band  and $i$-band ($\lambda_\mathrm{eff}$\,=\,7672~\AA) filters, plus a narrow-band H$\alpha$ filter; IPHAS was later complemented by the VST Photometric H$\alpha$ Survey \citep[VPHAS+;][]{drew2014} in the Southern Galactic Plane. In the near-infrared (NIR), the Two Micron All Sky Survey \citep[2MASS;][]{2MASS} mapped the entire celestial sphere in $J$ ($\lambda_\mathrm{eff}$\,=\,1.25~$\mu$m), $H$ ($\lambda_\mathrm{eff}$\,=\,1.65~$\mu$m), and $K_S$ ($\lambda_\mathrm{eff}$\,=\,2.16~$\mu$m), reaching a depth of 14.3~mag in $K_S$; similar filters were employed for the UKIRT Infrared Deep Sky Survey \citep[UKIDSS;][]{lawrence2007} to map an area of 7000~deg$^2$ to a depth of $K$$\sim$18.3~mag. The wide, deep and uniform spatial coverage brought in by such surveys dramatically enhanced our ability to resolve stellar populations across the Milky Way, hence bringing in first detailed maps of the structure of our Galaxy. 

In the field of young star clusters, extensive, dedicated IR campaigns performed notably with {\it Spitzer}, together with more general surveys such as 2MASS, enabled to reveal the variety of cluster morphologies, that bear the imprints of the dynamics of cluster formation in the molecular clouds (see \citealp{allen2007} for a review). However, investigations of young, embedded clusters, especially in the optical, have so far been limited to distances of {\small $\lesssim$} 1~kpc \citep[e.g.,][]{messineo2009, luhman2012}, and are often incomplete in the lower stellar mass regimes. And yet, low-mass, M-type stars ($M_\star$$\sim$0.1--0.5~$M_\odot$) are estimated to account for over 50\% of the total stellar population in the Galactic field and its young embedded clusters {\citep[e.g.,][]{offner2014}}; complete censuses of young ({\small $\lesssim$} 10~Myr), low-mass stellar populations in different environments are therefore crucial to address key issues in star formation and early evolution, like the modes and duration of star cluster formation, the initial mass function, or the timescales of protoplanetary disk evolution and dispersal \citep[e.g.,][]{briceno2007}. Remarkable progress in this respect has recently been warranted by the first set of surveys conducted with the Panoramic Survey Telescope and Rapid Response System (Pan-STARRS1; \citealp{chambers2016}). Pan-STARRS1 imaged the entire Northern Sky, {above} a declination of $\delta$$\sim$$-30^\circ$, in five broad-band filters ($g_{P1}$,$r_{P1}$,$i_{P1}$,$z_{P1}$,$y_{P1}$) which encompass the spectral window from $\sim$4000 to 10500~\AA. The survey, which included multiple pointings per observed field per filter over a span of four years, reached a depth of $g_{P1}$$\sim$23.3, $r_{P1}$$\sim$23.2, and $i_{P1}$$\sim$23.1 in the stacked images; these values theoretically allow simultaneous $g_{P1}$,$r_{P1}$,$i_{P1}$ detection of very young ($\sim$1~Myr), low-reddening stars at distances of 1--2~kpc down to masses of $\sim$0.05--0.1~$M_\odot$. 

In the next decade, a whole new scenario will be opened with the Large Synoptic Survey Telescope (LSST; \citealp{ivezic2008}). Starting from the early 2020s, LSST will survey the Southern Sky ($\delta$$\leq$$+10^\circ$) in six filters ($u$,$g$,$r$,$i$,$z$,$y$), modeled after the SDSS system, that extend from $\sim$3200 to 11000 \AA. Each portion of sky will be imaged repeatedly over the scheduled ten-year duration of the LSST campaign, with an expected single-epoch depth of $r$$\sim$24.5 and a coadded depth of $r$$\sim$27.5 at the end of the project. The wealth and homogeneity of photometric data that is being assembled and that will emerge over the next years will enable a first detailed census of pre-main sequence (PMS) stellar populations across the Galactic plane. A most important point, so far largely overlooked, is that photometric searches for young, distant clusters find a most favorable domain in the M-type stellar regime. Indeed, while early-type stars exhibit little luminosity evolution in the transition from the PMS to the zero-age main sequence (ZAMS), M-type PMS stars are significantly brighter than their early-main sequence (MS) counterparts \citep[e.g.,][]{siess2000}. As a consequence, if M-type PMS stars and field dwarfs are mingled in a region of sky, the MS population will fall below the detection threshold at much closer distances than the PMS population; M-type PMS stars will therefore stand out with respect to M-type MS stars in deep photometric surveys \citep{damiani2018}.

\begin{figure}
\resizebox{\hsize}{!}{\includegraphics{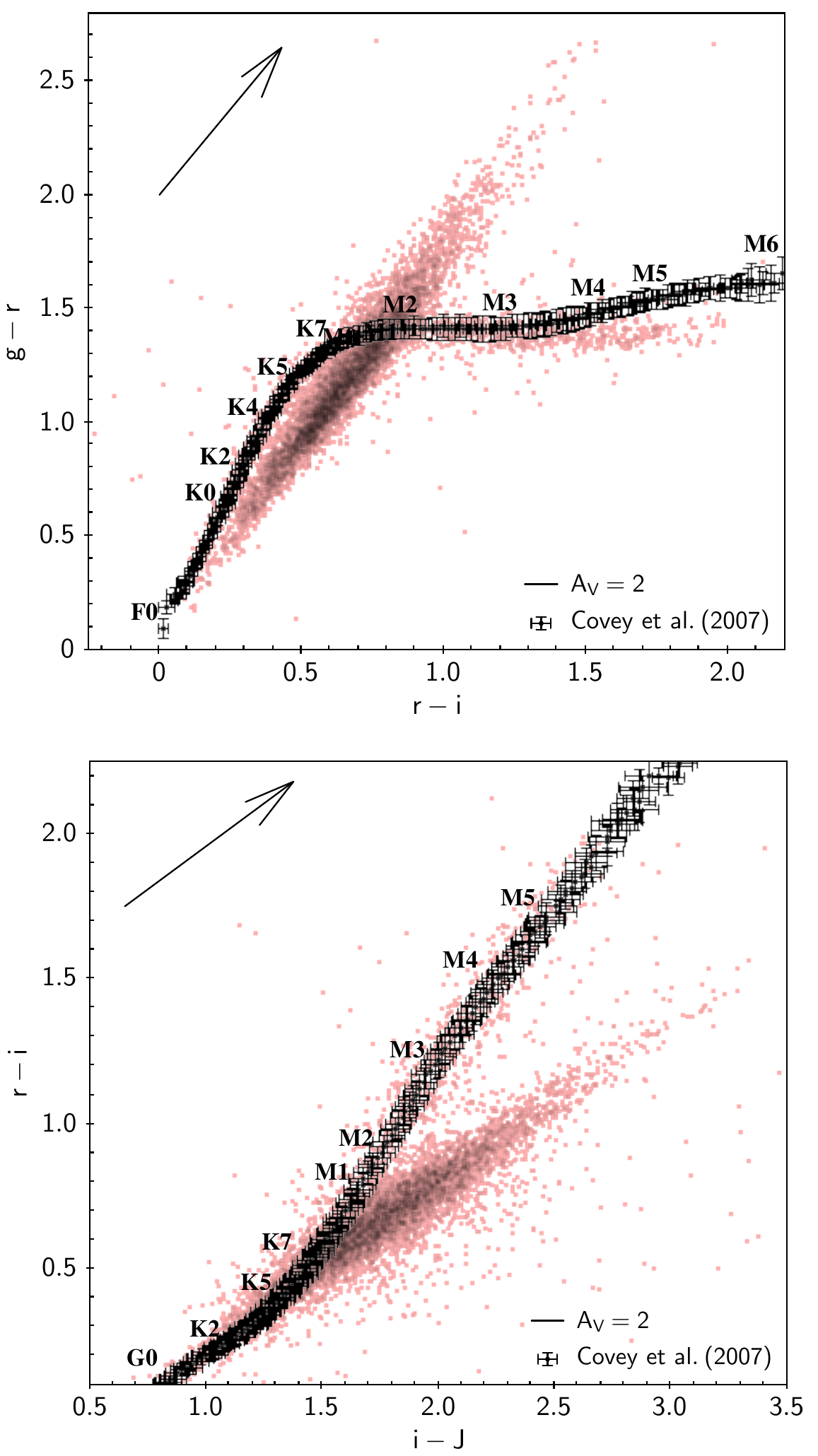}}
\caption{{\it Upper panel}: $g-r$ vs. $r-i$ color locus of stars extracted from a 1$^\circ$$\times$1$^\circ$ field around the NGC~2264 cluster (photometry from \citealp{venuti2014}). The expected color properties of MS dwarfs as a function of their spectral type, {and the associated uncertainties}, are {traced and} labelled following \citet{covey07} to guide the eye. {\it Lower panel}: $r-i$ vs. $i-J$ color locus for the same population of stars shown in the upper panel. The $J$-band photometry is extracted from 2MASS.}
\label{fig:gr_ri_iJ}
\end{figure}

Estimating the extinction A$_V$ suffered by individual objects is a major issue in spatial analyses of resolved stellar populations, since A$_V$ is a crucial parameter to determine the stellar properties ({$T_{\rm eff}$, $L_{\rm bol}$, and derived parameters}), and {also provides an indirect indicator} of the distance to the observed source. A classic photometric approach to determining A$_V$ consists in comparing the observed stellar colors to a reference intrinsic spectral type--color scale, assuming that the spectral types of the objects are known from independent (spectroscopic) measurements. However, this approach is subject to two limitations: i) spectroscopic surveys are significantly more time-consuming {than} wide-field photometric surveys, and can only be performed on small and/or nearby samples of stars; ii) the derived A$_V$ measurements are somewhat dependent on the specific reference scale used, as different models/calibrations can exhibit considerable offsets with respect to each other (see, e.g., Sect.~4.1 of \citealp{venuti2018}). As shown in Fig.\,\ref{fig:gr_ri_iJ}, mapping the stellar color locus in SDSS filters, possibly coupled with the $J$-band, offers a direct and empirical solution to this issue for low-mass stars: while early-type ($\leq$K7) stars exhibit intrinsic $g,r,i,J$ colors that progress with spectral type almost parallel to the reddening vector, the photospheric color sequence traced by M-type stars distinctly diverges from the reddening direction. This peculiar photometric behavior renders the ($r-i$, $g-r$) and the ($i-J$, $r-i$) color-color diagrams a powerful diagnostic tool to: i) identify and extract M-type stars from the bulk of sources detected in a given field; ii) infer a straightforward estimate of their individual A$_V$ by measuring the distance between their observed colors and the non-reddened envelope of the M-type color locus on the diagram along the reddening direction. 
Assuming a uniform reddening law along the cone of view, a stratification in A$_V$ across a given stellar population is a direct proxy to the range of distances within which stars belonging to the population under exam are distributed\footnote{This reasoning does not apply {across} a molecular cloud or an embedded star-forming site, characterized by more conspicuous and more rapidly spatially varying extinction than the surrounding areas.}. Therefore, deep $g,r,i,J$ photometry of M-type stars in a given field provides a direct means of probing the spatial structure of dust of the imaged region.  

\begin{figure}
\resizebox{\hsize}{!}{\includegraphics{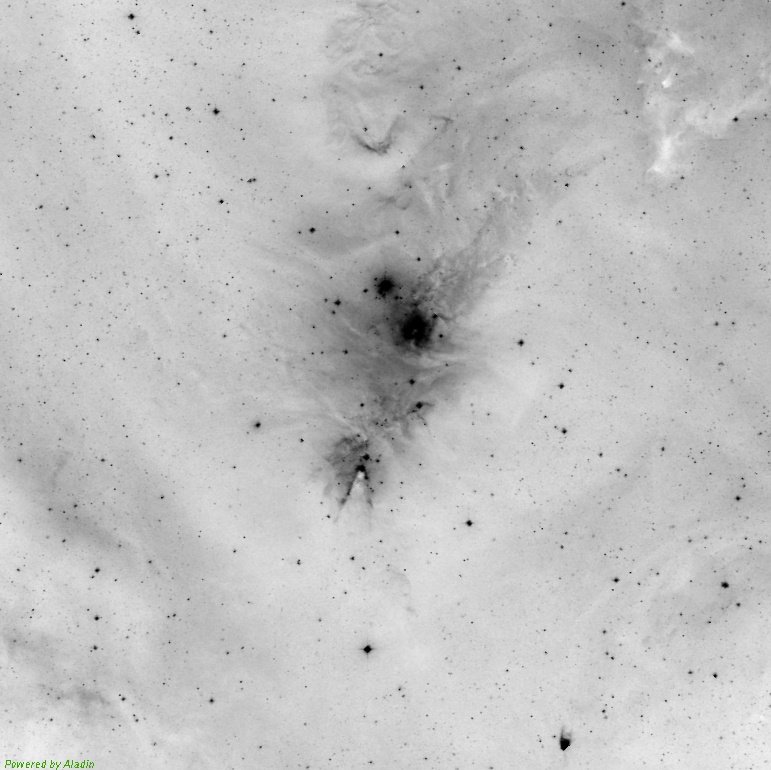}}
\caption{A Digitized Sky Survey (POSS-II) view of the field investigated in this study, from data collected in the red passband. North is up and East to the left.}
\label{fig:field}
\end{figure}

The relatively simple but very efficient conceptual approach outlined above was formalized and explored for the first time in \citet{damiani2018}. This method has enormous implications for the science of young star clusters in an era when extremely accurate photometric maps of the entire sky are taking shape. By combining the appropriate spatial and color diagrams of stars in a given field, without any additional information on their nature, this approach enables revealing young distant clusters by extracting their M-type population, more numerous with respect to earlier-type cluster members and favored for detection in deep surveys with respect to older field stars in the same spectral class. To test the performance and predictive capability of the method, we conducted a blind photometric study of the region around the $\sim$3-5~Myr-old cluster NGC~2264 \citep[e.g.,][]{dahm08}. Our investigation, reported in this paper, focused on a wide, 2$^\circ$$\times$2$^\circ$ field centered on the NGC~2264 cluster (R.A. = 06:41:00.0; Dec~= +09:38:20.0). {An image of the field, produced with the Aladin Sky Atlas \citep{ALADIN}, making use of data products from the Digitized Sky Survey \citep{lasker1996}, is presented in Fig.\,\ref{fig:field}}. We made use of the available, large-scale optical ($r$,$i$) and NIR ($J$) surveys to build an extensive photometric catalog for M-type stars detected in this area, and analyzed the gathered photometry to derive a reddening map of the region and to probe the nature of its stellar population. Other than the definition of the central pointing for our field, we carried out our study without assuming any prior knowledge on the presence of a young clustered population in the region, until the final assessment of our results. 

The paper is organized as follows. Sect.\,\ref{sec:data} describes the photometric catalog we assembled and used for the current study. In Sect.\,\ref{sec:results}, we illustrate the procedure adopted to identify M-type stars in the field, derive their reddening properties, and characterize the spatial structure of the region and the nature of its stellar population. In Sect.\,\ref{sec:discussion}, we discuss the impact of field contamination on our analysis, and compare our results to the information available in the literature for the NGC~2264 region, in order to assess the robustness of our selection method. Our conclusions are summarized in Sect.\,\ref{sec:conclusions}.

\section{Data selection} \label{sec:data}

\subsection{Optical {${r}$,${i}$} catalog} \label{sec:optical_cat}

\begin{figure}
\resizebox{\hsize}{!}{\includegraphics{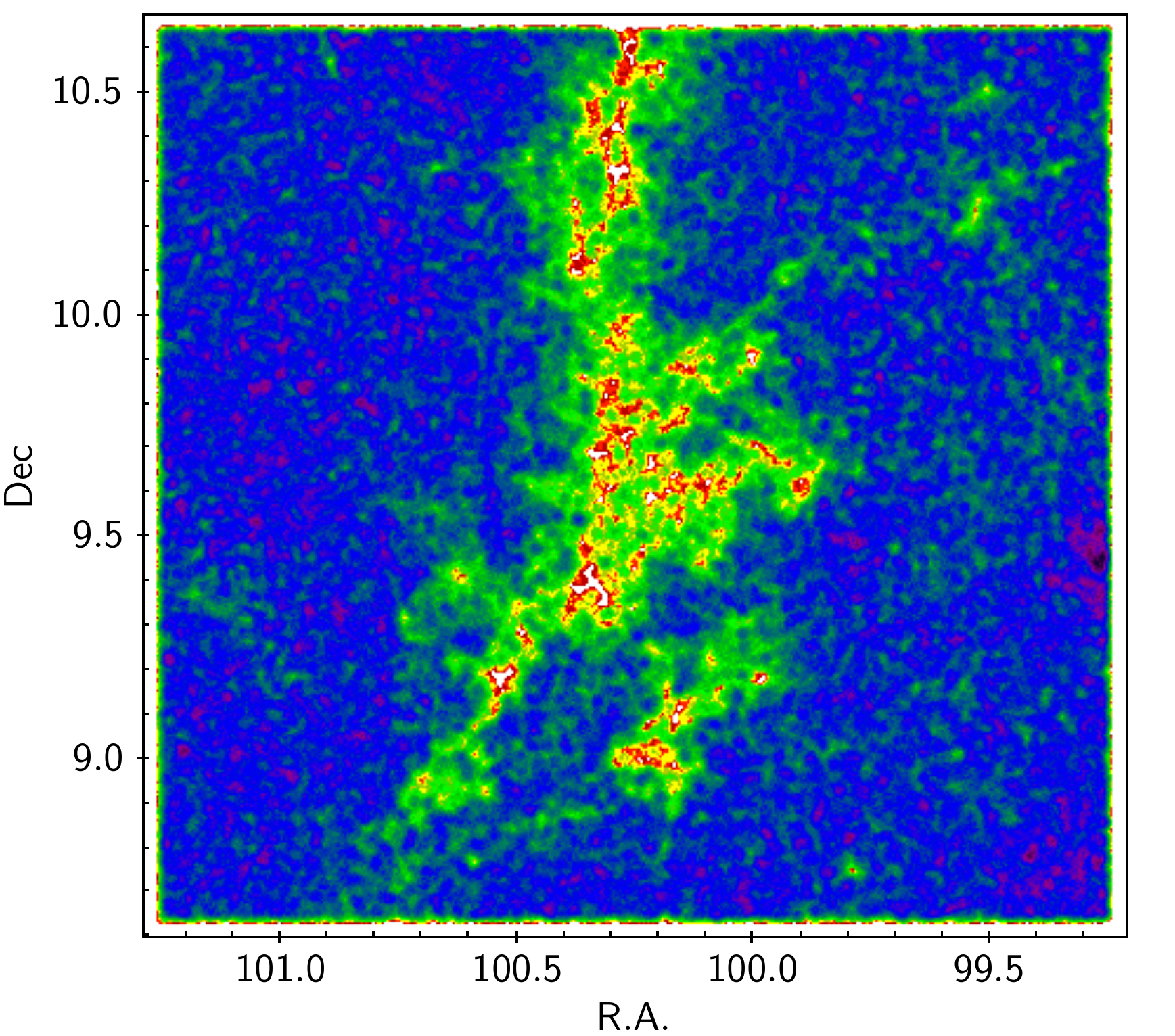}}
\caption{Number density map of the population of our catalog across the 2$^\circ$$\times$2$^\circ$ region investigated in this study (see text). Spatial coordinates (R.A. and Dec) are expressed in degrees ($^\circ$) in this and all subsequent diagrams. Objects densities are mapped according to a ``rainbow'' color scale, where purple and blue correspond to the highest number density, green indicates intermediate densities, and red corresponds to the lowest number density. North is up and East is left on the diagram.}
\label{fig:ra_dec_density}
\end{figure}

The $r$,$i$ point source catalog of the 4~deg$^2$ field investigated in this work was assembled with the goal to ensure a photometric coverage as robust, uniform, complete, and unbiased as possible across the entire area. We selected four deep photometric surveys as input sources for our catalog, as detailed below.
\begin{enumerate}
\item \textbf{\textit{\underline{CSI~2264}}}: the Coordinated Synoptic Investigation of NGC~2264 \citep{cody2014} provided the most extensive, multi-wavelength characterization of the NGC~2264 region available to date. The campaign included deep $u,g,r,i$ mapping performed at the Canada--France--Hawaii Telescope (CFHT) with the wide-field imager MegaCam \citep{venuti2014}. Simultaneous $r$-band and $i$-band detection was achieved for 20646 point sources in the magnitude range $r$$\sim$13.5--21.2, distributed over a 1$^\circ$$\times$1$^\circ$ area centered on (R.A., Dec) = (06:41:00.0, +09:38:20.0).
\item \textbf{\textit{\underline{Pan-STARRS1}}}: the stack photometry from the first data release (DR) of Pan-STARRS1 \citep{flewelling2016} covers, without gaps, the entire area investigated in this study. From the Pan-STARRS1 database we extracted all source detections located in our field of interest, subject to the conditions of i) possessing at least $r_{P1}$-band and $i_{P1}$-band photometry, and ii) being brighter than the mean 5$\sigma$ point source limiting sensitivities defined in \citeauthor{chambers2016} (\citeyear{chambers2016}; $r_{P1}$ =~23.2, $i_{P1}$~= 23.1). These criteria allowed us to retrieve 200602 sources with Pan-STARRS1 photometry in our field.
\item \textbf{\textit{\underline{IPHAS}}}: from the second and most recent DR of IPHAS (\citealp{barentsen2014}; {$r$$\sim$13--21.2, $i$$\sim$12--20}) we retrieved 138506 sources with $r$-band and $i$-band photometry in our field of interest. These sources are distributed across the entire region, except for two small patches, not mapped within IPHAS, located respectively in the north-west and in the south of our field; the total area of these two gaps is about 6.5\% of the 4~deg$^2$.
\item \textbf{\textit{\underline{SDSS}}}: SDSS photometric data for the region of interest to this study were released as part of DR7 \citep{SDSSDR7}. From the SDSS archive we extracted a photometric catalog of objects located within a radius of 60' around the coordinates defined earlier, and retained those entries which satisfied the following conditions: i) being flagged as ``point sources'', as opposed to ``extended sources''; ii) being a detection in both $r$ and $i$; iii) being brighter than the magnitude limits for 95\% detection repeatability for point sources defined for the survey ($r$ = 22.2, $i$ = 21.3). This procedure allowed us to select 118823 sources, distributed effectively across 75\% of our field, due to discontinuous area coverage.
\end{enumerate}
The final optical catalog of the region was built by cross-matching the four catalogs listed above. The cross-correlation was performed with TOPCAT \citep{TOPCAT}, using a matching radius of 1~arcsec and a Join Type ``2 not 1'', in order to avoid duplicates. Sources were retained preferentially from the input catalogs in the order shown in the above list. Prior to merging the individual catalogs, a cleaning procedure was performed on each of them in order to reject spurious or problematic detections, as described in the following.
\begin{itemize}
\item Regarding the CFHT/MegaCam catalog obtained as part of the CSI~2264 campaign, a description of the photometry extraction and validation procedure was presented in \citet{venuti2014}. Readers are referred to that paper for further details.
\item For the Pan-STARRS1 catalog, the rejection of bad sources was performed in four steps, taking into account several parameters and quality flags associated with each object in the Pan-STARRS1 database (see \citealp{flewelling2016}), in particular the number of valid single-epoch $r_{P1}$-band ($nr$) and $i_{P1}$-band ($ni$) detections, the individual flag values embedded in the $objInfoFlag$ bitmask, and the photometric errors associated with the measurements ($rMeanPSFMagErr \equiv err_\mathrm{r}$, $iMeanPSFMagErr \equiv err_\mathrm{i}$). At first, all catalog entries with $nr$=0 or $ni$=0 (i.e., sources without any good single-epoch detections in $r_{P1}$ or $i_{P1}$, despite having an associated photometry measurement from the stacked images) were rejected. Then, all sources flagged upon being affected by astrometric or parallax issues, or extended, or with poor-quality photometry, as coded in the $objInfoFlag$ value, were rejected. Afterwards, we computed the difference between the listed PSF photometry ($m_{PSF}$) and Kron photometry\footnote{The methodology to derive Kron magnitudes was introduced in \citet{kron1980} to characterize the photometry of extended sources (galaxies). Large discrepancies between the PSF and Kron magnitudes derived for a given source are therefore indicative of extended objects (see \citealp{farrow2014}).} ($m_{Kron}$) for each object, and rejected all sources with $m_{PSF}-m_{Kron}$ larger by more than 3$\sigma$ than the average difference computed across the entire sample. Finally, we sorted the remaining sources into groups based on the values of the $nr$ and $ni$ parameters; for each $nr$- and $ni$-group, we built the cumulative distribution respectively in $err_\mathrm{r}$ and $err_\mathrm{i}$, and discarded objects with ``anomalous'' photometric uncertainty, defined as lying below the 0.15th percentile or above the 99.85th percentile. This selection was performed individually for each $nr$- and $ni$-group, rather than collectively on the full sample, because the uncertainty associated with the photometric measurements naturally decreases with increasing number of detections.
\item For the IPHAS catalog, we applied a cut on the $pStar$ parameter (i.e., the probability that the extracted source is a star) to select {\it bona fide} point sources and discard likely extended sources. Namely, all objects with $pStar < 0.05$ were rejected. No further selection on the photometric error was performed in this case as all retained sources had associated uncertainties within the allowed range for Pan-STARRS1 data.
\item SDSS sources were filtered based on the associated photometric errors. The full sample of Pan-STARRS1 sources retained was used as reference to define the allowed uncertainty ranges in $r$ and in $i$; all objects with photometric errors in $r$ or in $i$ outside the allowed ranges were rejected.
\end{itemize}
The final optical catalog encompasses 185754 sources, ranging in magnitude from $\sim$12 to 22.5 (with a completeness limit of $\sim$21.5 in $r$ and 21 in $i$). The majority of stars in the catalog (83.4\%) were retrieved from Pan-STARRS1 observations, followed by CSI~2264--CFHT/MegaCam (11.1\%), IPHAS (3.2\%), and SDSS (2.3\%). Figure~\ref{fig:ra_dec_density} illustrates the spatial distribution of the population of the assembled catalog across our 4~deg$^2$ field. 
As can be seen, the number of sources detected is not uniform across the field: the outer regions of the area investigated exhibit the highest number densities of objects, while the concentration of stars in our catalog is lower in the central regions of our field, {where some nebulosity can be observed on the field picture in Fig.\,\ref{fig:field}}.

\subsection{$J$-band catalog} \label{sec:J_phot}

To provide a $J$-band counterpart to targets in our optical catalog, we used the UKIDSS catalog and the 2MASS catalog, as detailed below. 
\begin{enumerate}
\item \textbf{\textit{\underline{UKIDSS}}}: the NGC~2264 region was surveyed during the UKIDSS campaign as part of its Galactic Plane Survey (GPS) component {(depth $J$$\sim$19.9)}. From the UKIDSS/GPS database we extracted all sources matching our field, subject to the condition of being flagged as ``O'' (O.K.) or ``V'' (variable) in the classification of \citet{king2013}\footnote{Other flag values in the classification scheme of \citet{king2013} indicate non-stellar objects or issues with the photometry extraction from UKIRT images.}, when available.
\item \textbf{\textit{\underline{2MASS}}}: sources were extracted from the 2MASS database {(sensitivity $J$$\sim$15.8)} upon the condition of being assigned a quality flag ``A'' at least for the $J$-band photometry.
\end{enumerate}
Using the two input catalogs listed above, we could assign a $J$-band magnitude to 99.3\% of targets in our optical catalog, built as described in Sect.\,\ref{sec:optical_cat}. Following the same approach used to clean the optical catalog, we built the cumulative distribution in $J$-band photometric uncertainties ($err_\mathrm{J}$) across our sample, and rejected sources with $err_\mathrm{J}$ outside the $-3\sigma$-level to $+3\sigma$-level range. The final $r,i,J$ catalog assembled for this study therefore comprises 185205 objects.

\subsection{${g}$,${r}$,${i}$ {catalog}} \label{sec:gri_cat}

{As mentioned in Sect.\,\ref{sec:introduction}, M-type stars exhibit distinctive properties with respect to earlier-type stars both in $r,i,J$ and $g,r,i$ colors. We therefore adopted the same procedure outlined in Sect.\,\ref{sec:optical_cat} to assemble a $g,r,i$ catalog of the region from CFHT/MegaCam, Pan-STARRS1, and SDSS data (whereas IPHAS does not include the $g$-band). We decided to consider the $r,i,J$ and $g,r,i$ sets separately, rather than their intersection, because different bands are affected by different depths and completeness limits, as summarized in Table~\ref{tab:phot_range_completeness}. In particular, the catalog involving the $g$-band is significantly more shallow, and encompasses about one-third fewer sources (123370), than the $r,i,J$ catalog. In addition, the sequence of M-type stars on the ($r-i,\,\, g-r$) diagram only emerges for spectral types of M2 and later; instead, M-type stars can be identified over earlier-type stars on the ($i-J,\,\, r-i$) diagram already at the M0--M1 subclass. To ensure homogeneity and increase the statistical robustness of our study, in the following we will refer to the $r,i,J$ photometry as our primary tool of investigation, and will use the $g,r,i$ catalog, analyzed in a similar but independent fashion, to corroborate our results.
\begin{table}
\caption{Photometric ranges ($0.5^{th}-99.5^{th}$ percentiles) and completeness limits for the $r,i,J$ and $g,r,i$ catalogs of our field.}
\label{tab:phot_range_completeness}
\centering
\begin{tabular}{l| c| c}
\hline\hline
 & $r,i,J$ & $g,r,i$ \\
\hline
 $g$-band range (mag) & -- & 14.0--22.6\\
$r$-band range (mag) & 13.7--22.3 & 13.3--21.7\\
$i$-band range (mag) & 13.3--21.5 & 13.0--21.0\\
$J$-band range (mag) & 11.9--19.6 & -- \\
$g$-band completeness (mag) & -- & 22.0\\
$r$-band completeness (mag) & 21.5 & 20.5\\
$i$-band completeness (mag) & 20.5 & 19.5\\
$J$-band completeness (mag) & 18.5 & -- \\
\hline
\end{tabular}
\end{table}
}

\subsection{Photometric calibration} \label{sec:phot_calib}

To remove small systematic effects between the various SDSS-based photometric systems adopted for the surveys listed in Sect.\,\ref{sec:optical_cat}, we recalibrated the individual photometric datasets to the SDSS system. The photometric calibration was performed statistically, taking as reference the optical catalog from the CSI~2264 CFHT/MegaCam survey, which had been previously recalibrated to the SDSS system, as reported in \citet{venuti2014}. To derive the recalibration equations for Pan-STARRS1 and IPHAS photometry, we selected the sample of objects common to the CFHT/MegaCam catalog and, in turn, the Pan-STARRS1 and IPHAS catalogs; we then built ($m_\mathrm{catalog}-m$) vs. (${m'-m''}$) diagrams for the samples of common objects, where $m_\mathrm{catalog}$ is the photometry to be recalibrated, $m$ is the corresponding SDSS filter, and {${m'-m''}$ is the SDSS color of the objects used as reference for the calibration in the $m$ filter}. Calibration equations, accounting for both magnitude offsets and color effects, were derived from these diagrams via error-weighted linear least-squares fitting \citep{bevington}. Pearson's and Kendall's correlation test routines, implemented in R with the $cor.test()$ function, were applied to the data to ascertain that a significant correlation is detected. At the end of the procedure, we derived the following calibration equations for Pan-STARRS1 and IPHAS photometry:
\begin{equation} \label{eqn:panstarrs}
\begin{cases}
{g_{\scriptscriptstyle P1} - g = -0.0027 - 0.1821 \cdot (g-r)} \\
r_{\scriptscriptstyle P1} - r = 0.01155 - 0.0328 \cdot (r-i) \\
i_{\scriptscriptstyle P1} - i = 0.02776 + 0.00419 \cdot (r-i) 
\end{cases}
\end{equation}
\begin{equation} \label{eqn:iphas}
\begin{cases}
r_{\scriptscriptstyle IPHAS} - r = -0.10320 - 0.0253 \cdot (r-i) \\
i_{\scriptscriptstyle IPHAS} - i = -0.33614 - 0.04607 \cdot (r-i) 
\end{cases}
\end{equation}

No correction was applied to the $J$-band photometry, because the 2MASS and UKIDSS photometry sets for common targets in our field appear well correlated with negligible offset for $J$\,$>$\,11. The $J_\mathrm{2MASS}$ vs. $J_\mathrm{UKIDSS}$ sequence deviates from the equality line at the bright magnitude end. This effect could be due to saturation effects; however, it does not impact our analysis, since the focus of this work is on the fainter part of the stellar population.

\subsection{Properties of the final catalog} \label{sec:limiting_mag}

\begin{figure}
\resizebox{\hsize}{!}{\includegraphics{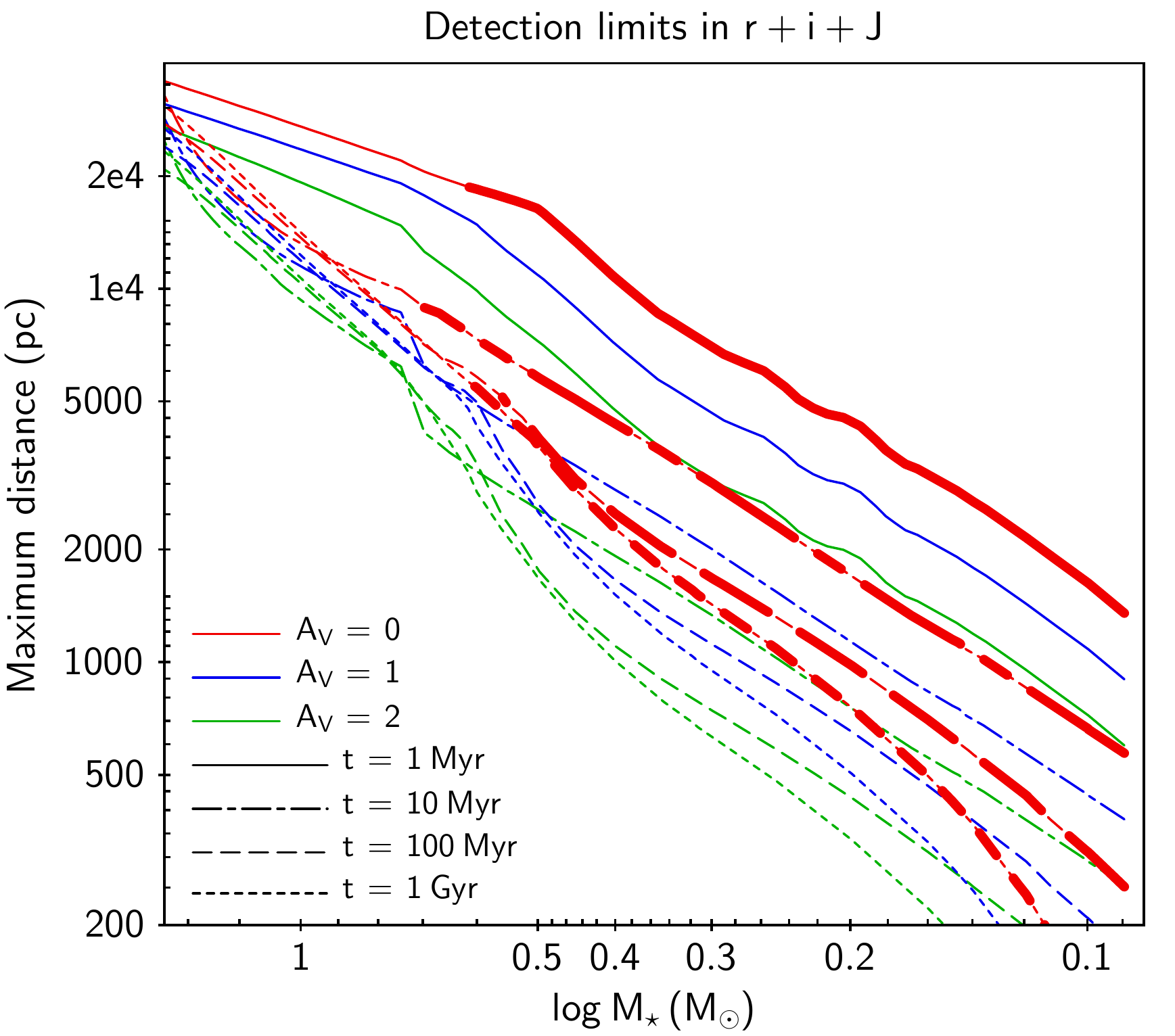}}
\caption{Maximum distance at which simultaneous $r,i,J$ detection in our catalog is expected for stars aged 1~Myr (solid line), 10~Myr (dash-dotted line), 100~Myr (dashed line), or 1~Gyr (dotted line), as a function of their A$_V$ (red = 0, blue = 1, green = 2) and mass. The curves were built by taking the absolute $r,i,J$ magnitudes predicted for stars of different mass and age in {\citeauthor{marigo2017}'s (\citeyear{marigo2017}) isochrones}, and using them as calibration to convert the limiting magnitudes of our catalog (see {Table~\ref{tab:phot_range_completeness}}) into limiting distances as a function of stellar mass, age, and A$_V$. {A thicker stroke marks the mass range of M-type stars (${T_{\rm eff} < 4000}$\,K) predicted by the models at the various ages, shown for illustration purposes on the A$_V$\,=\,0 curves.}}
\label{fig:riJ_detection_depth}
\end{figure}

{The photometric ranges covered by the bulk of objects in our final $r,i,J$ (185205 sources) and $g,r,i$ (123370 sources) catalogs}, assembled as described in the previous sections, {are reported in Table~\ref{tab:phot_range_completeness}. The boundary values of these ranges were defined as the $0.5^{th}$ and the $99.5^{th}$ percentile levels in magnitude in each filter for each catalog; the associated completeness limits were estimated by plotting the logarithmic density number of objects as a function of magnitude, and locating the point where the trend starts to deviate from a straight line at the faint magnitude end. {These values indicate the typical depth achieved with the assembled catalog across the field, although the estimated completeness limits can exhibit local variations of order 0.5~mag, as a result of differential crowding and nebulosity.}}

In Fig.\,\ref{fig:riJ_detection_depth}, we illustrate the limiting distances within which we expect to be able to detect stars of different mass, age and reddening in our field, considering the limiting magnitudes of our ${r,i,J}$ catalog in the three filters simultaneously. As discussed in Sect.\,\ref{sec:introduction} and shown in the diagram, very young stars are expected to be predominant among sources detected at large distances. This is especially true for M-type stars {(${T_{\rm eff} < 4000}$\,K, corresponding to $M_\star$\,{\small $\lesssim$}\,0.6--0.7~$M_\odot$ at an age of 1--10 Myr, and to $M_\star$\,{\small $\lesssim$}\,0.5--0.6~$M_\odot$ at ages $\sim$100--1000 Myr)}: as depicted in Fig.\,\ref{fig:riJ_detection_depth}, our limiting magnitudes enable simultaneous $r,i,J$ detection of $\sim$1~Myr-old, late-M stars suffering little to moderate extinction down to distances of 1 to 5~kpc, while early-M stars in the same age and extinction group will be detected down to distances of 5 to 10--15~kpc. Limiting distances 2 to 10 times smaller are instead expected for simultaneous $r,i,J$ detection of M-type stars older than 100~Myr. As already discussed earlier, the PMS/MS contrast for source detection at large distances becomes progressively less favorable for spectral types earlier than M, and the various sets of curves for stars of different ages in Fig.\,\ref{fig:riJ_detection_depth} tend to converge to similar values of limiting distances for $M_\star > 1.4\,\,M_\odot$ (spectral type\,$\sim$\,K3 at an age of a few Myr). {Our $r,i,J$ catalog enables measuring A$_V$ up to $\sim$7~mag for early-M, $\sim$1~Myr-old stars (M$_\star$\,$\sim$\,0.5~$M_\odot$) at a distance of 1~kpc; the limiting A$_V$ is about 4--5~mag for $\sim$M3 stars (M$_\star$\,$\sim$\,0.35~$M_\odot$) at the same age and distance, and reduces to $\sim$1.5~mag for late-M stars (M$_\star$\,$\sim$\,0.1~$M_\odot$). At the same distance, the largest A$_V$ that could be measured for a 1~Gyr-old, early-M star, according to the limiting magnitudes of our optical catalog, amounts to $\sim$4~mag; a $\sim$M3 star could be detected until A$_V$$\sim$1.5~mag, whereas late M-type stars could not be detected.}

\begin{figure}
\resizebox{\hsize}{!}{\includegraphics{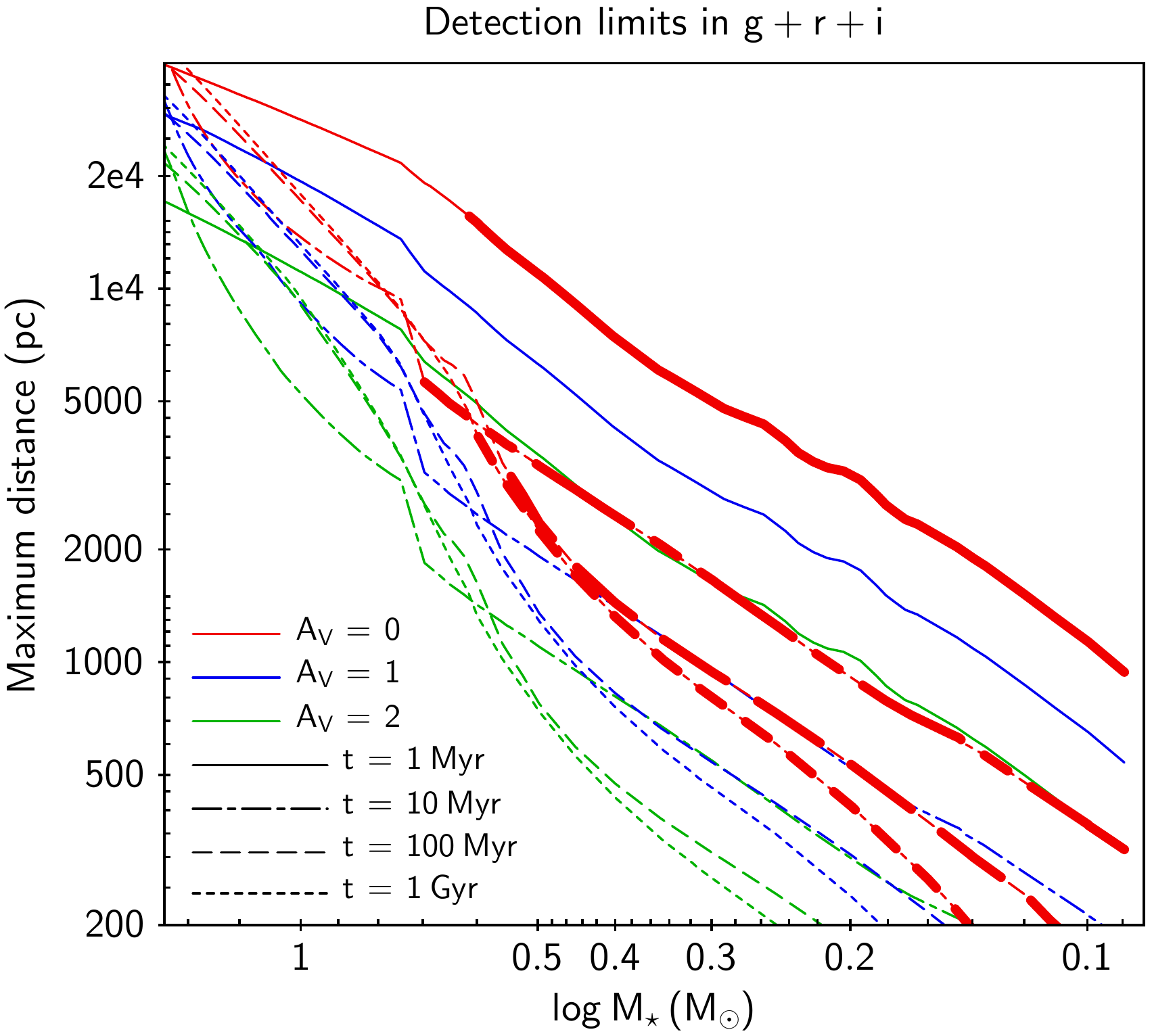}}
\caption{Same diagram as in Fig.\,\ref{fig:riJ_detection_depth}, but for our ${g,r,i}$ catalog (Sect.\,\ref{sec:gri_cat}).}
\label{fig:gri_detection_depth}
\end{figure}

{A similar analysis of expected limiting distances for simultaneous ${g,r,i}$ detection of stars in our catalog, depending on their mass, age and reddening, is illustrated in Fig.\,\ref{fig:gri_detection_depth}. Here, the effects of extinction are more conspicuous, resulting in a more shallow sky coverage compared to what achievable with the $r,i,J$ catalog. At a distance of 1~kpc, our $g,r,i$ limiting magnitudes would enable detecting 1~Myr-old, M0--M1 stars down to A$_V$$\sim$4--5~mag, M3 stars down to A$_V$$\sim$3~mag, and M5 stars down to A$_V$$\sim$1~mag. Assuming the same distance of 1~kpc, 1~Gyr-old, early M-type stars could be detected down to A$_V$$\sim$2~mag, and no M-type stars later than $\sim$M3 could be detected.}

\section{Results} \label{sec:results}

\subsection{Selection of M-type stars and derivation of A$_V$}

\subsubsection{The $r-i$ vs. $i-J$ diagram} \label{sec:iJ_ri}

\begin{figure}
\resizebox{\hsize}{!}{\includegraphics{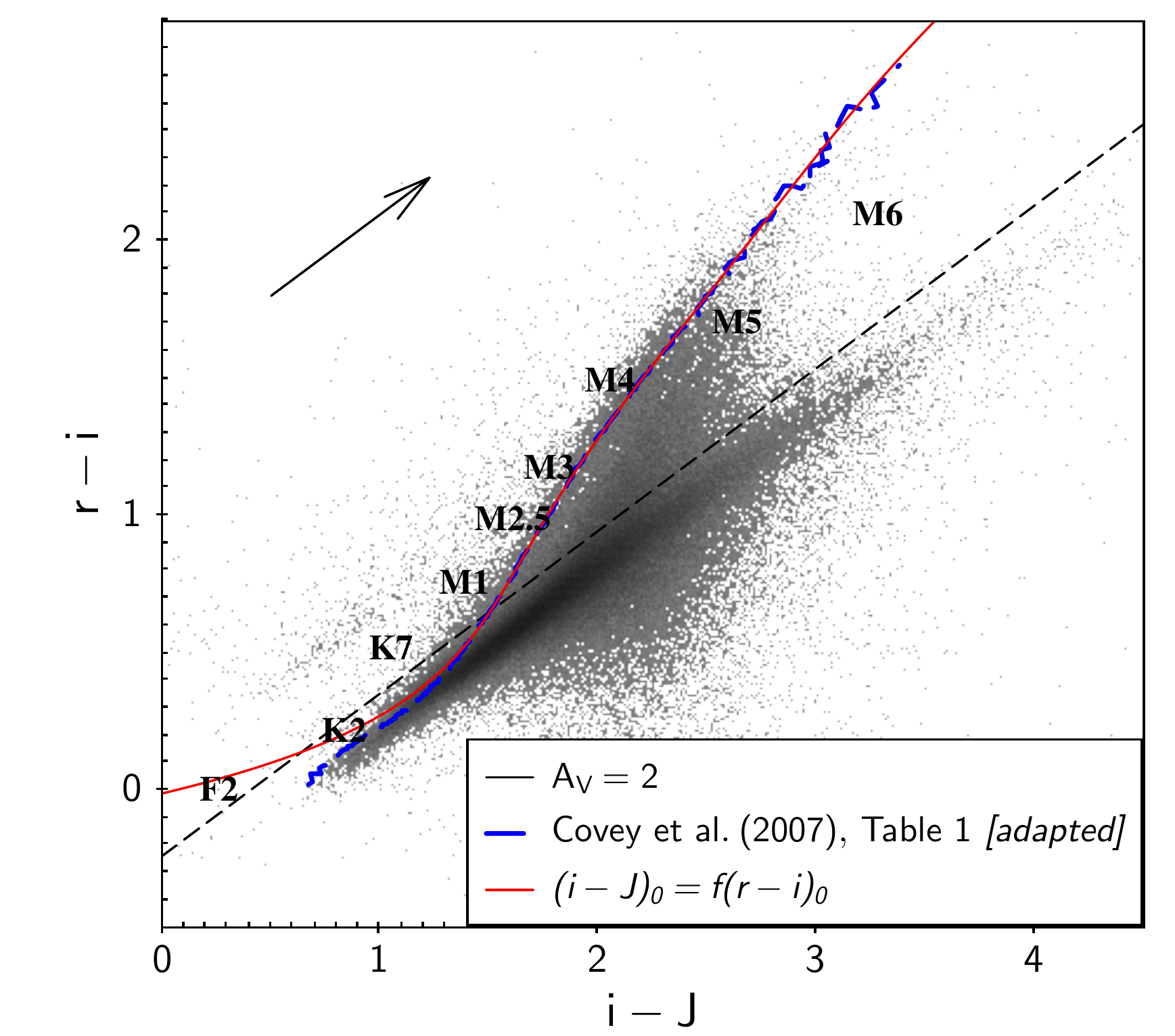}}
\caption{$r-i$ vs. $i-J$ color-color diagram for objects in the catalog built as in Sect.\,\ref{sec:data}. The reddening vector is traced using the $A_\lambda/A_V$ reported in Sect.\,\ref{sec:iJ_ri}. The black dashed line corresponds to the empirical threshold traced by hand, parallel to the reddening vector, to separate the color locus of earlier-type stars (nearly parallel to the reddening vector) from that of M-type stars. The color sequence traced in blue, adapted from \citet{covey07}, is shown as a guidance to locate stars of different spectral types on the diagram. The red line, traced as a polynomial fit to the blue curve on the diagram, corresponds to the reference sequence adopted in this study to measure A$_V$ for M-type stars (see text).}
\label{fig:iJ_ri}
\end{figure}

Figure~\ref{fig:iJ_ri} depicts the color properties of stars in our catalog on the $r-i$ vs. $i-J$ diagram.
To trace the effects of extinction on the colors in Fig.\,\ref{fig:iJ_ri}, we made use of the SVO Filter Profile Service\footnote{http://svo2.cab.inta-csic.es/svo/theory/fps/} \citep{rodrigo2012}, where the A$_\lambda$/A$_V$ coefficients for an extensive collection of photometric systems, surveys and instrument filters are reported. {These are calculated using the extinction law by \citet{fitzpatrick1999}, modified following \citet{indebetouw2005} in the infrared.} The A$_\lambda$/A$_V$ coefficients relevant to this study are reported in Table~\ref{tab:Av_coefficients}.
\begin{table}
\caption{Extinction coefficients $A_\lambda/A_V$ for the filters and surveys listed in Sect.\,\ref{sec:data}.}
\label{tab:Av_coefficients}
\centering
\begin{tabular}{l | c | c | c | c}
\hline\hline
 & ${A_g/A_V}$ & $A_r/A_V$ & $A_i/A_V$ & $A_J/A_V$ \\
\hline
CFHT/MegaCam & 1.20 & 0.87 & 0.67 & \\
Pan-STARRS & 1.19 & 0.89 & 0.67 & \\
IPHAS (INT) & & 0.88 & 0.65 & \\
SDSS & 1.23 & 0.89 & 0.68 & \\
2MASS & & & & 0.31\\
UKIRT & & & & 0.30\\
\hline
\end{tabular}
\end{table}
The final A$_\lambda$/A$_V$ used for the analysis for a given filter were computed as the mean of the corresponding values listed in Table~\ref{tab:Av_coefficients}; these average coefficients, reported in Eq.\,\ref{eqn:A_lambda}, describe well the trend traced by the bulk of objects below the black dashed line in Fig.\,\ref{fig:iJ_ri}.
\begin{equation} \label{eqn:A_lambda}
\begin{aligned}
{A_g/A_V = 1.207} \\
A_r/A_V = 0.883 \\
A_i/A_V = 0.668 \\
A_J/A_V = 0.305 
\end{aligned}
\enskip \Rightarrow \enskip
\begin{aligned}
{A_{g-r}/A_V = 0.324} \\
A_{r-i}/A_V = 0.215 \\
A_{i-J}/A_V = 0.363
\end{aligned}
\end{equation}

Two main stellar loci can be distinguished on the color diagram in Fig.\,\ref{fig:iJ_ri}. The first, populated by stars of spectral type earlier than K7, develops almost parallel to the reddening vector shown in the upper left part of the diagram. Little-extincted, K-type stars populate the lower-left part of the color distribution, starting at ($i-J$, $r-i$)$\sim$(0, 0.8); more extincted sources in the same spectral range extend the intrinsic K-type color locus, along the reddening direction, beyond the knee that marks the transition between K-type stars and M-type stars on the empirical color sequence dashed in blue. The second stellar color locus, populated by M-type stars, develops above the color locus of earlier-type stars, and is tilted with respect to the reddening vector. To separate the two loci, we traced a boundary line running parallel to the reddening vector above the bulk of K-type and earlier stars, as shown in Fig.\,\ref{fig:iJ_ri}; we then selected as M-type stars all objects located above this threshold line.

\subsubsection{Measuring A$_V$ for M-type stars in the field} \label{sec:Av_meas}

The empirical expectation from the color distribution of M-type stars in Fig.\,\ref{fig:iJ_ri} is that non-reddened (foreground) stars should be located along the left-hand envelope of the locus, of which the shape can be well reproduced with the SDSS stellar color sequence tabulated in \citet{covey07}. The positions of stars rightward of this envelope can be described as the combination of their intrinsic colors on the zero-reddening sequence plus a displacement along the reddening direction, the extent of which provides a measurement of their A$_V$.

If the colors of non-reddened stars were known without uncertainties, we would expect them to define a narrow sequence which would fall exactly on the left-hand edge of the observed color locus. In practice, the photometric uncertainties associated with the measured stellar colors, if unrelated and randomly distributed, will determine a symmetric widening of the empirical zero-reddening color sequence, both toward bluer and redder colors, by an amount corresponding to the typical uncertainty on $i-J$ for M-type stars on the diagram ($\sigma_{i-J}$$\sim$0.04~mag). As a consequence, if ($i-J$)$_0$ are the intrinsic colors of the sources in the absence of extinction, the leftmost edge of the color distribution will statistically correspond to the $(i-J)_0 - \sigma_{i-J}$ color sequence.

To take the above consideration into account, we first derived an analytical description for the left-hand envelope of the M-stars color locus on Fig.\,\ref{fig:iJ_ri} by fitting a polynomial curve to \citeauthor{covey07}'s (\citeyear{covey07}) color sequence, positioned onto the left~edge of the main datapoint distribution above the threshold line on the diagram. We then shifted the curve by 0.04~mag toward larger \mbox{$i-J$}, to statistically remove the effect of photometric uncertainties on the location of the zero-reddening color sequence. The final analytic curve to describe the sequence of intrinsic colors of M-type stars, that we adopted as reference to measure their A$_V$, is shown in red on the diagram in Fig.\,\ref{fig:iJ_ri}.

The final sample of objects for which we could derive an estimate of A$_V$ comprises 24089 sources, located above the black dashed line and to the right of the red curve in Fig.\,\ref{fig:iJ_ri}. For these stars, A$_V$ was computed by solving numerically the following set of equations:
\begin{equation} \label{eqn:riJ_Av_eq}
\left\{
\begin{aligned}
(r-i)_0 &= (r-i)_{obs} - 0.215 \cdot A_V \\
(i-J)_0 &= (i-J)_{obs} - 0.363 \cdot A_V \\
(i-J)_0 &= f(r-i)_0
\end{aligned}
\right.
\end{equation}
where ($r-i$)$_{obs}$ and ($i-J$)$_{obs}$ are the observed colors for a given object, ($r-i$)$_0$ and ($i-J$)$_0$ are its intrinsic colors, $f$ is the function that relates ($i-J$)$_0$ to ($r-i$)$_0$ for M-type stars, and the factors before A$_V$ are the coefficients that relate A$_V$ to A$_{r-i}$ or A$_{i-J}$ following Eq.\,\ref{eqn:A_lambda}.

{A similar procedure was adopted to select the sample of M-type stars and derive their A$_V$ from the $r-i$ vs. $g-r$ diagram. A total of 4277 sources were extracted for the analysis of A$_V$ from $g,r,i$ photometry. A$_V$ estimates were then derived by solving a similar set of equations as as in (\ref{eqn:riJ_Av_eq}), but with $g-r$ replacing $i-J$, and the $A_\lambda/A_V$ coefficients changed accordingly. }

\subsubsection{A reddening map of the NGC~2264 region}

\begin{figure}
\resizebox{\hsize}{!}{\includegraphics{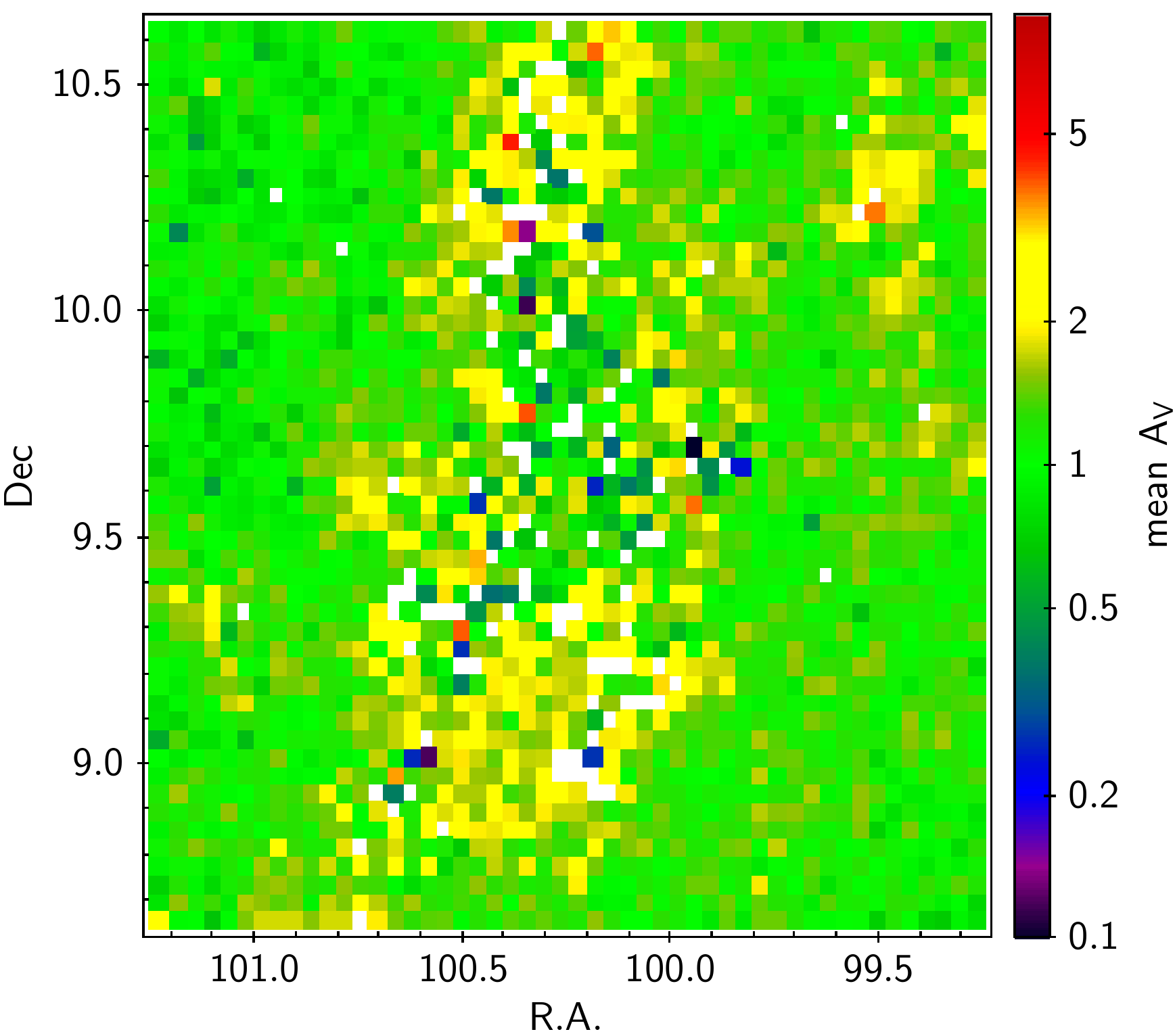}}
\caption{Map of the average A$_V$ measured for M-type stars across the region. Each of the 2500 {2.4$'$$\times$2.4$'$} boxes that constitute the map is colored according to the mean A$_V$ computed across the population of M-type stars projected onto that box. The reference color--A$_V$ scale used is shown in the side axis. Colors are assigned only to boxes containing at least two M-stars in our sample, and with an internal spread in A$_V$ which is no more than 3\,$\sigma$ larger than the typical dispersion in A$_V$ measured among boxes containing the same number of sources within the map.}
\label{fig:mean_Av_map}
\end{figure}

The A$_V$ distribution derived for our sample of M-type stars {from Fig.\,\ref{fig:iJ_ri}}, selected as described in Sects.~\ref{sec:iJ_ri} and \ref{sec:Av_meas}, is centered on A$_V$\,$\sim$\,1.2~mag; about 25\% of sources have A$_V$\,$<$\,0.7~mag, and about 25\% have A$_V$\,$>$\,1.7~mag. Fig.\,\ref{fig:mean_Av_map} shows an average A$_V$ map of our field.
While the computation of the average A$_V$ can be locally affected by differential population density across the field or by classification outliers, some interesting global indications can be deduced from the map. The A$_V$ distribution is not uniform across the region: typical extinction values tend to be lower in the outer parts of the field, whereas typically higher values of A$_V$ are registered toward the center of the field. Also, many of the points with the lowest concentrations of sources or with the strongest local variations in A$_V$, marked in white on the map in Fig.\,\ref{fig:mean_Av_map}, are projected onto the central regions of the field. 

\begin{figure}
\resizebox{\hsize}{!}{\includegraphics{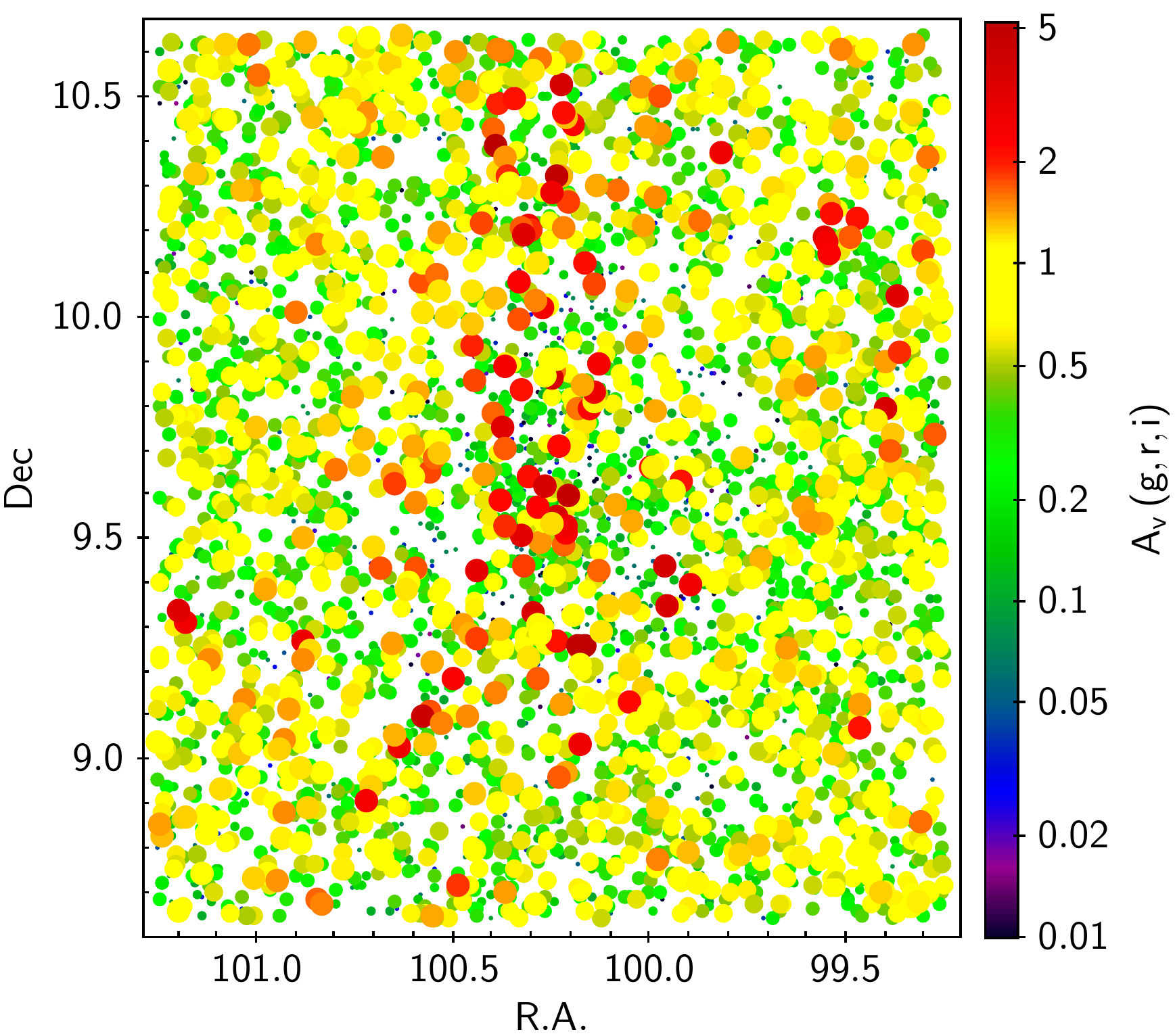}}
\caption{Spatial map of M-type stars selected from the (${r-i}$, ${g-r}$) diagram as a function of their measured A$_V$. The color of the symbols varies with A$_V$ as shown on the side axis of the diagram. The size of the symbols is scaled with A$_V$ in five groups: ${A_V < 0.1}$ (smallest dots); ${0.1 \leq A_V < 0.2}$; ${0.2 \leq A_V < 0.5}$; ${0.5 \leq A_V < 1}$; ${A_V \geq 1}$ (largest dots).}
\label{fig:Av_gri_map}
\end{figure}

{Since the number density of M-type stars selected from the (${r-i}$, ${g-r}$) diagram is about six times smaller than that of our $r,i,J$ sample, we did not build an averaged extinction map from the $g,r,i$ A$_V$ estimates. Instead, we examined the spatial distribution of the M-stars from the $g,r,i$ sample as a function of their A$_V$, as illustrated in Fig.\,\ref{fig:Av_gri_map}. A comparison between Figs.\,\ref{fig:mean_Av_map} and \ref{fig:Av_gri_map} shows that the two diagnostics yield a very similar picture of reddening across the field. While the two sets of A$_V$ measurements exhibit {a tendency for estimates from $r,i,J$ colors to be larger than those from $g,r,i$ colors,} and a scatter of $\sim$0.5~mag among individual common sources (similar to what found in earlier studies which compared different A$_V$ diagnostics; e.g., \citealp{cauley2012}, \citealp{venuti2018}), the relative A$_V$ scale indicates in both cases higher extinction toward the central part of the field, and lower extinction in the outer regions. A very close match in morphology can be observed between the higher-extinction area (colored in yellow) in the averaged A$_V$ map in Fig.\,\ref{fig:mean_Av_map}, and the distribution of point sources with the largest A$_V$ values (colored in red) in Fig.\,\ref{fig:Av_gri_map}.}

\subsection{The A$_V$ structure of the NGC~2264 field} \label{sec:Av_structure}

\begin{figure}
\resizebox{\hsize}{!}{\includegraphics{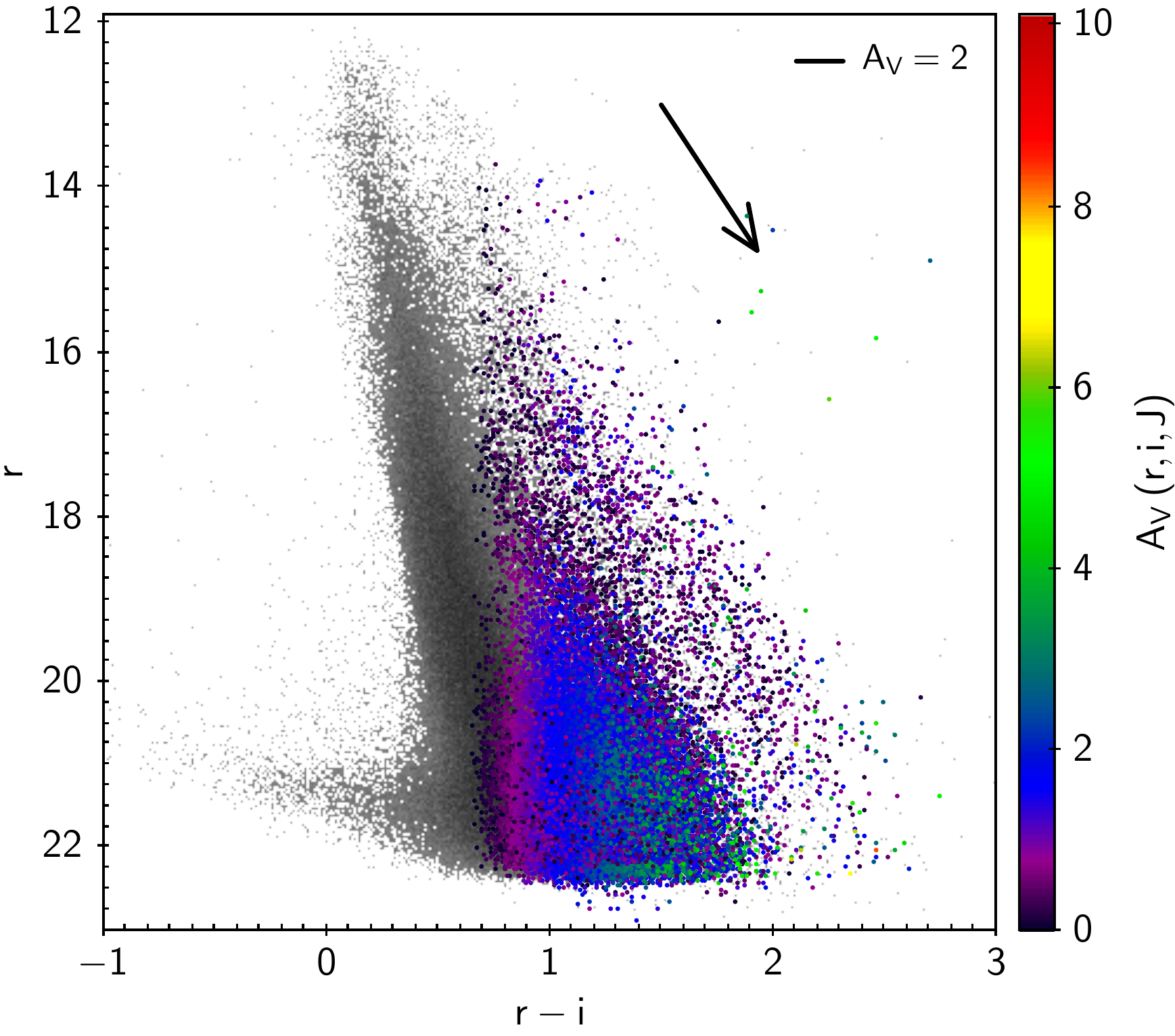}}
\caption{($r-i$, $r$) color-magnitude diagram populated by the 185205 sources in our catalog (see Sect.\,\ref{sec:data}). M-type stars in the field for which we could derive an A$_V$ estimate (see Sect.\,\ref{sec:Av_meas}) are highlighted as colored points; colors are scaled according to the A$_V$ measurements, as shown to the right of the diagram.}
\label{fig:ri_r_CMD_Av}
\end{figure}

Fig.\,\ref{fig:ri_r_CMD_Av} illustrates the distribution of sources in our sample on the $r$ vs. $r-i$ diagram. The main field locus is clearly delineated as a densely populated region between $r-i$\,$\sim$\,0.1 and 2. However, a separate sequence of objects, spanning most of the magnitude range and running along the redder side of the main field locus, can be also observed on the diagram. Another feature of the datapoint distribution in Fig.\,\ref{fig:ri_r_CMD_Av} is the tail of objects that develops on the bluer side of the main field locus at the faint magnitude end. This is likely a residue of the catalog cleaning procedure on the $r$-band filter; at any rate, it does not impact our analysis, since this small group of outliers is well distinct from the locus of M-type stars, highlighted on the diagram, which are the focus of our study.

\begin{figure*}
    \centering
\begin{subfigure}{0.31\textwidth}
  \includegraphics[width=\linewidth]{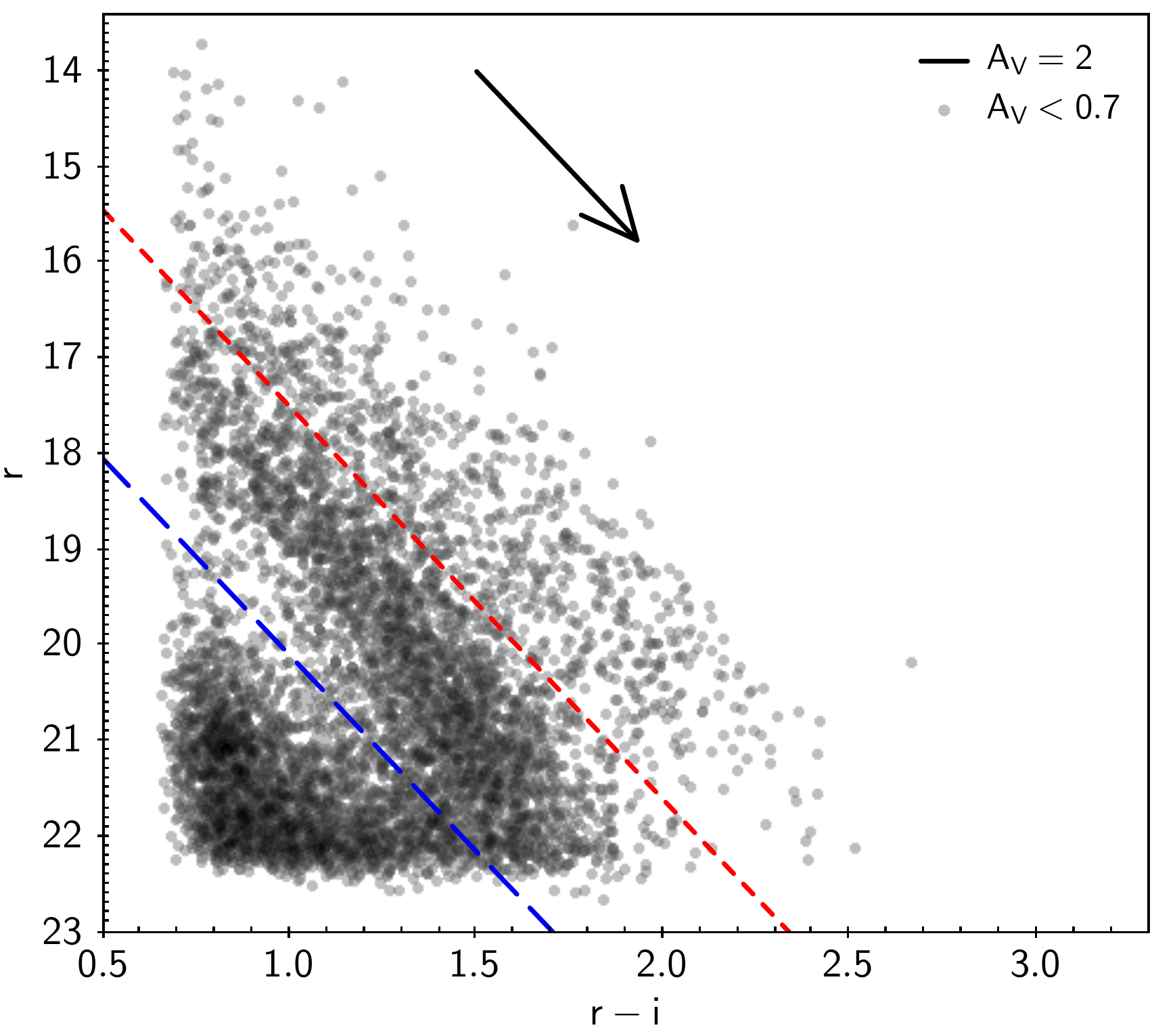}
  \caption{$r$ vs. $r-i$ for M-stars with $A_V < 0.7$.}
  \label{fig:a}
\end{subfigure}\hfil
\begin{subfigure}{0.31\textwidth}
  \includegraphics[width=\linewidth]{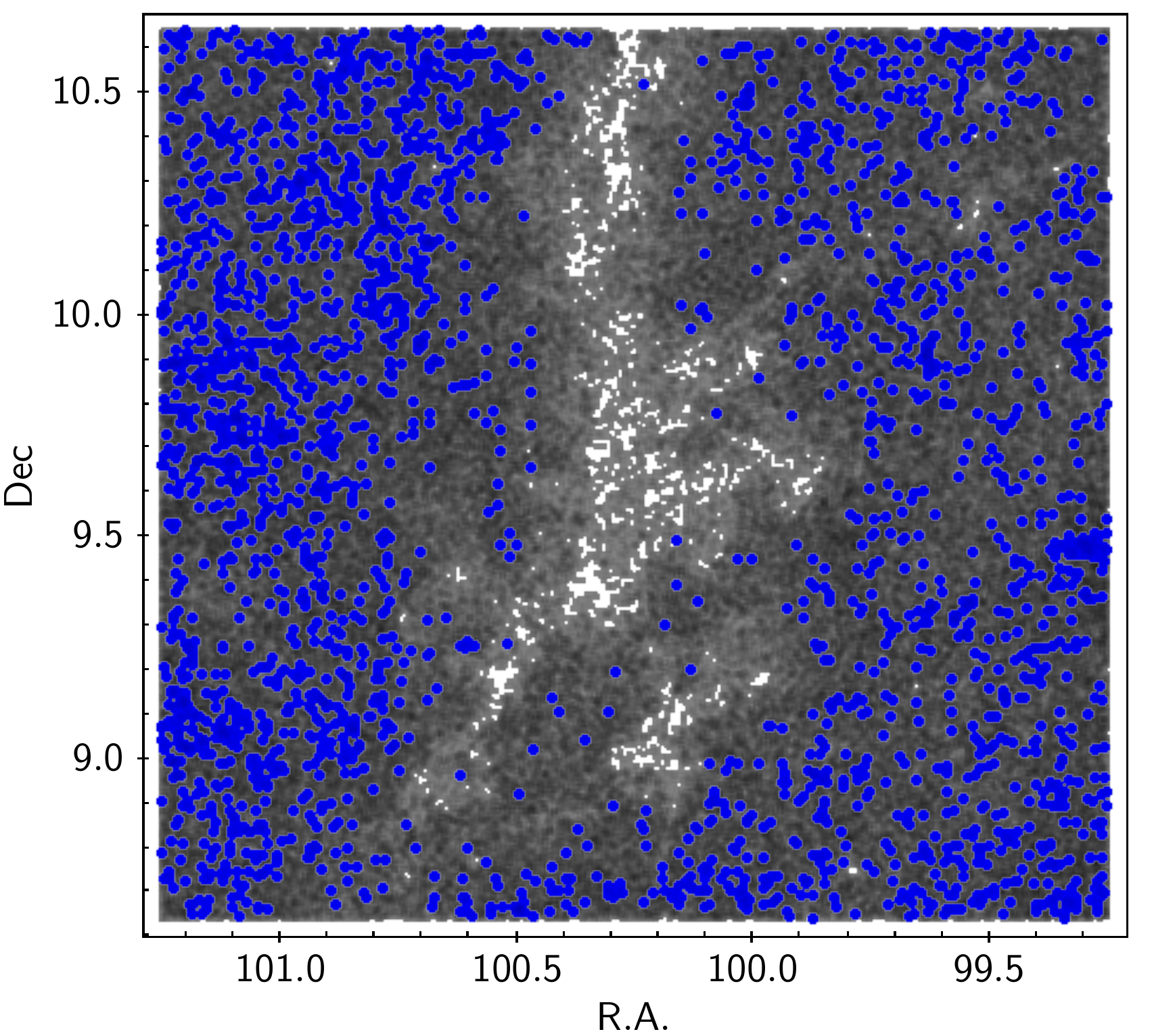}
  \caption{Map of objects below the blue line in (a).}
  \label{fig:b}
\end{subfigure}\hfil
\begin{subfigure}{0.31\textwidth}
  \includegraphics[width=\linewidth]{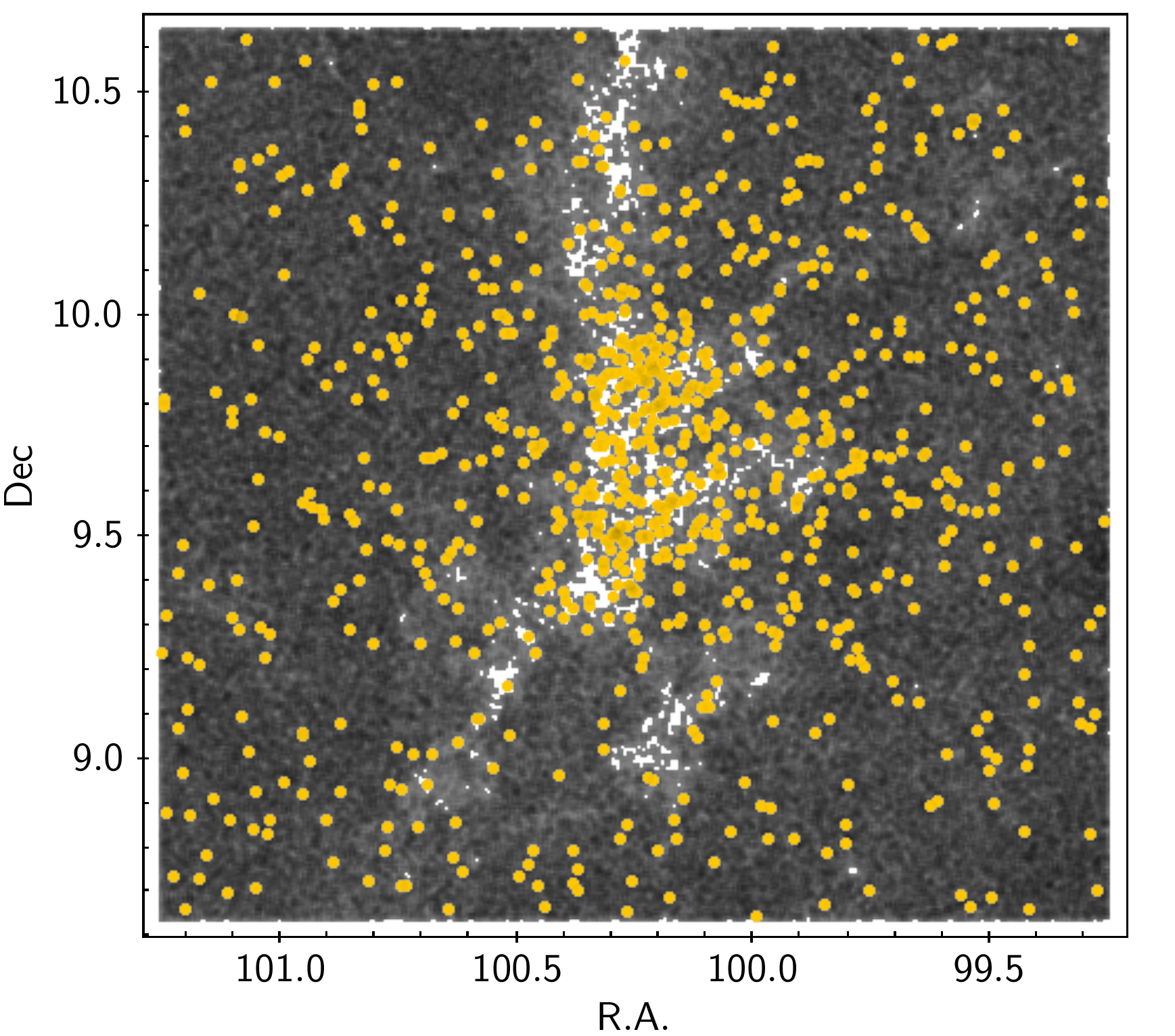}
  \caption{Map of objects above the red line in (a).}
  \label{fig:c}
\end{subfigure}

\medskip
\begin{subfigure}{0.31\textwidth}
  \includegraphics[width=\linewidth]{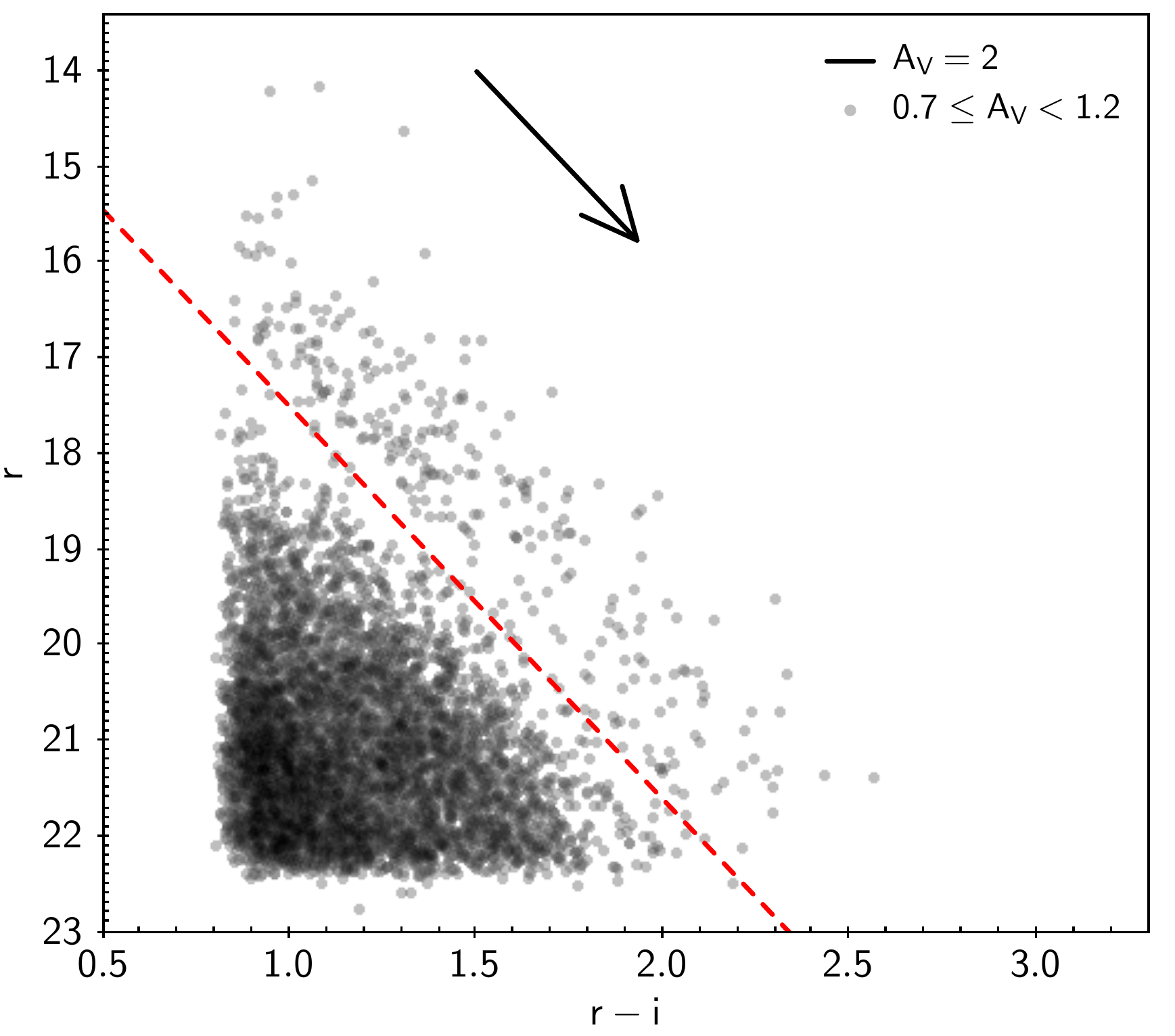}
  \caption{$r$ vs. $r-i$ for M-stars with $A_V = 0.7-1.2$.}
  \label{fig:d}
\end{subfigure}\hfil
\begin{subfigure}{0.31\textwidth}
  \includegraphics[width=\linewidth]{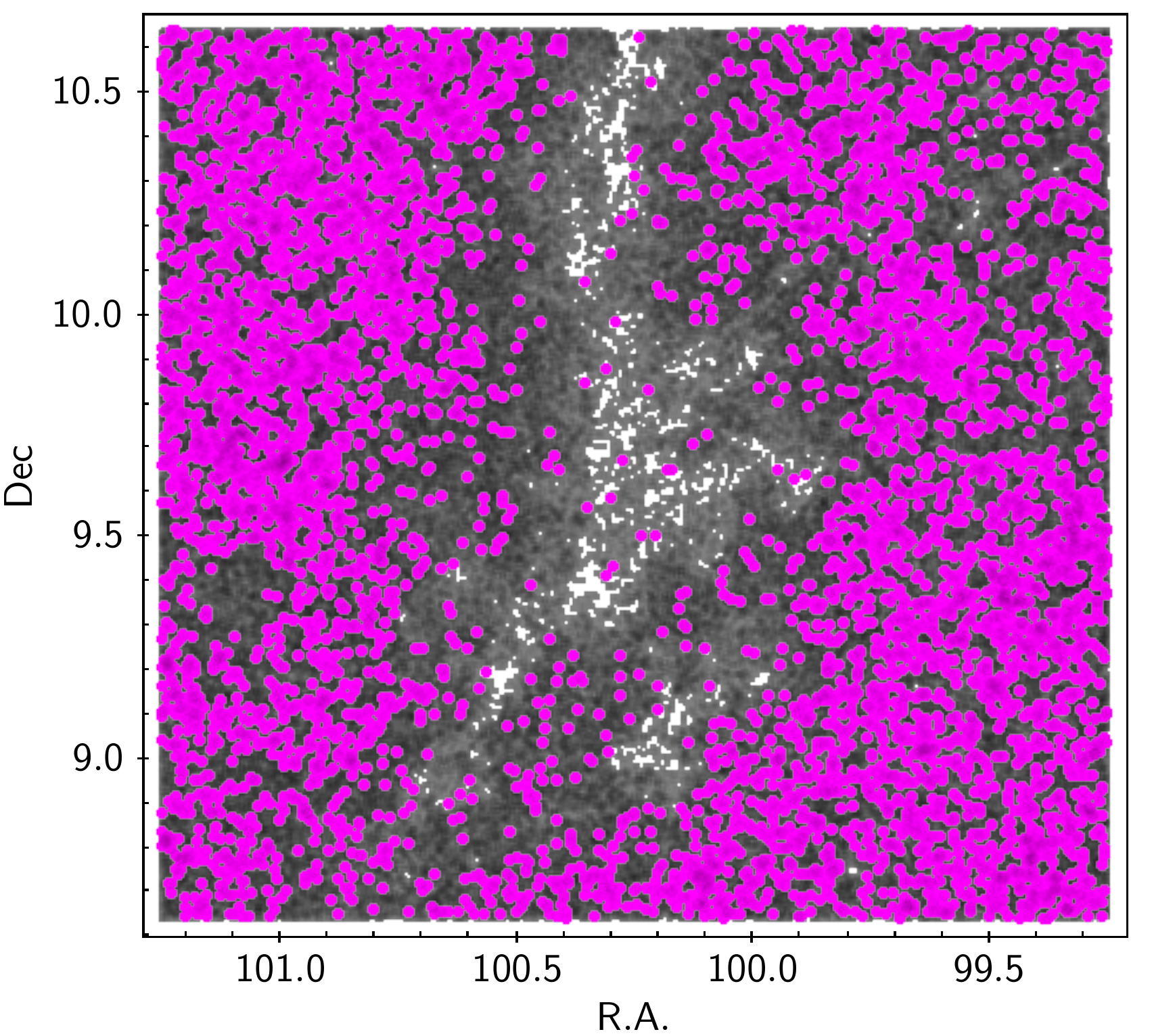}
  \caption{Map of objects below the red line in (d).}
  \label{fig:e}
\end{subfigure}\hfil
\begin{subfigure}{0.31\textwidth}
  \includegraphics[width=\linewidth]{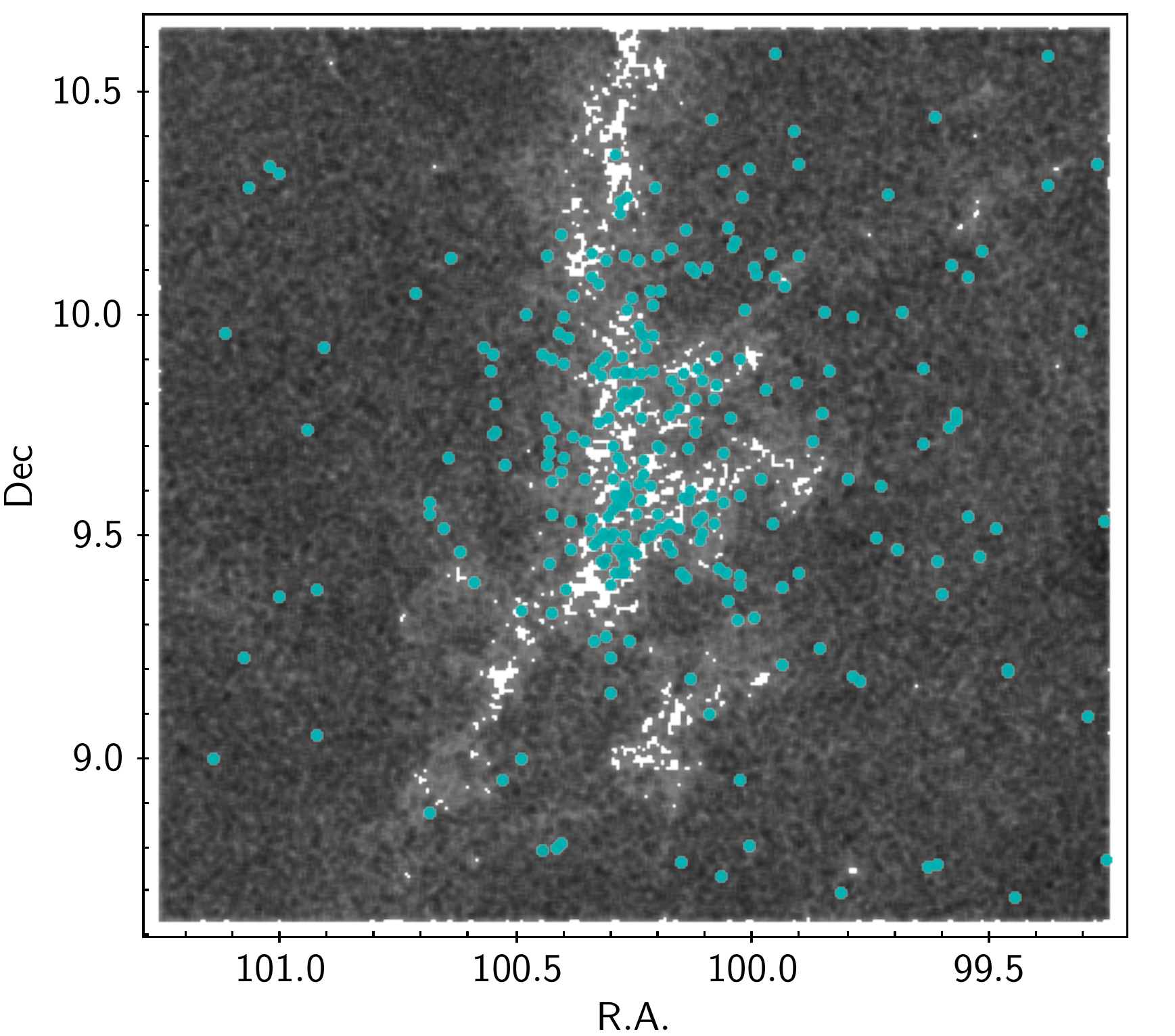}
  \caption{Map of objects above the red line in (d).}
  \label{fig:f}
\end{subfigure}

\medskip
\begin{subfigure}{0.31\textwidth}
  \includegraphics[width=\linewidth]{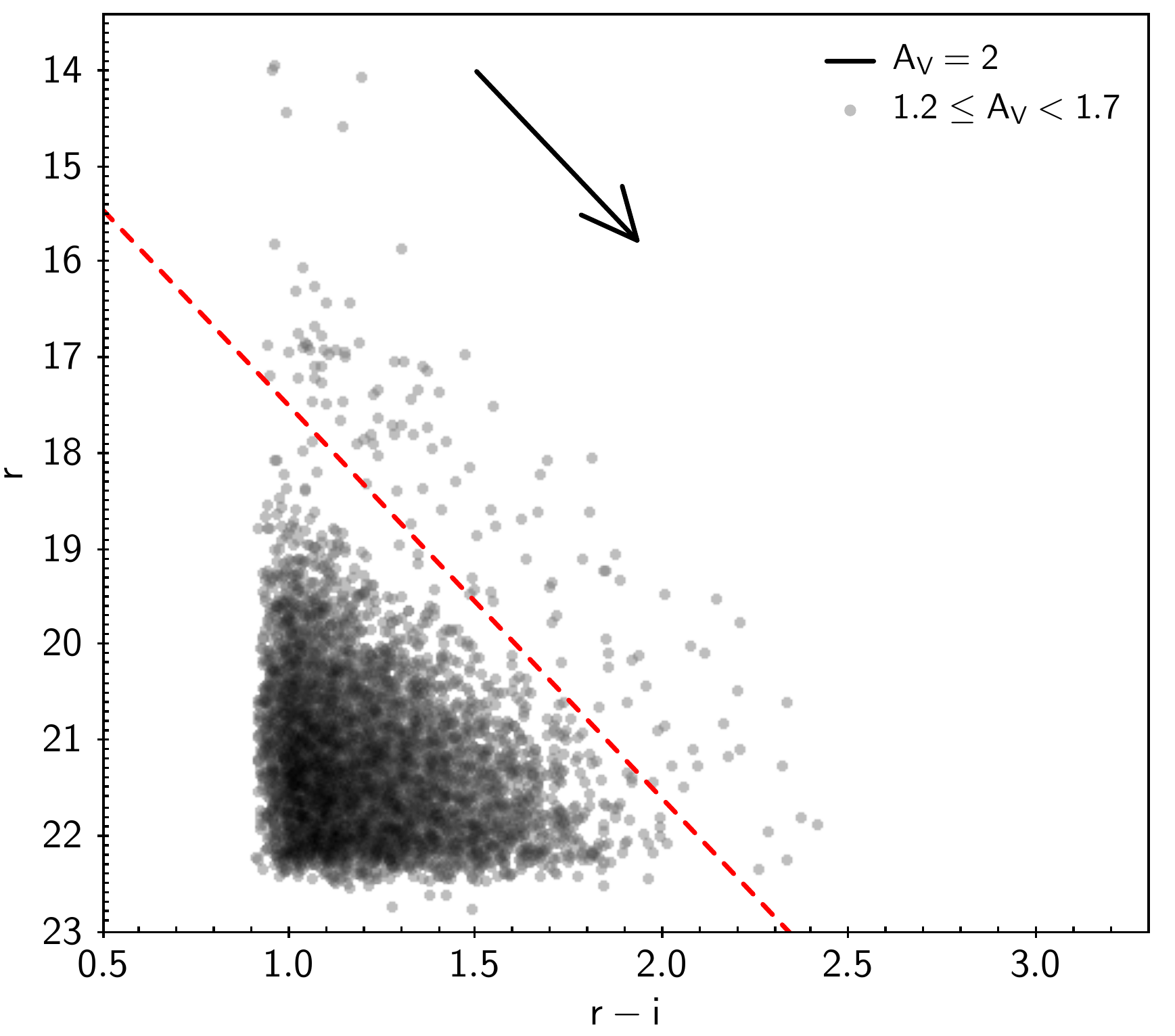}
  \caption{$r$ vs. $r-i$ for M-stars with $A_V = 1.2-1.7$.}
  \label{fig:g}
\end{subfigure}\hfil
\begin{subfigure}{0.31\textwidth}
  \includegraphics[width=\linewidth]{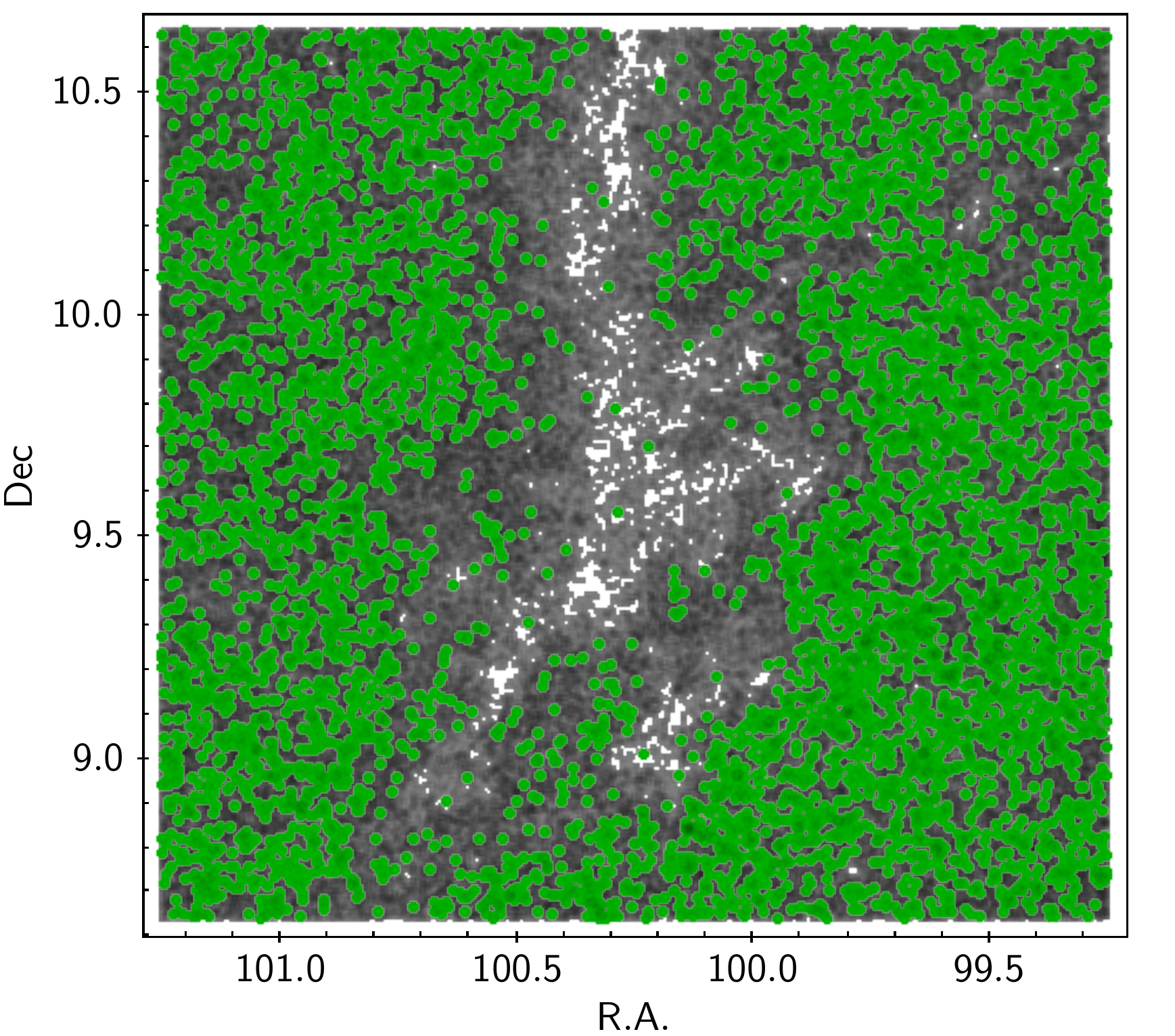}
  \caption{Map of objects below the red line in (g).}
  \label{fig:h}
\end{subfigure}\hfil
\begin{subfigure}{0.31\textwidth}
  \includegraphics[width=\linewidth]{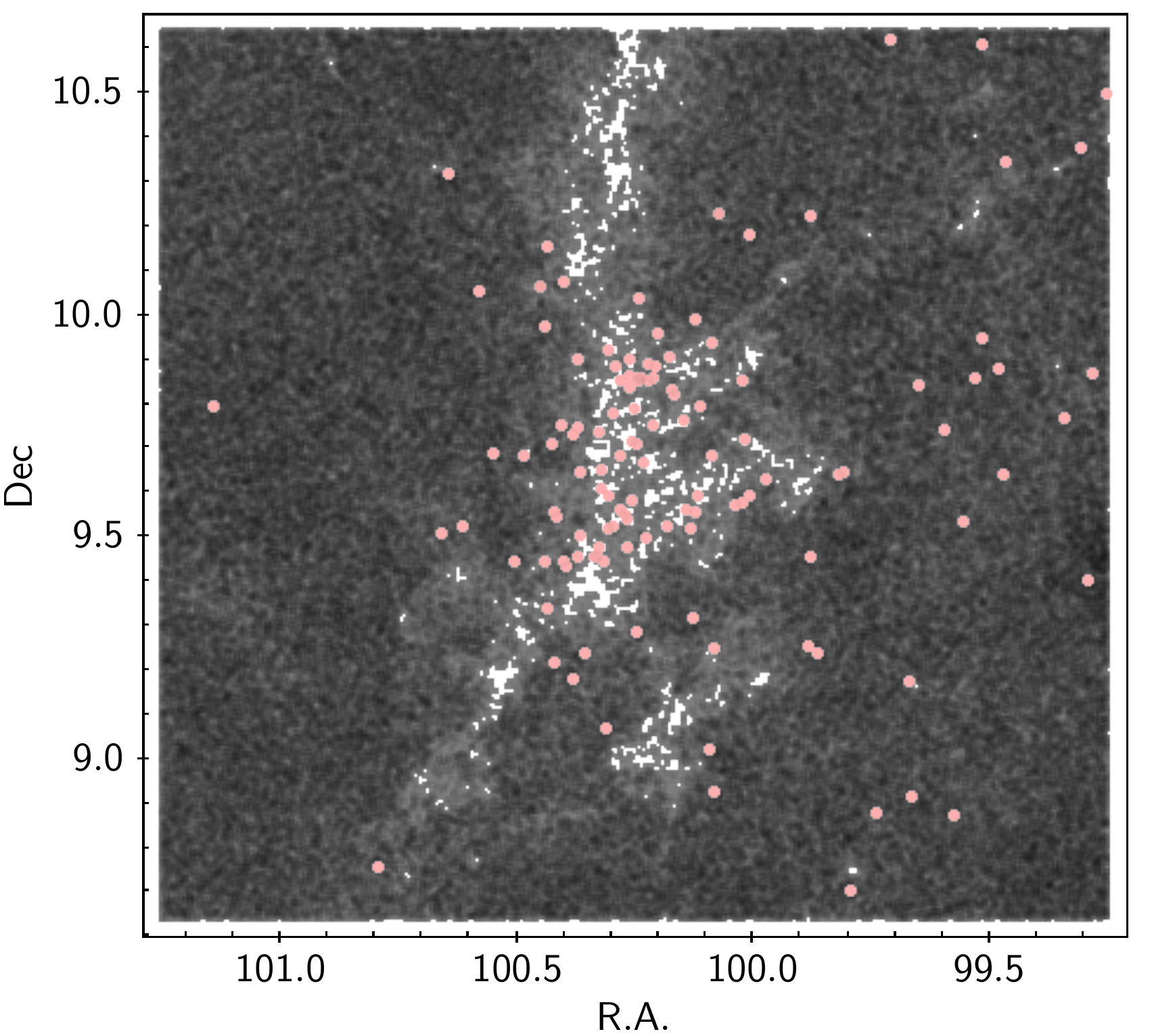}
  \caption{Map of objects above the red line in (g).}
  \label{fig:i}
\end{subfigure}

\medskip
\begin{subfigure}{0.31\textwidth}
  \includegraphics[width=\linewidth]{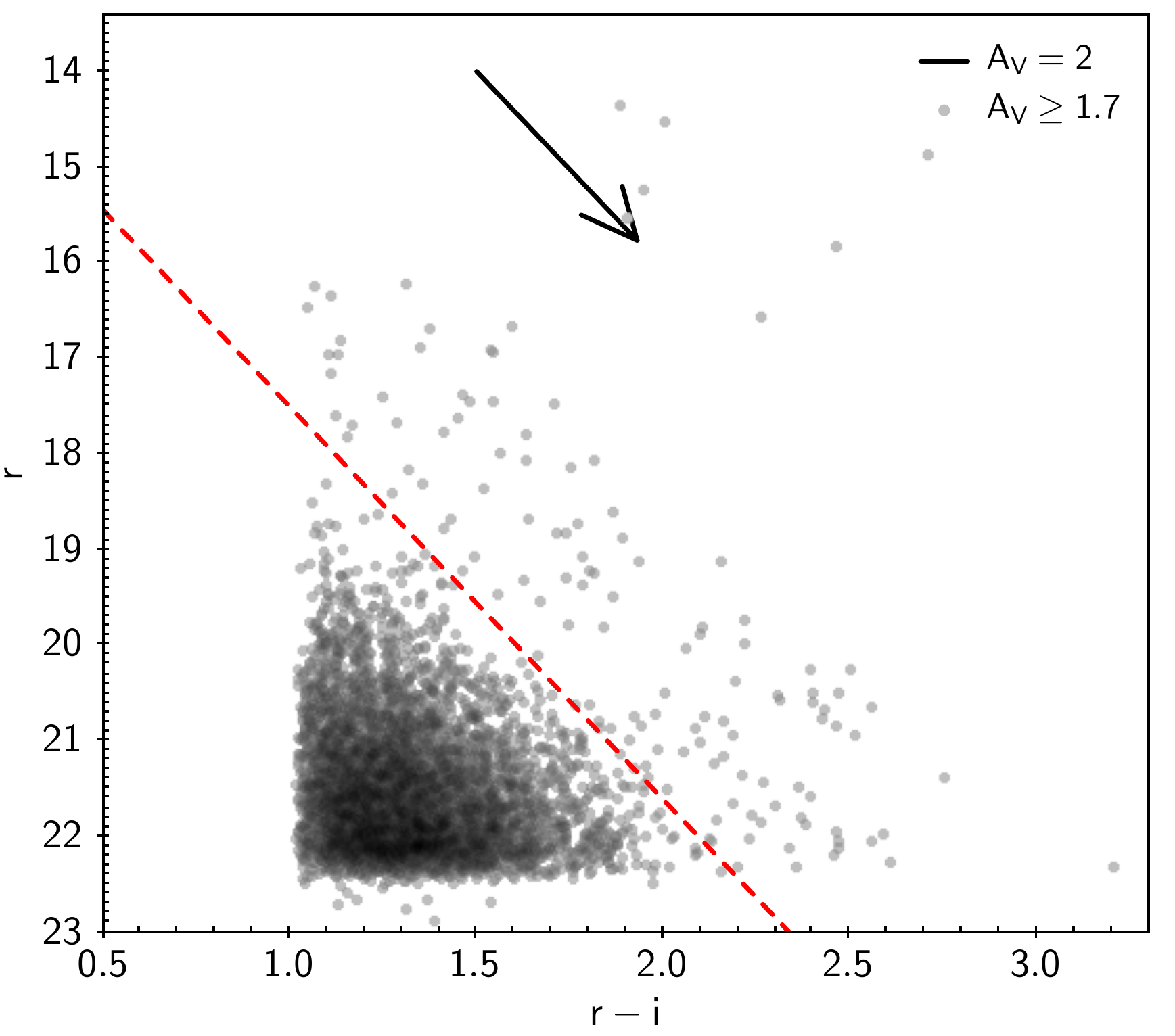}
  \caption{$r$ vs. $r-i$ for M-stars with $A_V \geq 1.7$.}
  \label{fig:j}
\end{subfigure}\hfil
\begin{subfigure}{0.31\textwidth}
  \includegraphics[width=\linewidth]{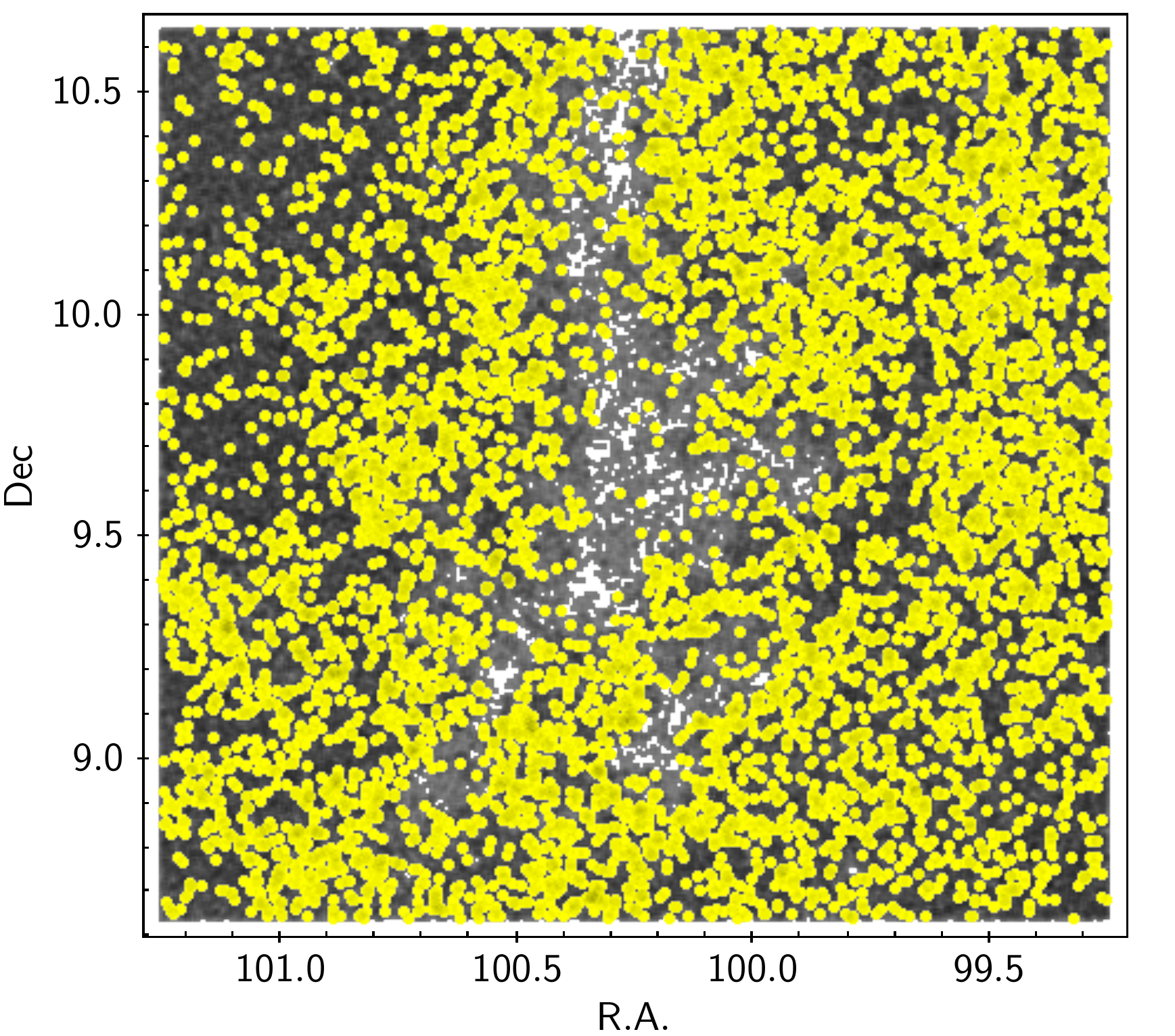}
  \caption{Map of objects below the red line in (j).}
  \label{fig:k}
\end{subfigure}\hfil
\begin{subfigure}{0.31\textwidth}
  \includegraphics[width=\linewidth]{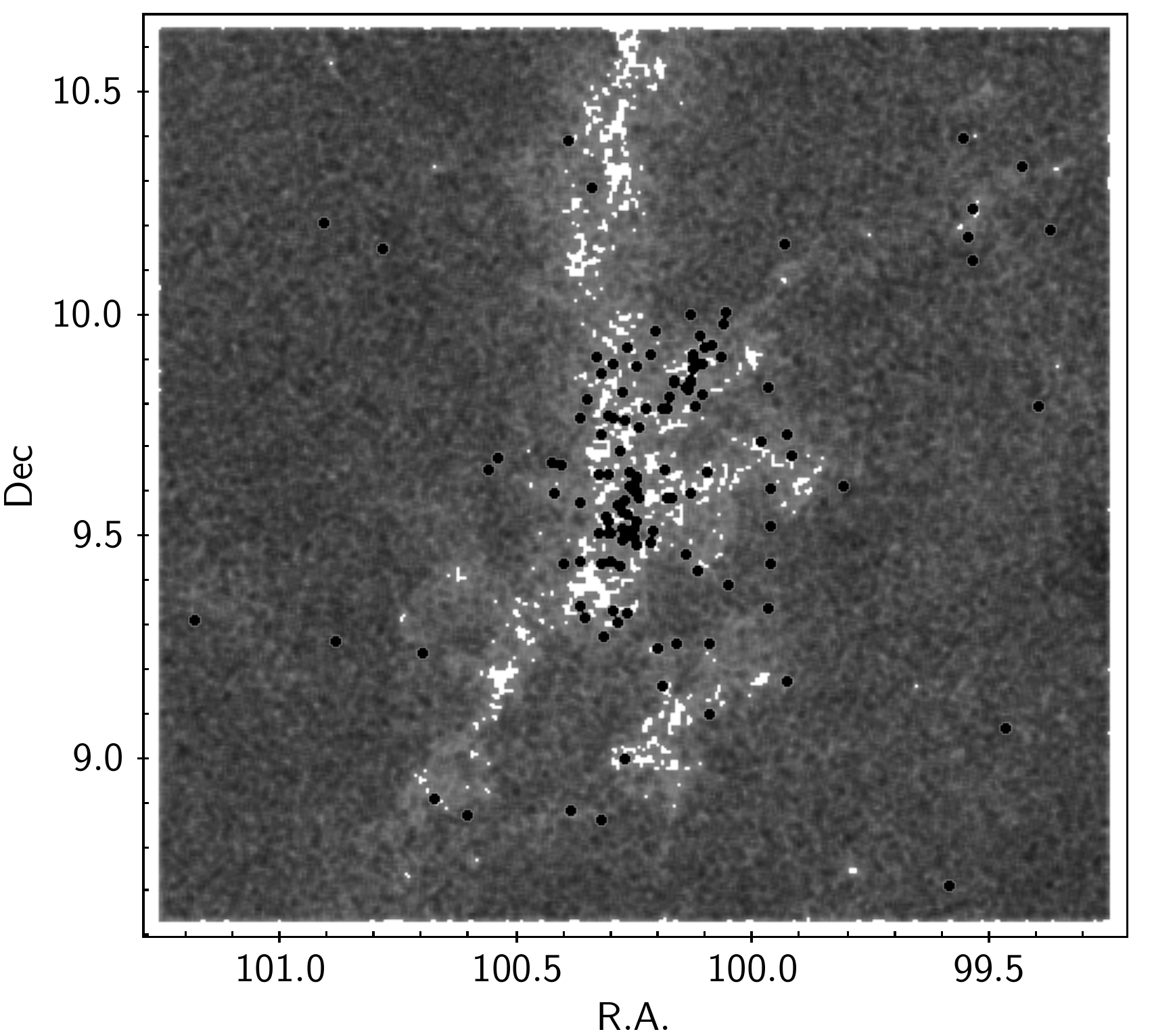}
  \caption{Map of objects above the red line in (j).}
  \label{fig:l}
\end{subfigure}

\caption{Spatial distribution of the different A$_V$ subgroups extracted from the ($r-i$, $r$) diagram, overplotted to the map of the entire population of our catalog (in gray on the various R.A., Dec diagrams).}
\label{fig:images}
\end{figure*}

To better investigate the nature of the M-type population of the region and identify potential subpopulations, we performed a detailed analysis of the A$_V$ structure of the region. We sorted our sample of objects, {extracted from the (${i-J}$, ${r-i}$) diagram in Fig.\,\ref{fig:iJ_ri}}, into four A$_V$ groups, selected in such a way as to ensure that each group contains about one fourth ($\sim$6020) of the total M-type population of the field: A$_V$\,$<$\,0.7~mag; 0.7\,$\leq$A$_V$\,$<$\,1.2; 1.2\,$\leq$A$_V$\,$<$\,1.7; A$_V$\,$\geq$\,1.7~mag. We then examined the distribution of each of these four groups of objects on the ($r-i$, $r$) diagram, together with their spatial distribution across the region. Results are illustrated in Fig.\,\ref{fig:images} and discussed in the following sections.

\subsubsection{Nature of the low-reddening population ($A_V < 0.7$)} \label{sec:lowAv}

Fig.\,\ref{fig:a} illustrates the color properties of objects from the first quartile in A$_V$ ($A_V < 0.7$) on the $r$ vs. $r-i$ diagram. The points do not form a continuous distribution on the diagram; rather, low-A$_V$ M-type stars appear to split into three distinct loci. About 40\% of the sources in this group populate a compact, triangular locus in the lower-left corner of the diagram; a small strip with a clear dearth of points separates this locus from an intermediate sequence of objects, that contains another 45\% of the sources in this A$_V$ group. A smaller fraction of objects ($\sim$15\%) defines instead a third locus, running above the intermediate sequence of objects on the diagram, and well distinct from the latter at the faint magnitude end, although partially merging with the intermediate population locus at the bright magnitude end.

\begin{figure}
\resizebox{\hsize}{!}{\includegraphics{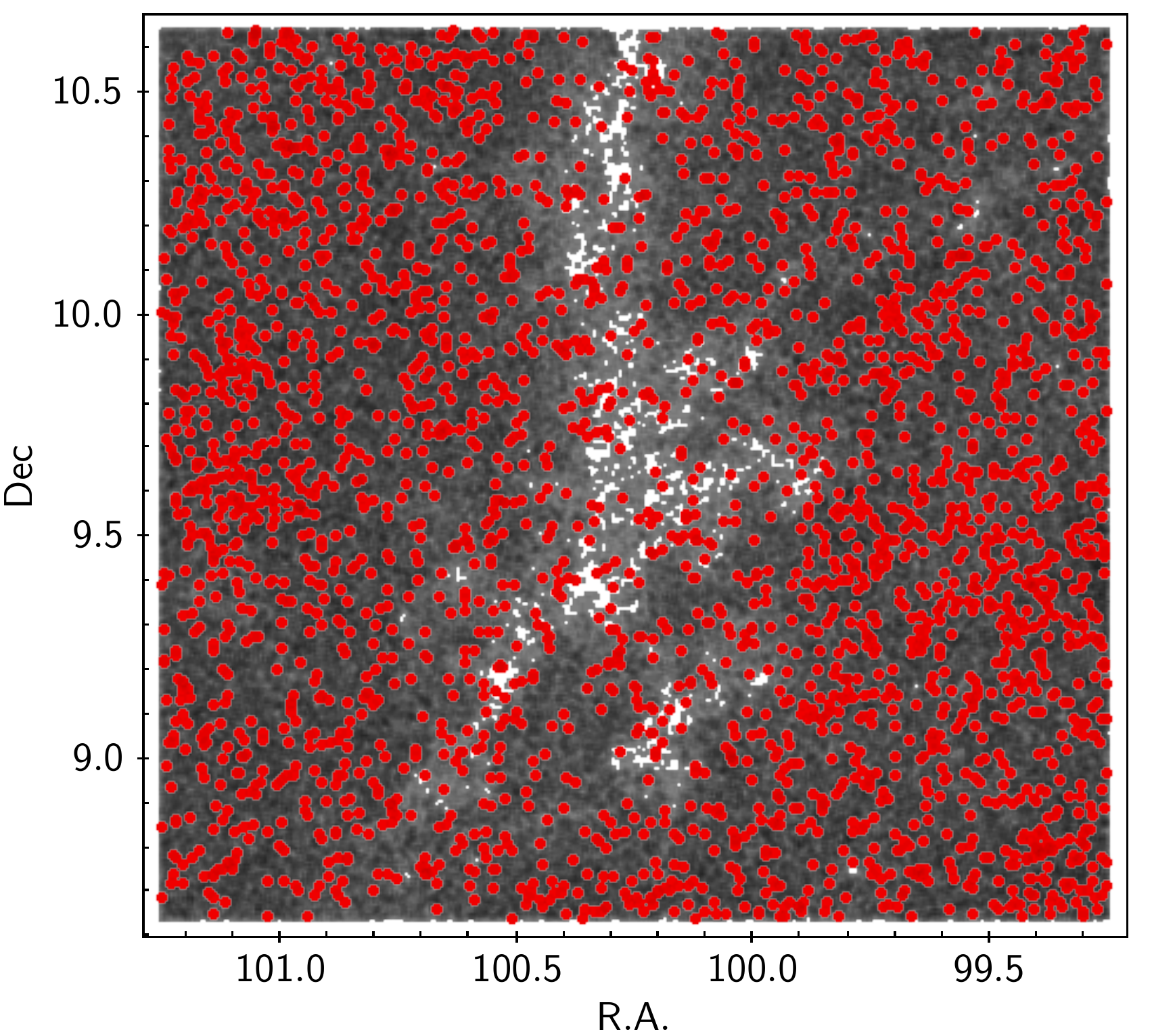}}
\caption{Spatial distribution of objects from the intermediate color-magnitude locus in Fig.\,\ref{fig:a} (red points), projected onto a map of the entire population of our catalog (gray dots).}
\label{fig:radec_Av_0_07_2}
\end{figure}

To investigate the nature of the distinct stellar loci, we traced two boundary lines, parallel to the reddening vector on the diagram and crossing the gaps between two neighboring loci; we then extracted the population of each of the three resulting areas on Fig.\,\ref{fig:a} separately, and examined their spatial distribution. A visual inspection suggests that the three subpopulations extracted from Fig.\,\ref{fig:a} are associated with different spatial properties. Objects from the lower locus on the ($r-i$, $r$) diagram populate the outer areas of our field, whereas the central part of the region is devoid of such sources (Fig.\,\ref{fig:b}); conversely, objects from the upper locus on the ($r-i$, $r$) diagram in Fig.\,\ref{fig:a} are predominantly concentrated toward the center of the field, albeit with some dispersion of objects across the outer field areas (Fig.\,\ref{fig:c}). Interestingly, objects from the intermediate locus in Fig.\,\ref{fig:a} exhibit a rather uniform spatial distribution across the whole field, as illustrated in Fig.\,\ref{fig:radec_Av_0_07_2}.

\begin{figure}
\resizebox{\hsize}{!}{\includegraphics{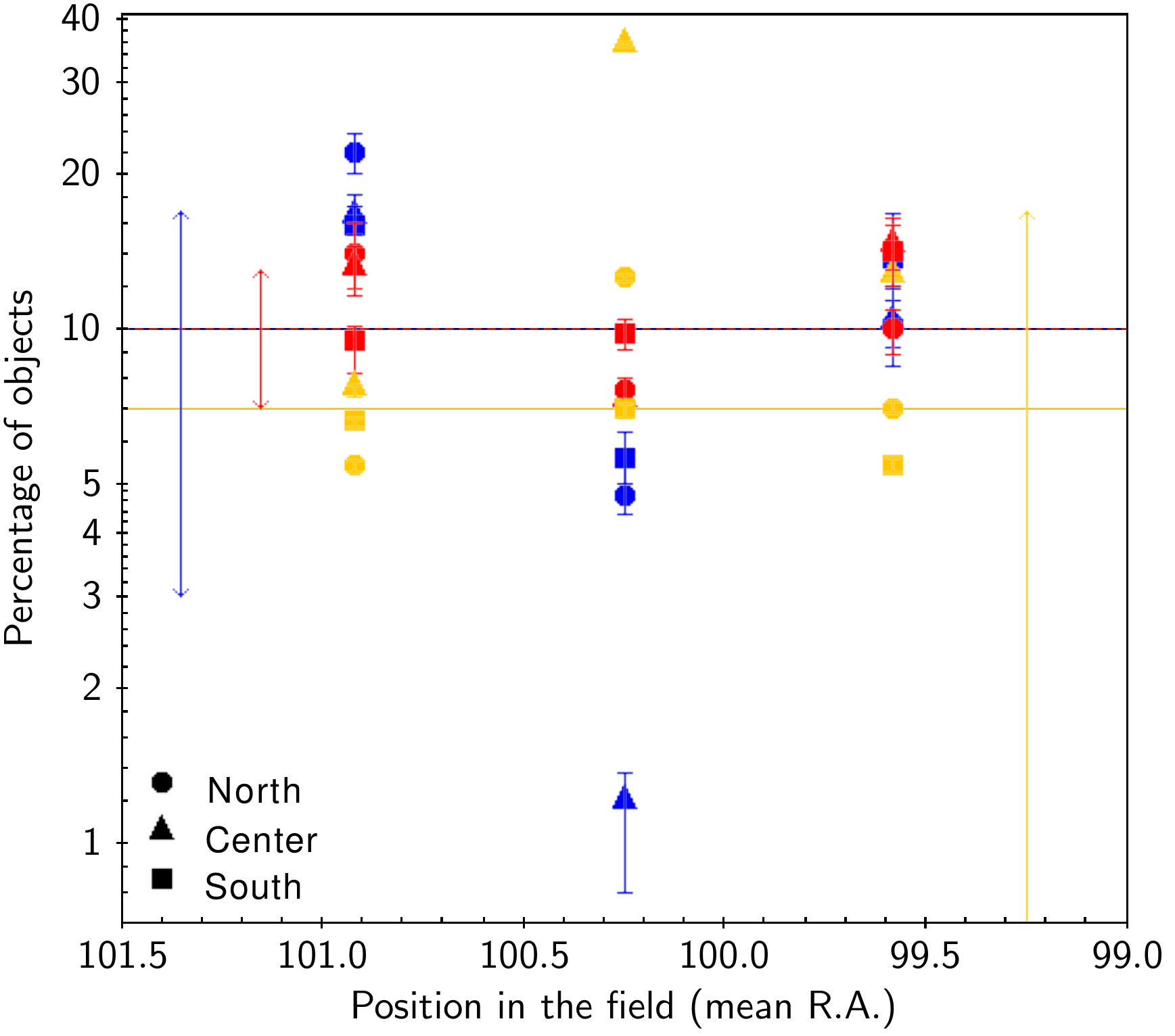}}
\caption{Percentage of objects from each of the three subpopulations extracted from Fig.\,\ref{fig:a} which falls onto each of the nine subfields defined within our region (see text). Following the color convention in Figs.\,\ref{fig:images} and \ref{fig:radec_Av_0_07_2}, blue symbols are referred to the lower population from Fig.\,\ref{fig:a} (below the blue dashed line), red symbols are referred to the intermediate population from Fig.\,\ref{fig:a} (between the blue dashed line and the red dotted line), and yellow symbols are referred to the upper population from Fig.\,\ref{fig:a} (above the red dotted line). The $x$-axis coordinates correspond to the mean R.A. position of the subfield from which the corresponding percentages were extracted. The shape of the symbols identifies the Dec coordinate of the subfield from where the corresponding percentages were extracted: circle for subfields in the northern part of the region (mean Dec = 10.3056$^\circ$), triangle for subfields in the central part of the region (mean Dec = 9.6389$^\circ$), and square for subfields in the southern part of the region (mean Dec = 8.9722$^\circ$). Error bars associated with the percentage values take into account the photometric uncertainties on datapoints in Fig.\,\ref{fig:a}, which may determine individual objects to shift from one area to the neighboring one on the diagram. Horizontal lines mark the median percentage of objects from each of the three subpopulations in Fig.\,\ref{fig:a} across the nine subfields; the double arrows mark the dispersion of values around the typical percentage (i.e., the extent of the spatial variation in the distribution of objects from the corresponding subpopulation across the field).}
\label{fig:radec_lowAv_distr}
\end{figure}

To provide some quantitative support to the visual analysis outlined in the above paragraph, we divided our field (Fig.\,\ref{fig:ra_dec_density}) into a grid of 3$\times$3 equal squares; we then computed what fraction of each of the three populations extracted from Fig.\,\ref{fig:a} falls within each of the nine subfields. The results of this test are illustrated in Fig.\,\ref{fig:radec_lowAv_distr}. If the three subpopulations were distributed uniformly across the field, we would expect to find about 11\% of objects belonging to each of them into each of the nine subfields that we defined in our region, with only small variations in number from one subfield to the other. As shown in Fig.\,\ref{fig:radec_lowAv_distr}, the average distribution across the field of the lower and the intermediate populations from Fig.\,\ref{fig:a} is overall consistent with the expected fraction per subfield in the case of a uniform distribution. However, while for the intermediate population the individual percentages measured in each subfield (red dots) exhibit little fluctuation around this typical value (indicated by the vertical double arrow), the behavior of the lower population is significantly more spatially variable across the region (blue dots). The strongest departure from the typical trend occurs in the central subfield, where hardly a few objects from this subpopulation are found. The upper population in Fig.\,\ref{fig:a} exhibits the most non-uniform spatial distribution: a dearth of objects from this group is observed at any location within the region but in the central subfield, which contains alone 40\% of the subpopulation members. This strongly spatially varying behavior is illustrated by the yellow double arrow in Fig.\,\ref{fig:radec_lowAv_distr}, which marks the dispersion in value of single-subfield percentages with respect to the median percentage measured for the upper population across the whole field. To summarize this analysis more quantitatively, each of the nine subfields defined in our region contains typically 10\% of objects from the intermediate population in Fig.\,\ref{fig:a}, with an rms scatter of 3\% among individual subfields (i.e., 10\% $\pm$ 3\%). These percentages become $(10 \pm 7)$\% for the lower population, and $(7 \pm 10)$\% for the upper population.

As a further check on the respective spatial distributions of the three subpopulations with $A_V < 0.7$, we applied the minimum spanning tree (MST) test, as formulated by \citet{prim1957}. To ensure a uniform sampling of the three populations and account for the different number of objects contained in each of them, we randomly picked 100 stars from each subpopulation, and computed the minimum path length required to spatially connect each of the 100 points in a network without closed loops. We then iterated this procedure 100\,000 times, in order to build an MST distribution for each subpopulation from Fig.\,\ref{fig:a}. A MST of $13.4 \pm 0.4$ was derived when sampling the intermediate population, of which the spatial distribution is illustrated in Fig.\,\ref{fig:radec_Av_0_07_2}; this value is very similar to the one ($13.3 \pm 0.4$) found when applying the same procedure to the entire group of objects with $A_V < 0.7$, irrespective of their location on Fig.\,\ref{fig:a}. Noticeably lower values were derived for the lower ($12.1 \pm 0.5$) and for the upper ($11.9 \pm 0.5$) populations in Fig.\,\ref{fig:a}; this {difference, while not conclusive,} supports the view of a more wide-spread spatial distribution for the intermediate population than for the other groups of objects, although it does not enable distinguishing between the spatial properties of the lower population and the upper population (Figs.\,\ref{fig:b} and \ref{fig:c}, respectively).

\subsubsection{Nature of the reddened population ($A_V \geq 0.7$)}

Figs.\,\ref{fig:d}, \ref{fig:g}, and \ref{fig:j} illustrate the photometric properties of progressively more extincted M-stars in the field ($0.7 \leq A_V < 1.2$, $1.2 \leq A_V < 1.7$, and $A_V \geq 1.7$, respectively). A clear separation can be observed, on each panel, between the bulk of the field population, grouped in a triangular locus in the lower-left quarter of the $r$ vs. $r-i$ diagram, and a more sparse sequence of objects, which develops above the main population locus, nearly parallel to its right-hand edge and to the reddening vector on the diagram. A straight line, parallel to the reddening vector and delimiting the transition between the densely populated field locus and the upper sequence of objects, was applied uniformly in every A$_V$ group to extract the two subpopulations and investigate their spatial properties. This line is the same as that shown in red in Fig.\,\ref{fig:a}.

As already observed for the upper vs. lower populations in Fig.\,\ref{fig:a} (see Sect.\,\ref{sec:lowAv}), objects located below the red line in Figs.\,\ref{fig:d}, \ref{fig:g} and \ref{fig:j} exhibit statistically distinct spatial properties from the sequence of objects located above the red line on the same diagrams. The former are predominantly distributed across the outer areas of our field, and are substantially absent in the innermost regions (see Figs.\,\ref{fig:e}, \ref{fig:h}, and \ref{fig:k}); on the contrary, sources that form the upper sequence of objects on the $r$ vs. $r-i$ diagram tend to be spatially concentrated in the center of the field. The same behavior is observed irrespective of the A$_V$ group we are considering, albeit with slight variations in number between the fractions of objects in the main population locus and in the upper sequence locus.

The analysis illustrated in this and in the previous section indicates that two distinct stellar populations are present in our field. A diffuse field population comprises about 94\% of the objects in our catalog; the remaining 6\% form a clustered, presumably young population, which stands out clearly as a separate sequence above the main field locus on the $r$ vs. $r-i$ diagram {(i.e., above the red line on the diagrams in Fig.\,\ref{fig:images})}, and is spatially gathered in the inner regions of the field. The clustered population appears to be associated with a concentration of obscuring material, {clearly seen in Fig.\,\ref{fig:field}}, and silhouetted on the spatial maps in Figs.\,\ref{fig:ra_dec_density} and \ref{fig:images}. This region is characterized by a higher average extinction than the surrounding areas (see Fig.\,\ref{fig:mean_Av_map}), {and the nebulosity presumably} filters out background stars. This explains the underdensity of objects observed in Fig.\,\ref{fig:ra_dec_density} at the location of our clustered population, and the corresponding lack of detected field stars with moderate extinction and spatially projected onto the center of the field, as shown in the middle panels of Fig.\,\ref{fig:images}. 

A somewhat different behavior is observed for field stars in the first A$_V$ quartile ($A_V < 0.7$~mag), which split into two photometric loci on the $r$ vs. $r-i$ diagram, with correspondingly different spatial properties (see Sect.\,\ref{sec:lowAv}). The nature of these two subgroups of objects can be interpreted in the light of the above considerations: objects located below the blue line in Fig.\,\ref{fig:a} are low-extinction background stars, and are therefore detected only in the outer regions of our field, not blocked by the central obscuring material; objects between the blue and the red lines in Fig.\,\ref{fig:a} are instead foreground stars, uniformly detected throughout the field. 

{At the time of the analysis, no parallax and proper motion information is available, from Gaia DR1, for any of the M-stars in our sample. Further tests on the astrometric properties of the various populations identified in the NGC~2264 region are deferred to a future work, making use of the wealth of information from the newest Gaia DR2.}

\subsection{A clustered population in the NGC~2264 region}

\subsubsection{Distance and age} \label{sec:dist_age}

The peculiar features of the $r$ vs. $r-i$ diagram for lower-extinction stars in our sample ($A_V<0.7$; Fig.\,\ref{fig:a}) allow us to provide some constraints on the distance and age of the clustered population.

\begin{figure}
\resizebox{\hsize}{!}{\includegraphics{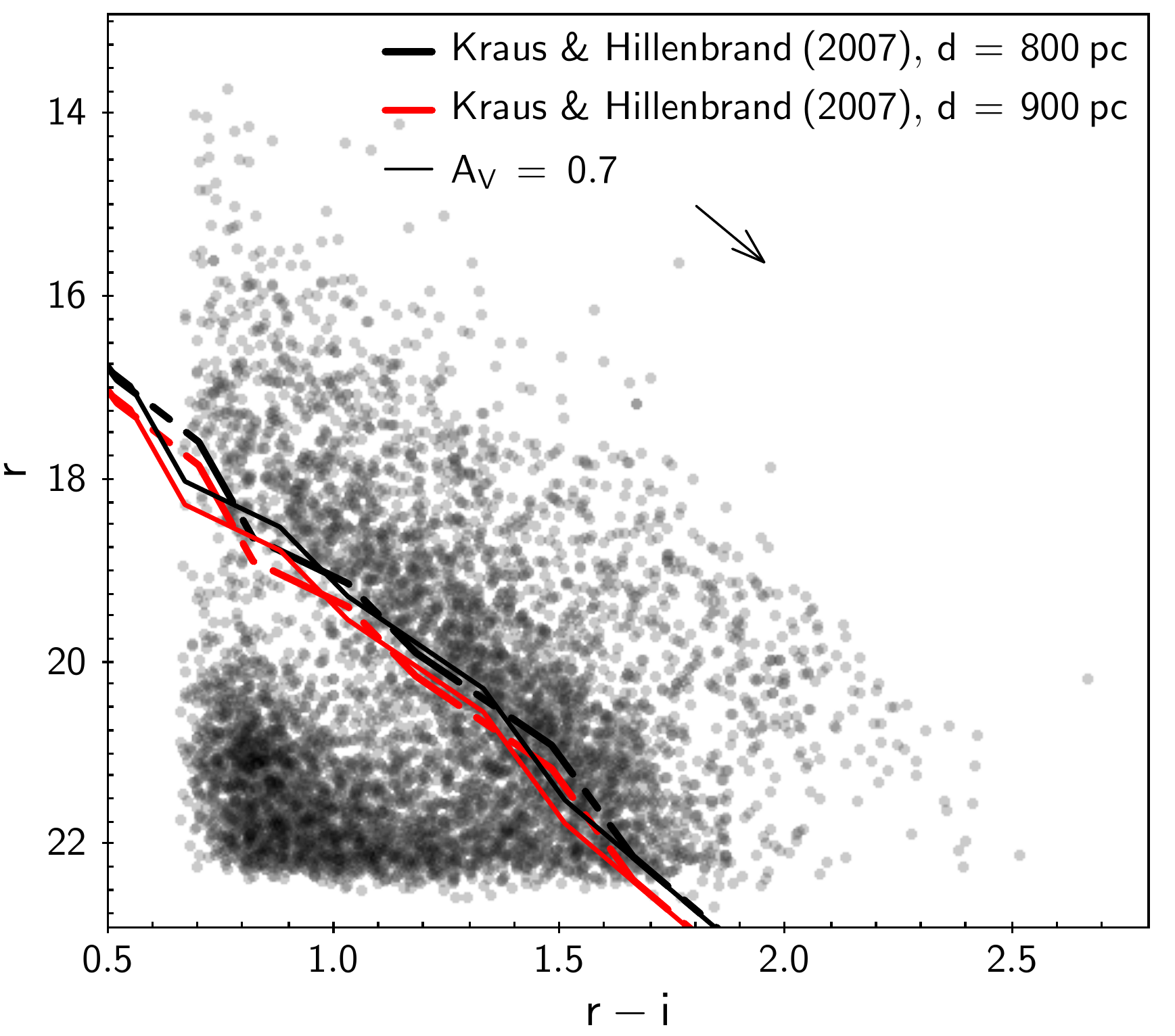}}
\caption{Same diagram as in Fig.\,\ref{fig:a}, with a superimposed color-magnitude sequence for dwarfs extracted from \citet{kraus07}, brought to a distance of 800~pc (black) and 900~pc (red), assuming $A_V=0$ (solid line) or $A_V=0.7$ (dash-dotted line), to reproduce the lower envelope of the foreground stars locus on the diagram.}
\label{fig:ri_r_CMD_Av_0_1_dist}
\end{figure}

In particular, the clear separation between foreground stars and background stars on the diagram enables a direct estimation of the distance to the obscuring material spatially associated with the clustered population. Such estimate of distance can be obtained by fitting the lower envelope of the foreground (intermediate) population on the diagram with a MS track; a similar approach was adopted by{, e.g., \citet{knude2001} to determine the distance to Lupus~2, or by} \citet{prisinzano2005} to determine the distance to NGC~6530. For the purposes of this analysis, we adopted the SDSS absolute magnitude sequence for dwarfs tabulated in \citet{kraus07}, and shifted it to tentative distances until we obtained a satisfactory match between the photometric sequence and the lower envelope of the upper part of the foreground stars locus on Fig.\,\ref{fig:a}, where the foreground-to-background gap is more marked. The results are illustrated in Fig.\,\ref{fig:ri_r_CMD_Av_0_1_dist}, and are substantially unaffected by extinction, since the color-magnitude M-dwarf sequence on the diagram runs almost parallel to the reddening vector.
The concentration of obscuring material which determines the discontinuity between the distributions of foreground stars and background stars in Fig.\,\ref{fig:a} appears to {begin} at a distance of roughly 800--900~pc.

\begin{figure}
\resizebox{\hsize}{!}{\includegraphics{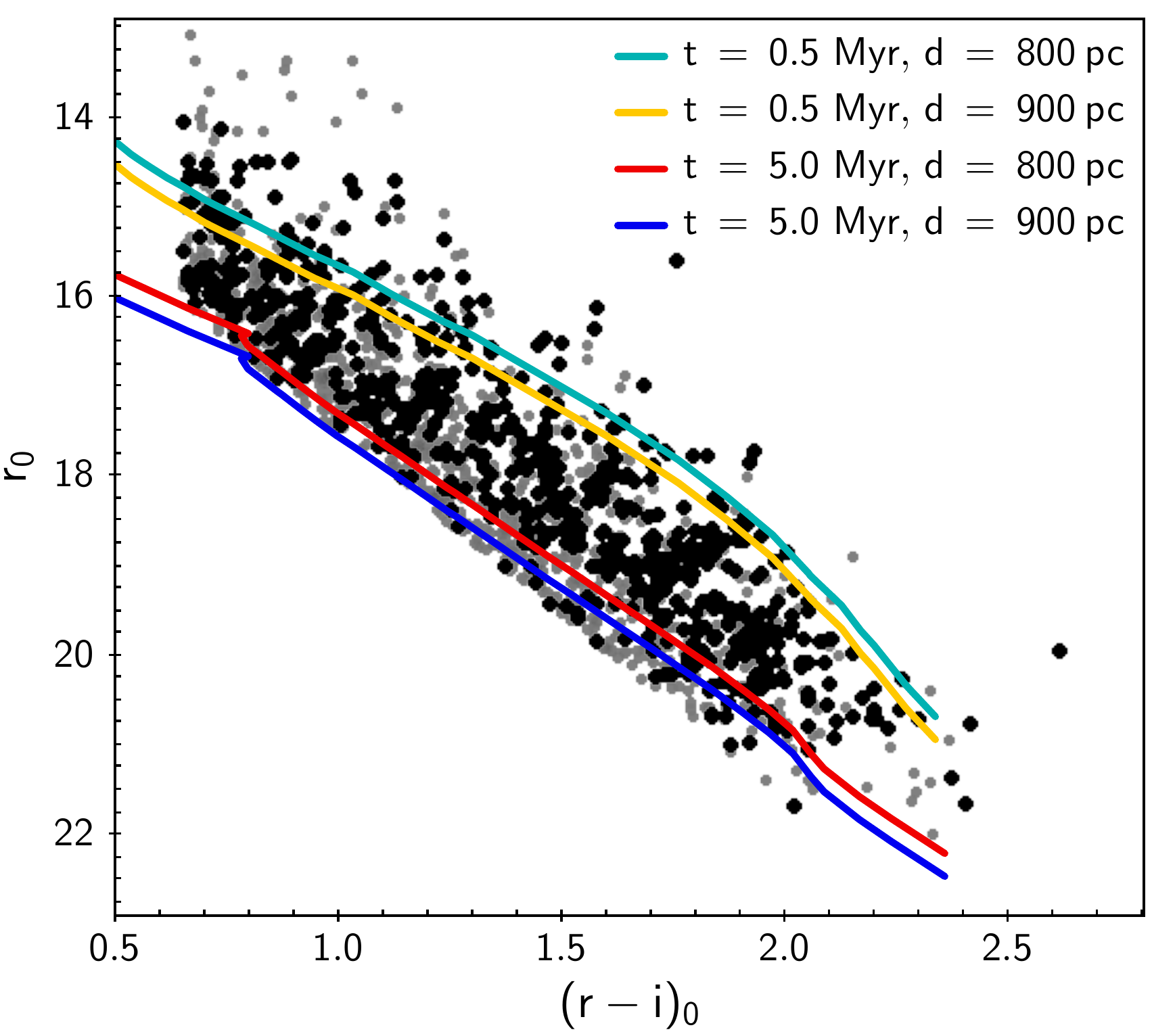}}
\caption{Dereddened $r$-band photometry and $r-i$ colors for M-type stars that fall in the clustered population locus (i.e., above the red line) on the diagrams shown in Fig.\,\ref{fig:images} (left panels). Sources extracted from the photometric clustered population locus and spatially projected onto the central part of the field are highlighted further as black points. {0.5-Myr and 5-Myr PMS model isochrones from \citet{marigo2017}}, brought to a distance of 800--900~pc, are overplotted to the datapoint distribution to mark a tentative age range for the clustered population.}
\label{fig:ri_r_CMD_seq_Baraffe15}
\end{figure}

Assuming that the distance derived above is a good approximation to the distance of the clustered population, we can use this estimate to tentatively constrain the age of the clustered population. To do so, we selected all objects located above the main field locus in the various $r$ vs. $r-i$ diagrams in Fig.\,\ref{fig:images} (left panels), and built a dereddened $r_0$ vs. $(r-i)_0$ diagram for the clustered population in the field. We then adopted {the grid of PARSEC--COLIBRI stellar isochrones \citep{marigo2017} in the SDSS system}, brought to a distance of 800--900~pc, to identify the range in ages covered by the sequence of clustered objects on the color-magnitude diagram. This comparison is illustrated in Fig.\,\ref{fig:ri_r_CMD_seq_Baraffe15}; the analysis suggests that the clustered population identified within our field may span an age range between {$\sim$0.5 and 5~Myr.}

\begin{figure}
\resizebox{\hsize}{!}{\includegraphics{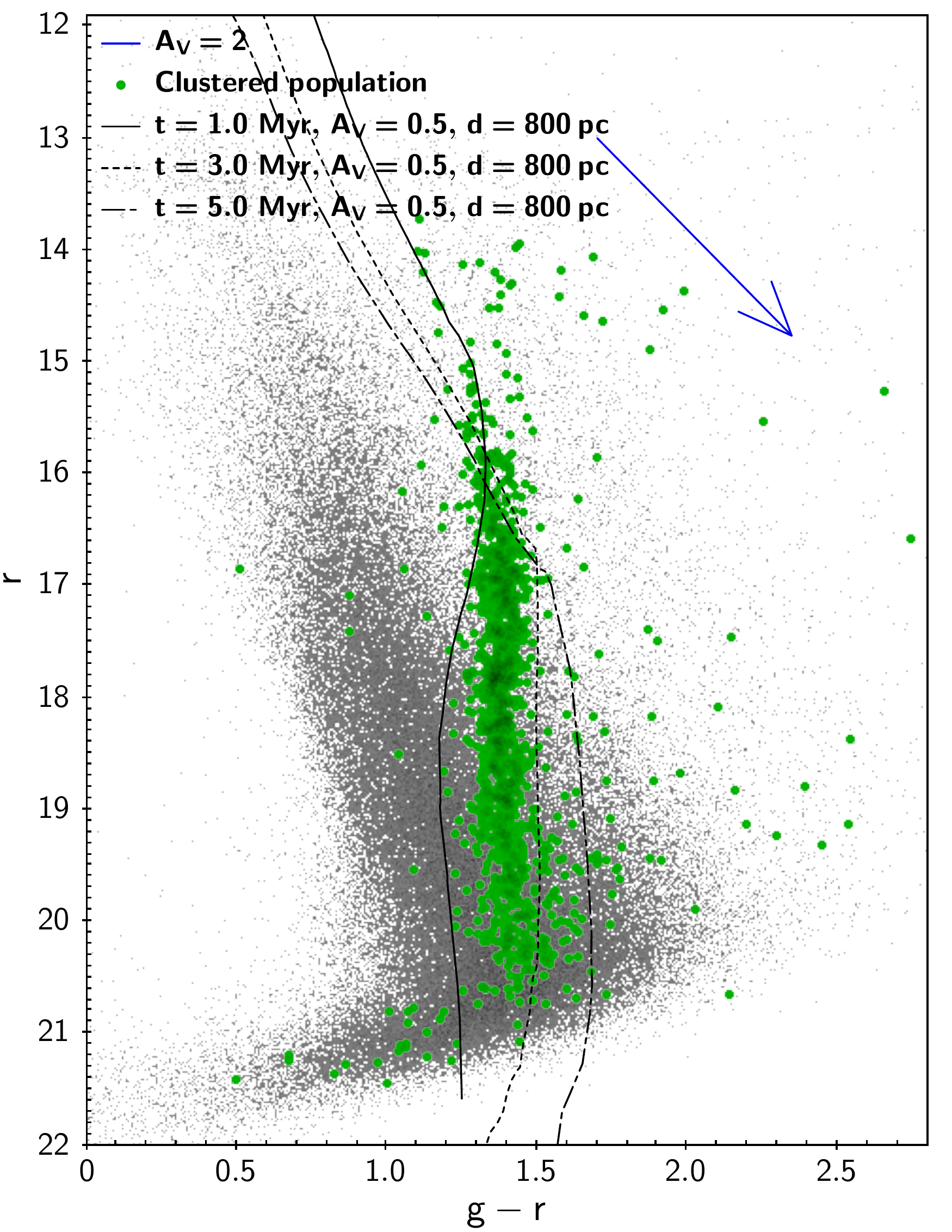}}
\caption{Field locus on the ${r}$ vs. ${g-r}$ diagram (gray). The sequence traced by the clustered population, extracted as described in Sect.\,\ref{sec:Av_structure}, is highlighted in green. 1-Myr, 3-Myr and 5-Myr model isochrones from \citet{marigo2017}, brought to a distance of 800~pc and reddened by an amount corresponding to ${A_V = 0.5}$ (typical A$_V$ measured for sources in the clustered population), are overlaid on the diagram.}
\label{fig:gr_r_CMD_seq_PARSEC}
\end{figure}

{The clustered population does not define a sequence of its own, parallel to the main field locus, on the (${g-r}$, ${r}$) diagram, but a nearly vertical sequence on ${g-r}$, reflecting the color properties for M-type stars observed in Fig.\,\ref{fig:gr_ri_iJ}. The ${r}$ vs. ${g-r}$ location of clustered objects (selected from the $r$ vs. $r-i$ diagram) is in broad agreement with the predicted loci traced by $\sim$1-Myr to 5-Myr isochrones from \citeauthor{marigo2017}'s (\citeyear{marigo2017}) compilation, brought to a distance of 800~pc (as inferred from the analysis in Fig.\,\ref{fig:ri_r_CMD_Av_0_1_dist}), and reddened by an amount corresponding to the typical A$_V$ ($\sim$0.5) measured across the clustered population from $r,i,J$ photometry. 

The analysis illustrated in Fig.\,\ref{fig:ri_r_CMD_seq_Baraffe15} may be affected by the presence of unresolved binaries, mapped on the color-magnitude diagram as a single object with magnitude $m = m_{primary} - 2.5 \log{\left(1+\frac{F_{secondary}}{F_{primary}}\right)}$. To test the possible effect on the age range derived earlier, we simulated a population of unresolved binaries, randomly selected across our clustered population, assuming a binary fraction between 0.15 and 0.5, and secondary-to-primary mass ratios between 0 and 1 and distributed following a power-law distribution with index between 0.2 and 1 \citep{kraus2012, duchene2013}. For each object selected as an unresolved binary, we then artificially corrected the apparent photometry, starting from the observed magnitudes and the randomly assigned mass ratio, and assuming that the magnitude difference between secondary and primary is related to the mass ratio of the system following the predicted mass--magnitude trend along the PMS theoretical isochrones. Results from this test suggest that the presence of unresolved binaries may impact the location of both the upper and lower envelopes of the photometric locus of our clustered population in Fig.\,\ref{fig:ri_r_CMD_seq_Baraffe15} with respect to the isochrone grid. At the upper end, unresolved binaries interpreted as intrinsically brighter sources may determine an artificial widening of the sequence of objects toward lower ages by $\sim$0.5--1~Myr. Similarly, unresolved binaries may artificially lift the lower envelope of the datapoint distribution to younger apparent ages; the shift in age is of the order of a few (1--2 to 4--5) Myr, and appears to depend on the actual binary fraction in the population.

We also note that the age range inferred for the population is somewhat dependent on the model tracks used for the comparison with the photometric locus on Fig.\,\ref{fig:ri_r_CMD_seq_Baraffe15}. A different, although overlapping, age range between $\sim$3 and 20~Myr would be deduced from the SDSS-calibrated PMS model isochrones of \citet{baraffe2015}. We adopted here the grid of isochrones from \citet{marigo2017}, rather than that from \citet{baraffe2015}, because the latter fail to reproduce the clustered population locus on the (${g-r}$, ${r}$) diagram in Fig.\,\ref{fig:gr_r_CMD_seq_PARSEC}. Stellar variability may represent another source of uncertainty on the age range spanned by the cluster population: as reported in \citet{venuti2015}, a day-to-year $r$-band variability of up to $\sim$0.4~mag may be expected for $\sim$3~Myr-old stars, translating to a vertical shift of 1--2~Myr with respect to the grid of PARSEC isochrones on the ($r-i$, $r$) diagram.
}

\subsubsection{Spatial structure} \label{sec:structure}

\begin{figure*}
\centering
  \includegraphics[width=\textwidth]{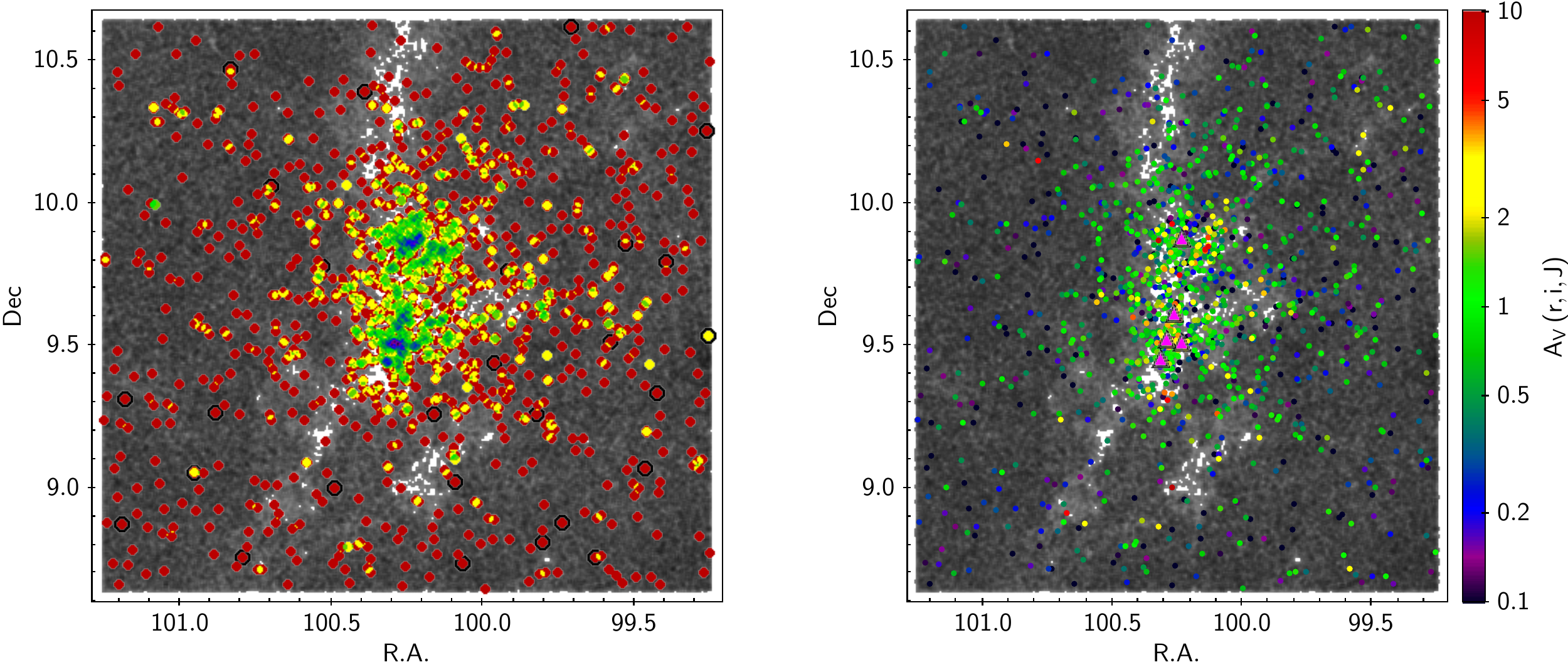}
  \caption{{\it Left panel}: Spatial density map of the clustered population extracted from Fig.\,\ref{fig:images}, overplotted to the map of the complete catalog population (gray dots). Colors progress with local number density of objects from red (lowest) to blue/purple (highest) following a rainbow scale. Points encircled in black are possible giant contaminants (see Sect.\,\ref{sec:contaminants}). {\it Right panel}: $A_V$ distribution of objects highlighted in the left panel, after removing the giant contaminants. Colors are scaled according to A$_V$ as shown in the side axis. The field population is shown in gray in the background. Fuchsia triangles mark the locations with highest density of objects from the clustered population (in blue on the left panel diagram).}
\label{fig:radec_cluster_Av_density}
\end{figure*}

A spatial distribution and number density map of the clustered population identified in our analysis is presented in Fig.\,\ref{fig:radec_cluster_Av_density} (left panel). Objects are predominantly grouped in two main clumps, distributed vertically in the innermost regions of our field. The core of the northern clump, where the concentration of objects from the clustered population is highest, occurs at (R.A., Dec) $\sim$ (100.23, 9.865). The most populated locations in the southern clump occur at (R.A., Dec) $\sim$ (100.29, 9.51), (100.23, 9.5), and (100.315, 9.445). An additional, densely populated, elongated structure is observed at (R.A., Dec) $\sim$ (100.26, 9.6), above the main southern clump. A comparison of the density map with the spatial A$_V$ distribution of members of the clustered population, shown in Fig.\,\ref{fig:radec_cluster_Av_density} (right panel), indicates that the areas where the clustered population is most represented tend to be associated with higher extinction than the surrounding areas. {The median A$_V$ measured across the clustered population is $\sim$0.5~mag, well within the range of measurable A$_V$ predicted from the analysis in Fig.\,\ref{fig:riJ_detection_depth}; conversely, the cutoff in A$_V$ for detected sources with large extinction ($\sim$5--8~mag) is likely affected by the limiting magnitudes of our optical catalog.} A fraction of objects from the clustered population distribute around the two main clumps of stars described earlier, forming a wide-spread halo that extends across the entire field, as illustrated in Fig.\,\ref{fig:radec_cluster_Av_density} (left panel).

\section{Discussion} \label{sec:discussion}

\subsection{Contamination by giants and field stars to the selected clustered population} \label{sec:contaminants}

A certain degree of contamination by foreground field stars is expected on the photometric selection of the clustered population. This is evident from Fig.\,\ref{fig:a}, where a partial overlap can be observed between the locus of the clustered population (well determined from Figs.\,\ref{fig:d}, \ref{fig:g}, and \ref{fig:j}) and the locus of foreground stars. An additional source of contamination to the clustered population locus is represented by reddened giant stars in the field, with photometric properties that would place them in the brighter mag regime but at redder colors than the range spanned by field stars (see, for instance, the trail of points located at $r < 16$ and $r-i > 1$ in Fig.\,\ref{fig:ri_r_CMD_Av}). \citet{covey2008} simulated SDSS/2MASS observations to map the respective loci of giant stars and field dwarfs on a $J$ vs. $J-K_S$ diagram, and to identify the region of the diagram where giants are predominant over the dwarf population. Following their example, we combined 2MASS and UKIDSS photometry to build the ($J-K_S$, $J$) diagram of the population of our field, and discarded as likely giant contaminants all sources brighter than $J = 12$ and redder than $J-K_S = 0.5$. This procedure allowed us to identify 27 probable giant contaminants in our selected clustered population. The spatial distribution of these 27 sources with respect to the clustered population is highlighted in Fig.\,\ref{fig:radec_cluster_Av_density} (left panel); as can be seen, these are mostly dispersed in the outer regions of the field. We note that this analysis only enables to identify the brightest giants; a fraction of giant stars is expected to be mingled with the field dwarf locus (see Fig.\,5 of \citealp{covey2008}).

\begin{figure}
\resizebox{\hsize}{!}{\includegraphics{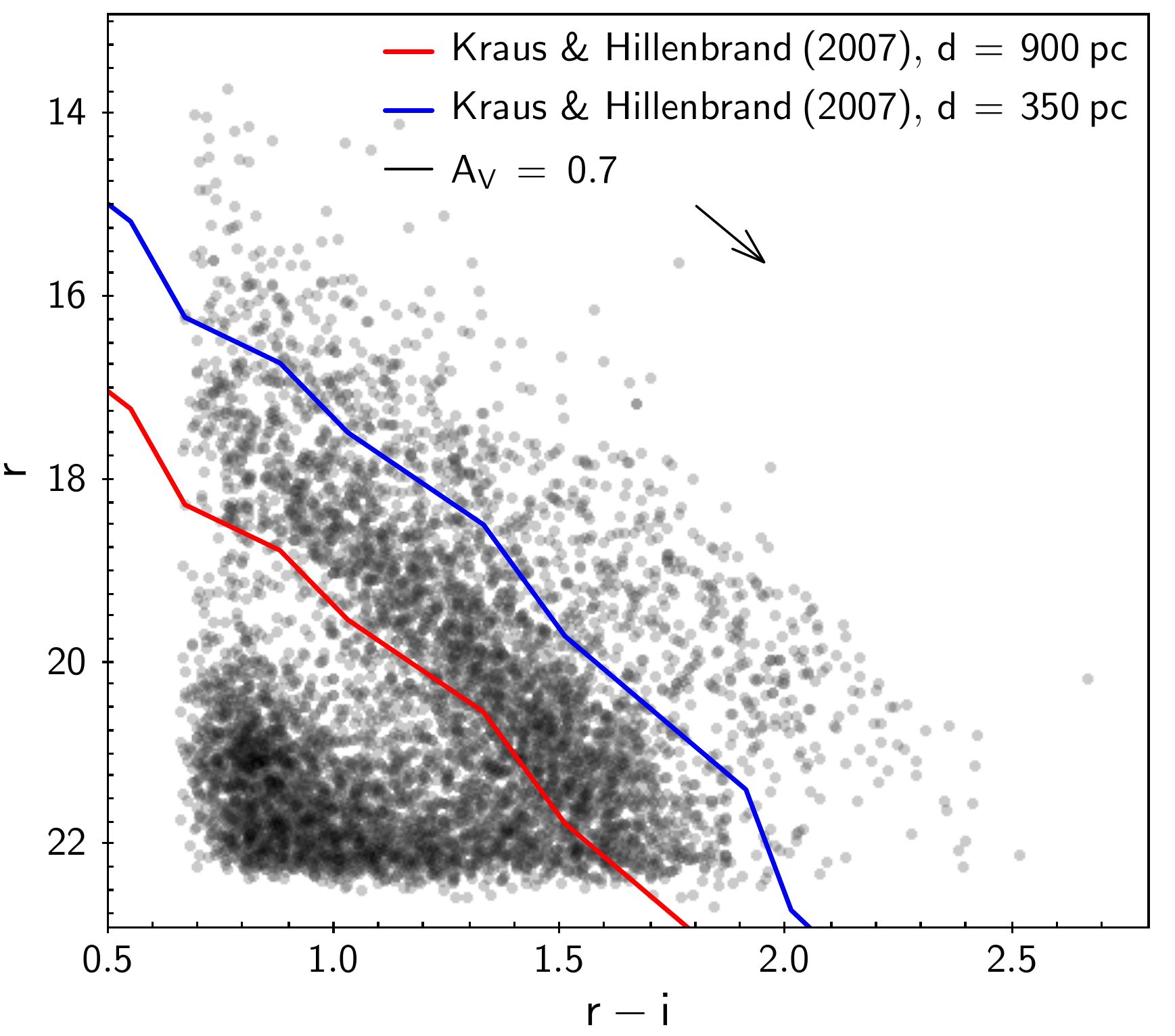}}
\caption{($r-i$, $r$) diagram for M-type stars with $A_V < 0.7$ in our field, with the theoretical color-magnitude sequence from \citet{kraus07} brought to a distance of 350~pc (blue) and 900~pc (red) to delimit the range of distances within which foreground stars can be distinguished from the clustered population along the line of sight based on their photometric properties.}
\label{fig:ri_r_d1_d2}
\end{figure}

To derive a statistical estimate of the number of field contaminants expected among our clustered population members, we proceeded as follows. We used the SDSS spectral type-absolute magnitude sequences tabulated in \citet{kraus07}, already adopted to derive the distance to the obscuring material, to infer an estimate of the minimum distance at which foreground stars and members of the clustered population can be distinguished. This minimum distance was obtained by shifting the theoretical color-magnitude sequence of dwarfs until it matched the upper envelope of the foreground stars locus on the $r$ vs. $r-i$ diagram for objects with $A_V < 0.7$, as shown in Fig.\,\ref{fig:ri_r_d1_d2}.
The comparison between the observed field locus and the color-magnitude sequence on the diagram indicates that clustered stars are photometrically mixed with MS foreground stars in the spatial volume between 0~pc and 350~pc along the line of sight. On the other hand, MS foreground stars located between 350~pc and 900~pc define a distinct photometric locus from that of the clustered population; they can therefore be used as a reference to measure the spatial density of foreground stars. We then counted the number of stars distributed between the two sequences in Fig.\,\ref{fig:ri_r_d1_d2} to evaluate the volume density $\rho_N$ of foreground stars geometrically located inside a square frustum having our field of view as base and height between 350~pc and 900~pc. In the assumption that foreground stars exhibit a uniform spatial distribution, the number of field contaminants expected among the extracted clustered population is then equal to $\rho_N$ times the volume of the section of the cone of view between 0 and 350~pc. This statistical assessment suggests that about 10\% ($\sim$138) of the clustered population objects selected may be field contaminants. If these contaminants are randomly distributed across the field, this result implies that we expect about 15 contaminants in each of the nine subfields defined in Sect.\,\ref{sec:lowAv} to evaluate the quantities in Fig.\,\ref{fig:radec_lowAv_distr}. This number amounts to about 20\% of the objects dispersed in the outer parts of the field in Fig.\,\ref{fig:radec_cluster_Av_density} (left panel); on the other hand, the corresponding contamination rate in the innermost region of the field, where the two main clumps of objects in Fig.\,\ref{fig:radec_cluster_Av_density} (left) are located, is only about 2\%. We therefore conclude that the contamination by foreground stars has no significant impact on our results regarding the presence and structure of a young clustered population in the center of our field. 
\begin{figure}
\resizebox{\hsize}{!}{\includegraphics{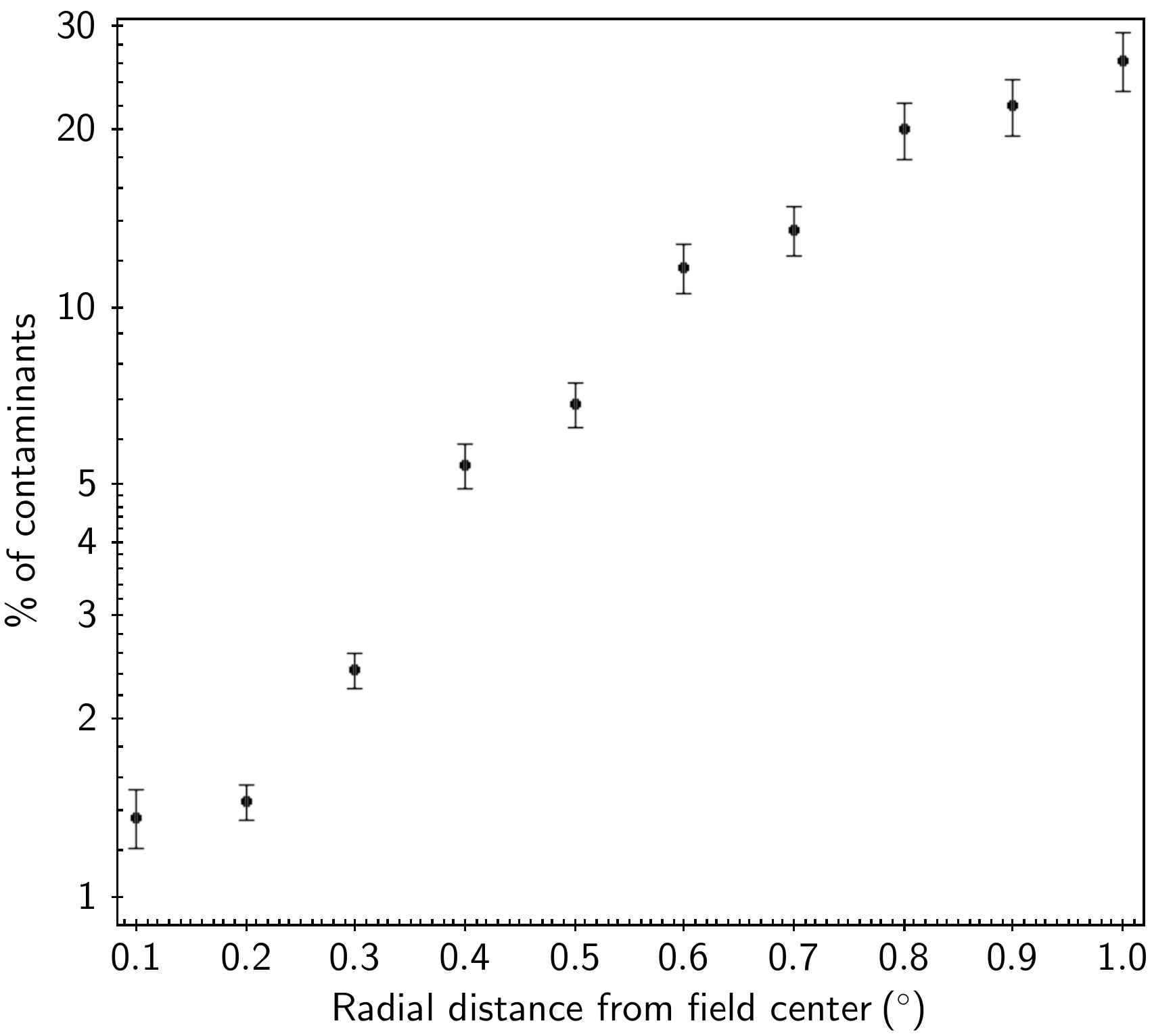}}
\caption{Percentage of expected field contaminants among clustered objects selected in annuli of increasing outer radius (reported on the $x$-axis) and fixed difference of 0.1$^\degree$ between outer and inner radii. The error bars follow from Poissonian error propagation on the number of clustered objects selected in each annulus.}
\label{fig:radius_contaminants}
\end{figure}
{The radial dependence of the rate of field contaminants expected among the selected clustered population, as a function of distance from the field center, is illustrated in Fig.\,\ref{fig:radius_contaminants}. The very low rate of contaminants, of merely a few percents, expected in the innermost field regions (where the bulk of the clustered population is located) increases to nearly 30\% in the outermost areas of the field.}

\subsection{Blind vs. informed survey: assessment of the method performance on the NGC~2264 cluster} \label{sec:csi2264_comp}

In this section, we compare the results of our blind study to the properties and census of the NGC~2264 cluster reported in the literature.

\subsubsection{Cluster structure}

The cluster is located at a distance of 760${\pm 90}$~pc \citep{sung97}, and exhibits an average age of $\sim$3--4~Myr and an internal age spread of several Myr \citep[e.g.,][]{venuti2018}. These values are overall consistent with the estimates of distance and age reported in Sect.\,\ref{sec:dist_age}. Several studies have shown that the NGC~2264 cluster has a hierarchical structure, with multiple clumps. Based on optical and infrared observations, \citet{teixeira2006} and \citet{sung08, sung09} identified multiple subclusters within the region: the S~Mon region, in the northern part of the cluster, which develops around the O-type binary star S~Mon, and the Cone Nebula region, in the southern part of the cluster, which envelops the most embedded parts of NGC~2264 (the Spokes subcluster and Cone (C), the core of the Cone region). In a recent study \citep{venuti2018}, we investigated the nature of these multiple NGC~2264 populations, and showed that they may be the results of distinct star formation episodes occurred within a time lapse of several Myr.
\begin{figure}
\resizebox{\hsize}{!}{\includegraphics{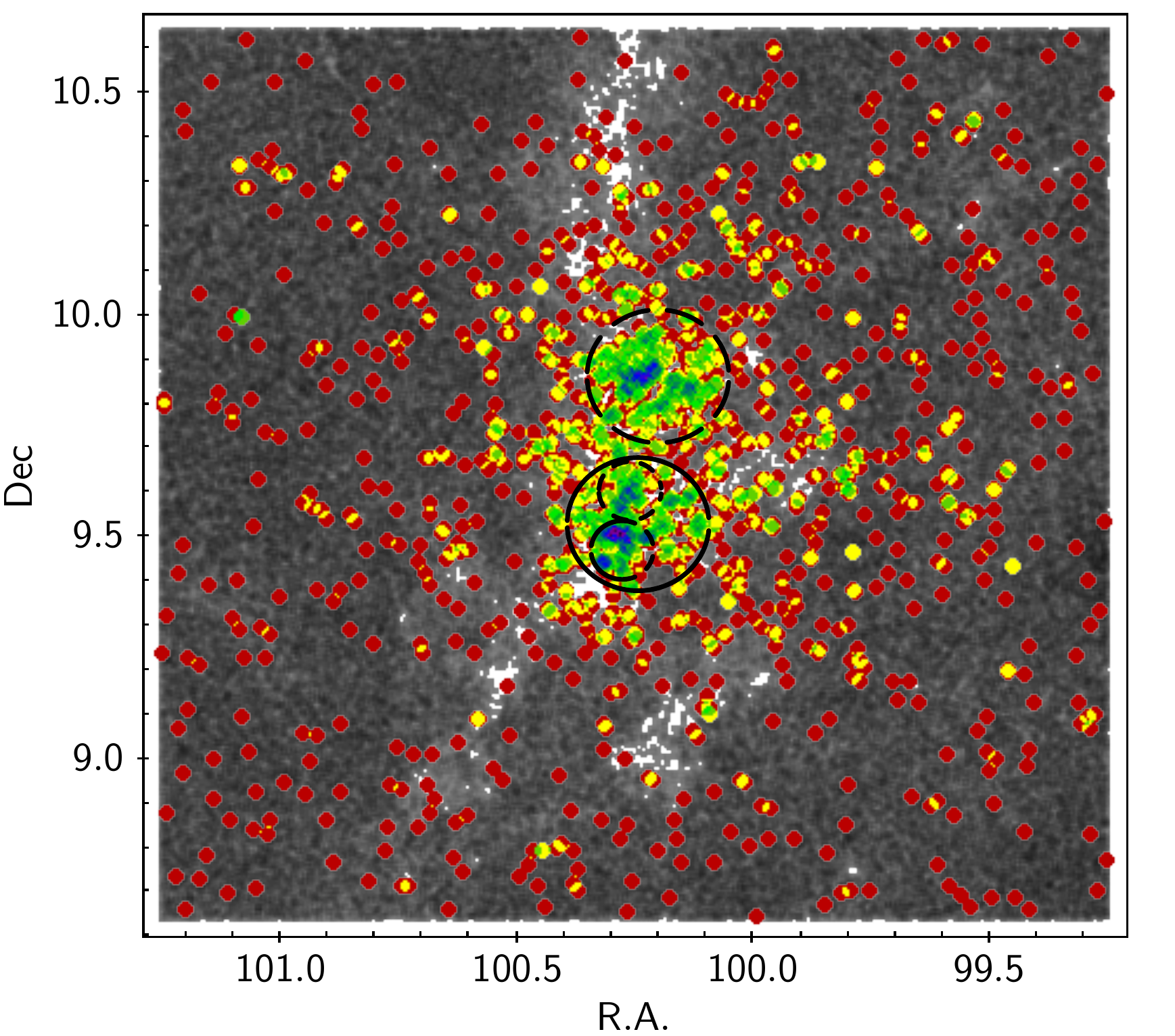}}
\caption{Spatial density map of the clustered population identified in this study (see Fig.\,\ref{fig:radec_cluster_Av_density} and Sect.\,\ref{sec:structure}), with contours indicating the NGC~2264 subregions defined in \citet{sung08, sung09}: the S~Mon subregion (dashed line), the Cone subregion (Cone\,(H), solid line), the Spokes subcluster (upper contour inside the Cone subregion), and the Cone core (Cone\,(C), lower contour inside the Cone subregion).}
\label{fig:radec_sequence_density_Sung}
\end{figure}
In Fig.\,\ref{fig:radec_sequence_density_Sung}, we compare the location of the NGC~2264 subregions defined in \citet{sung08, sung09} to the spatial density map of the clustered population identified in this study, already illustrated in Fig.\,\ref{fig:radec_cluster_Av_density} (left panel). The diagram exhibits an excellent agreement between the clumps and number density peaks outlined by the photometry-selected clustered population and the cataloged substructures of the NGC~2264 cluster. The contours delimiting the S~Mon and the Cone subregions encompass the two main clumps of objects in the center of our field; the highest-density regions in the southern clump are part of the Cone~(C) subcluster, while the Spokes subcluster spatially corresponds to the elongated high-density structure observed in the number density map of our clustered population in Fig.\,\ref{fig:radec_cluster_Av_density} (left). The population ratios found between the various subclusters in the center of the field are remarkably similar to those that can be deduced from Table~1 of \citet{venuti2018}. The S~Mon contour in Fig.\,\ref{fig:radec_sequence_density_Sung} encompasses about the same number of clustered stars as the external Cone contour (including the Spokes and Cone\,(C) subregions); moreover, the Cone\,(C) subregion contains about 1.6 times as many objects as the Spokes subregion.

\begin{figure}
\resizebox{\hsize}{!}{\includegraphics[width=\textwidth]{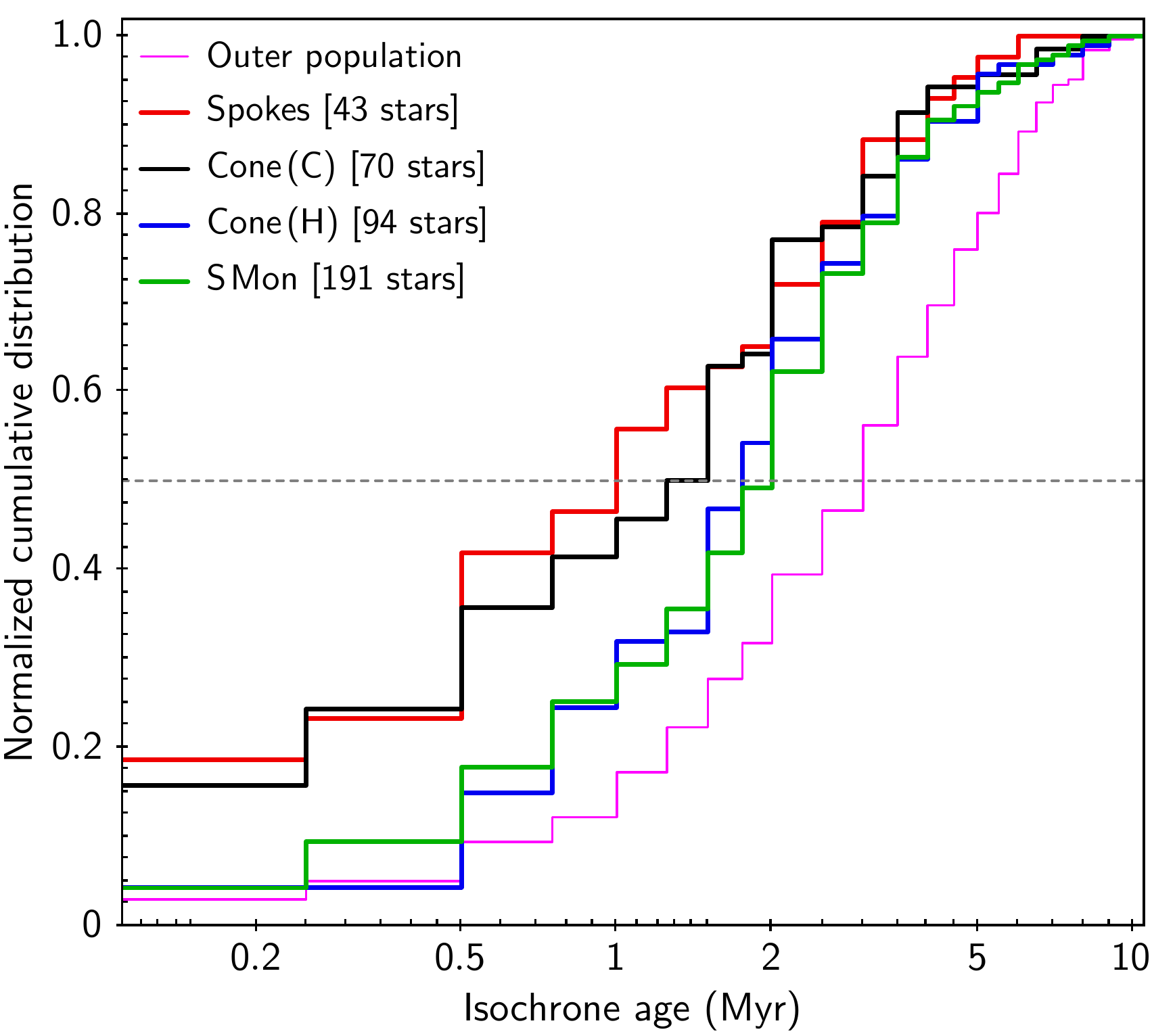}}
\caption{Normalized cumulative distribution in isochronal age of clustered stars from the various subregions highlighted in Fig.\,\ref{fig:radec_sequence_density_Sung} (red~= Spokes, black = Cone\,(C), blue = Cone\,(H), green = S~Mon). The number of stars included in each subregion is reported between square brackets. The underlying cumulative distribution in fuchsia corresponds to the population of M-type clustered stars located outside the S~Mon+Cone region. {\citeauthor{marigo2017}'s (\citeyear{marigo2017}) PARSEC-COLIBRI stellar isochrones were adopted to sort our clustered population of M-type stars into age groups, using a binning of 0.25~Myr below 2~Myr, a binning of 0.5~Myr between 2 and 8~Myr, and a binning of 1~Myr for later ages}. The gray dotted line indicates the position of the median value in each cumulative distribution.}
\label{fig:clustered_age_groups}
\end{figure}

\subsubsection{Age distribution of cluster members}

A detailed analysis of individual stellar ages for our clustered population was not attempted here; however, we did explore the respective statistical distributions in age of clustered stars projected onto the various NGC~2264 subregions. Namely, following \citet{venuti2018}, we selected all objects from our clustered population which are spatially projected inside one of the four subregions highlighted in Fig.\,\ref{fig:radec_sequence_density_Sung}: S~Mon (upper contour), Cone\,(H) (lower contour, excluding objects comprised within the two inner contours), Spokes (upper contour inside the Cone\,(H) subregion), and Cone\,(C) (lower contour inside the Cone\,(H) subregion). We then examined the distribution of each group of stars on the $r_0$ vs. $(r-i)_0$ diagram (Fig.\,\ref{fig:ri_r_CMD_seq_Baraffe15}) with respect to the grid of PMS theoretical isochrones from {\citet{marigo2017}}, brought to a distance of 800~pc. For each subpopulation, we counted the number of stars located between two successive isochrones; the results, normalized to the total number of objects in each subpopulation, are illustrated in Fig.\,\ref{fig:clustered_age_groups}. A quantitative comparison with earlier studies is hampered by the different mass range probed here (see, for instance, Appendix~A of \citealp{venuti2018}) and by the lack of additional information on the nature (e.g., accreting/non-accreting) of selected cluster members; nevertheless, some qualitative inferences can be extracted from Fig.\,\ref{fig:clustered_age_groups}. The Spokes and Cone\,(C) subpopulations exhibit a distinct age trend from Cone\,(H) and S~Mon, with a lower median age than that extracted from the cumulative distributions of the last two subregions. This suggests that the Spokes and Cone\,(C) subclusters have hosted the most recent star formation events within NGC~2264, while the Cone\,(H) and S~Mon subregions comprise on average older cluster members. These visual inferences are supported by a two-sample Kolmogorov-Smirnov (K-S) test \citep{numerical_recipes}, applied to each pair of subregions. As summarized in Table~\ref{tab:KS_age}, the cumulative distributions in age calculated for the S~Mon and Cone\,(H) subregions are statistically indistinguishable from each other; the same occurs for the age distributions pertaining to the Spokes and Cone\,(C) subregions, in spite of the fact that the median age associated with the former appears to fall in an earlier age bin than that extracted from the latter. On the contrary, the K-S test returns a very low probability that the age distribution of the Spokes or Cone\,(C) subpopulations might be extracted from the same parent distribution as that for the Cone\,(H) or S~Mon subpopulations. It is also interesting to note that the normalized cumulative distribution (in fuchsia on Fig.\,\ref{fig:clustered_age_groups}) for the more dispersed component of the M-type clustered population, i.e., stars located outside the S~Mon and Cone contours in Fig.\,\ref{fig:radec_sequence_density_Sung}, exhibits a markedly slower increase with age than observed for any of the inner NGC~2264 subregions. This indicates that the dispersed cluster population was formed earlier than cluster members projected onto the innermost regions of NGC~2264, and may have started to evolve away from the cluster cores. {While the absolute age coordinates are somewhat dependent on the specific model grid adopted, as discussed earlier, a very similar qualitative picture is derived when using the isochrone grid from \citeauthor{baraffe2015}'s (\citeyear{baraffe2015}) models to build the relative age ladder across the field. Similarly, the statistical inferences regarding the age ordering of the different cluster substructures remain unchanged if a fraction of unresolved binaries is assumed across the population (see Sect.\,\ref{sec:dist_age}), although the individual cumulative distributions in age would exhibit a lower starting level and a more slowly increasing trend along the age domain.} These results are consistent with previous studies of the star formation history of NGC~2264 \citep[e.g.,][]{sung2010, venuti2018}; future studies of the dynamics of cluster members may help discriminating between the scenarios of {\it in situ} formation or evaporation from the cluster center for the dispersed cluster population. The presence of a significant fraction of objects at older ages in the cumulative distributions for Spokes and Cone\,(C) may indicate multiple generations of stars, as well as mixing of different star populations along the line of sight.

\begin{table}
\caption{Probabilities, resulting from a two-sample K-S test, that the age distributions illustrated in Fig.\,\ref{fig:clustered_age_groups} for each NGC~2264 subregion, taken in pairs, are extracted from the same parent distribution.}
\label{tab:KS_age}
\centering
\begin{tabular}{l | c c c c}
\hline\hline
 & Spokes & Cone\,(C) & Cone\,(H) & S~Mon \\
\hline
Spokes &  & $\cdots$ & $\cdots$ & $\cdots$ \\
Cone\,(C) & {0.92} &  & $\cdots$ & $\cdots$ \\
Cone\,(H) & {0.018} & {0.05} &  & $\cdots$ \\
S~Mon & {0.011} & {0.019} & {0.99} & \\
\hline
\end{tabular}
\tablefoot{The table is symmetric with respect to its diagonal.
}
\end{table}

\subsubsection{Membership census}

To appraise the robustness of our object selection, we cross-correlated our list of M-type stars in the clustered population with the census of the NGC~2264 region provided by the CSI~2264 collaboration\footnote{https://irsa.ipac.caltech.edu/data/SPITZER/CSI2264/} \citep{cody2014}. The latter is the result of a comprehensive effort aimed at gathering a complete point source catalog of the field, with membership information from the diverse studies on the region available in the literature and a final membership flag assigned to each catalog entry (1~= ``very likely NGC~2264 member'', 2~= ``possible NGC~2264 member'', -1~= ``likely field object'', 0~= ``no membership information''). Among the 1369 sources that comprise our clustered population, 1203 have a counterpart in the CSI~2264 catalog within a radius of 1~arcsec{\footnote{For comparison, this value corresponds to the $\sim$0.06th percentile in the distribution of angular separations between each point source and its closest neighbor in the CSI~2264 and $r,i,J$ catalogs (the distribution peaks in both cases around an intra-source separation of $\sim$8~arcsecs).}}. The vast majority of them (95.5\%) were assigned a CSI~2264 membership flag which is consistent with the photometric association with the clustered population derived here (namely, 33.7\% have membership flag ``1'', 27.9\% have membership flag ``2'', and 33.8\% have membership flag ``0''). Only the remaining 4.5\% (54 stars) have a contrasting membership classification (i.e., field stars) in the CSI~2264 scheme. Sources matched between our clustered population catalog and the CSI~2264 catalog are effectively distributed over $\sim$3~deg$^2$ of the field in Fig.\,\ref{fig:radec_cluster_Av_density}; based on the analysis in Sect.\,\ref{sec:contaminants}, we would expect about 100 field contaminants in our selection of clustered stars across this area. The number of objects with discordant classification in the CSI~2264 catalog is therefore consistent with our statistical estimate of the rate of contaminants (a few tens of additional field contaminants may be found in the 33.8\% of objects with no membership status in the CSI~2264 catalog, or among those with more uncertain classification).

The 1369 sources contained in our clustered population also include 166 objects with no counterpart in the CSI~2264 catalog. This group of objects are predominantly located in the most external areas of the field shown in Fig.\,\ref{fig:radec_cluster_Av_density}, a dozen of parsecs away from the NGC~2264 cluster core, and therefore lying outside the region typically targeted in focused surveys. Based on the statistical contamination rate estimated in Sect.\,\ref{sec:contaminants}, we expect about 35 field contaminants among the 166 sources; therefore, our analysis suggests new statistical membership to the NGC~2264 cluster for about a hundred stars located at a distance of 10--15 pc from the cluster cores. A complete list of the 1369 M-type stars selected statistically as part of our clustered population, and of the relevant literature information, is reported in Table~\ref{tab:Mstars_NGC2264}.

\begin{table*}
\caption{Photometry and literature classification for the M-type stars with statistical membership to the NGC~2264 clustered population identified in this study.}
\label{tab:Mstars_NGC2264}
\centering
\resizebox{\textwidth}{!} {
\begin{tabular}{c c c c c c c c c c c c c}
\hline\hline
R.A. & Dec & $r$-mag & $i$-mag & $J$-mag & $r$-err & $i$-err & $J$-err & ($r,i$)\_source & $J$\_source & A$_V$ & {\small CSIMon-\#} & {\small Memb} \\
\hline
100.39517 & 9.83891 & 20.19 & 18.263 & 15.606 & 0.05 & 0.003 & 0.008 & CSI~2264 & UKIDSS & 0.205 & 000002 & 2\\
100.32146 & 9.89437 &	17.574 & 16.331 & 14.192 & 0.006 & 0.004 & 0.005 & {\small Pan-STARRS1} & UKIDSS & 0.933 & 000005 & 1\\
100.32758 & 9.90211 & 21.89 & 19.505 & 16.084 & 0.17 & 0.019 & 0.008 & {\small Pan-STARRS1} & UKIDSS & 2.301 & 000013 & 2\\
100.30493 & 9.91909 & 16.073 & 15.0328 & 12.9150 & 0.003 & 0.0006 & 0.0009 & CSI~2264 & UKIDSS & 1.577 & 000018 & 1\\
100.26904 & 10.01185 & 17.798 & 16.2571 & 13.931 & 0.008 & 0.0010 & 0.007 & CSI~2264 & UKIDSS & 0.527	& 000022 & 2\\
100.30178 & 9.91670 & 18.212 & 16.8474 & 14.676 & 0.010 & 0.0014 & 0.007 & CSI~2264 & UKIDSS & 0.565 & 000023 & 1\\
100.27161 & 10.00526 & 17.180 & 15.5069 & 13.052 & 0.005 & 0.0007 & 0.022 & CSI~2264 & 2MASS & 0.540 & 000026 & 2\\
100.36946 & 9.89951 & 19.95 & 18.098 & 15.328 & 0.03 & 0.008 & 0.004 & {\small Pan-STARRS1} & UKIDSS & 1.461 & 000032	& 1\\
100.35084 & 9.89526 & 19.26 & 17.318 & 14.594 & 0.04 & 0.008 & 0.005 & {\small Pan-STARRS1} & UKIDSS & 0.552 & 000039 & 2\\
100.43947 & 9.92845 & 18.314 & 16.7654 & 14.410 & 0.011 & 0.0013 & 0.007 & CSI~2264 & UKIDSS & 0.666 & 000042 & 1\\
\hline
\end{tabular}
}
\tablefoot{A full version of the Table is available in electronic form at the CDS. A portion is shown here for guidance regarding its form and content.\\
``$r$-mag, $i$-mag'' = photometry calibrated to the SDSS system as described in Sect.\,\ref{sec:phot_calib}.
``$r$-err, $i$-err, $J$-err'' = photometric uncertainties computed taking into account the instrumental errors and the calibration relationships.
``($r,i$)\_source'' = survey from where the original optical photometry for the corresponding object was extracted. Possible values are: CSI~2264, Pan-STARRS1, IPHAS, SDSS (see Sect.\,\ref{sec:optical_cat}).
``$J$\_source'' = survey from where the $J$-band photometry for the corresponding object was extracted. Possible values are: UKIDSS, 2MASS (see Sect.\,\ref{sec:J_phot}).
``CSIMon-\#'' = object identifier in the CSI~2264 master catalog, if present.
``Memb'' = membership flag associated with the corresponding source in the CSI~2264 catalog, if present. Possible values are: 1, 2, 0, -1 (see Sect.\,\ref{sec:csi2264_comp}).
}
\end{table*}  

\subsection{Control field}

\begin{figure}
\resizebox{\hsize}{!}{\includegraphics{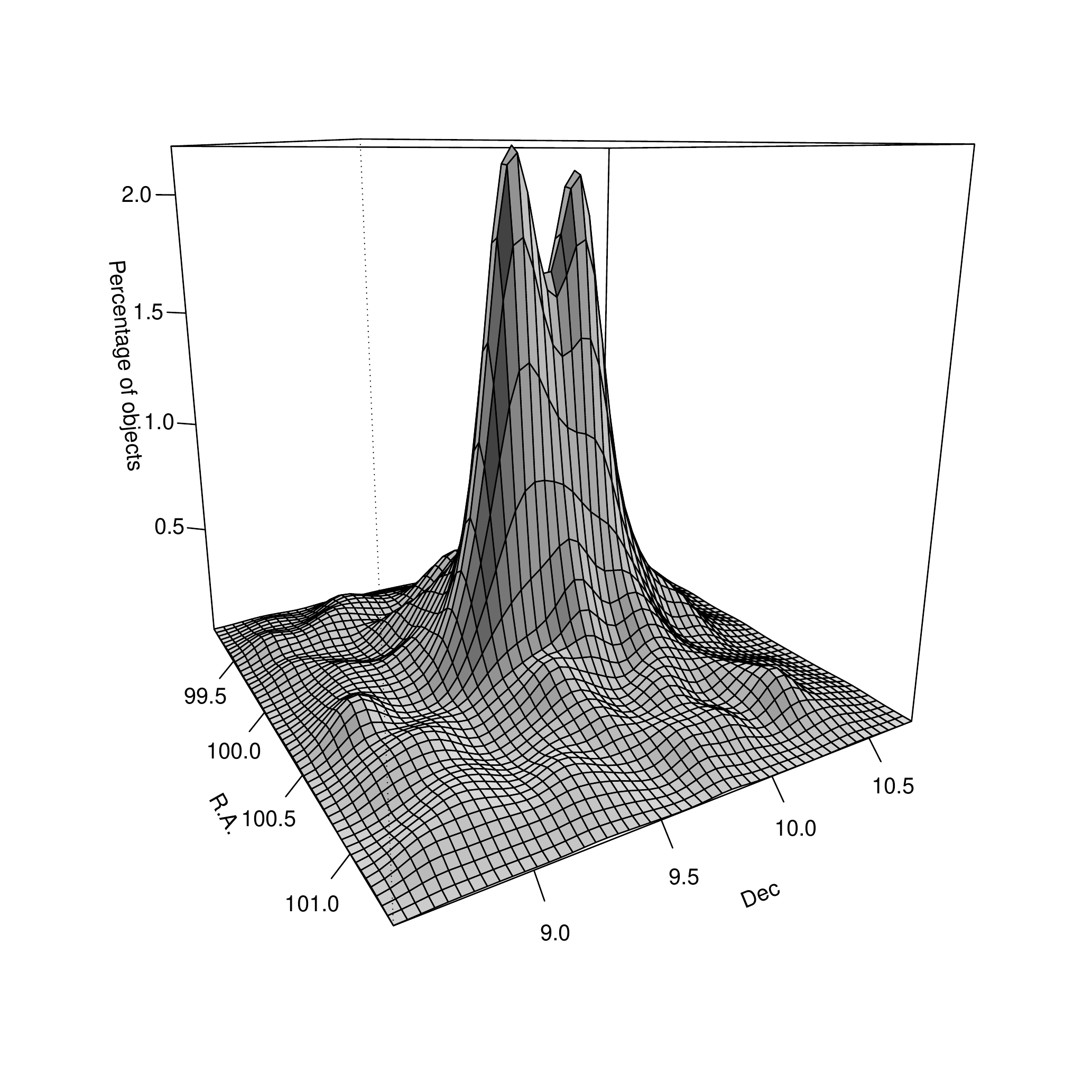}}
\caption{Smoothed two-dimensional density map of objects in the clustered population across the field.}
\label{fig:radec_density_curve}
\end{figure}

Fig.\,\ref{fig:radec_density_curve} illustrates a normalized two-dimensional density curve of the clustered population across the field, created by computing a binned two-dimensional kernel density estimate map with R\footnote{The map was built using the R function $bkde2D()$ with a bandwidth of 0.08$^\circ$, and plotted using the R function $persp()$.}. As already noted in the discussion around Fig.\,\ref{fig:radec_cluster_Av_density}, two main peaks in population are observed, corresponding to the northern (S~Mon) and southern (Cone) clumps of the cluster. The number density of objects decreases by half within a radius of $\sim$0.15$^\circ$; this implies that a significant fraction of the cluster develops within a radius of $\leq$2--2.5~pc from the cluster cores, while a tail of objects spreads out until the edges of our spatial domain, where the number density is about 50 times smaller than that in the peak. If our mapping is accurate, then we would expect to not find any evidence of the clustered population if we shifted our field of investigation to a neighboring area outside of the region depicted in Fig.\,\ref{fig:ra_dec_density}.

To test this point and as a further validation of our method of analysis, we selected a {$\sim$1.8~deg$^2$} control field, adjacent to our NGC~2264 field {in Galactic (l, b) coordinate space}. We then built an extensive $r,i,J$ point source catalog for the control field, using the {same databases employed for the science field with the exception of the CSI\,2264 CFHT catalog (limited to the central part of the science field)}, and following the same procedure detailed in Sect.\,\ref{sec:data}. The final catalog assembled for the control field encompasses {111410} sources; this corresponds to an average surface density of objects of $\sim${61900} stars/deg$^2$, slightly larger than the average population density of $\sim$46300 stars/deg$^2$ found in the science field{, possibly ensuing from the lack of nebulosity or molecular clouds, and less crowding}. {The photometric ranges covered by the selected control field population (${r = 13.4-23.6}$, ${i = 13.1-22.1}$, ${J = 11.7-19.8}$) and the completeness limits estimated for the control field catalog (${r \sim 22}$, ${i \sim 20.5}$, ${J \sim 18.5}$) are comparable to, although slightly deeper than, the properties of our science field catalog.} About {14.3}\% of the {111410} sources ({15881}) were selected as M-type stars following the same analysis illustrated in Sect.\,\ref{sec:iJ_ri} and Fig.\,\ref{fig:iJ_ri}. {This fraction is slightly larger than the fraction of stars extracted above the M-type boundary line} from the population of the science field (13\%). We applied the same numerical approach described in Sect.\,\ref{sec:Av_meas} to measure the individual A$_V$ of M-type stars in the control field, and then investigated its A$_V$ structure similarly to what described in Sect.\,\ref{sec:Av_structure}. The same A$_V$ groups were adopted to sort the M-type population of the control field, and boundary lines at the same locations as shown in the left panels of Fig.\,\ref{fig:images} were used to identify the ``counterparts'' of the ``field stars locus'' and of the ``upper sequence locus'' on the $r$ vs. $r-i$ diagrams for the various A$_V$ groups, and to inspect the spatial distribution of the corresponding populations.

\begin{figure*}
\centering
\includegraphics[width=\textwidth]{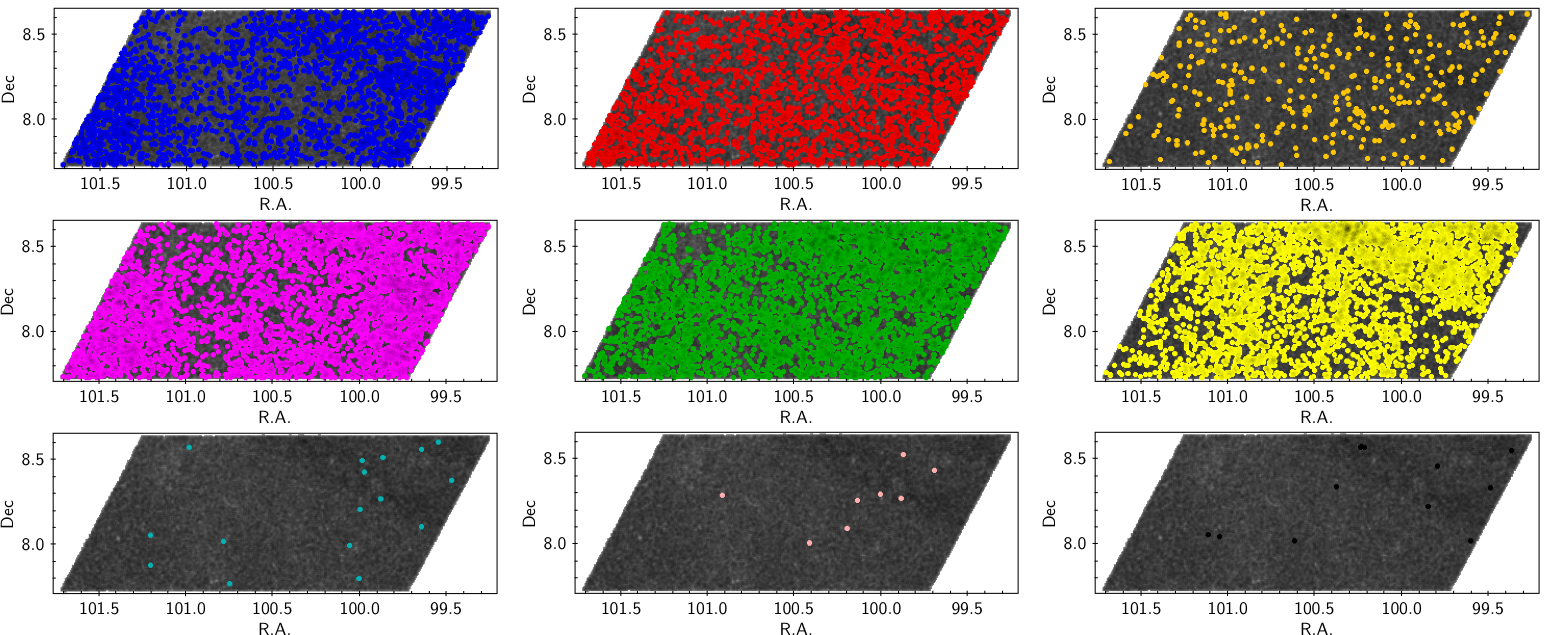}
\caption{Spatial distribution for M-type stars in the control field, sorted following the scheme in Fig.\,\ref{fig:images}. The upper panels, from left to right, depict $A_V < 0.7$ stars with the same photometric properties of the lower, middle, and upper populations in Fig.\,\ref{fig:a}. Middle and lower panels are populated respectively by ``field locus'' and ``sequence locus'' stars with $A_V$\,=\,0.7--1.2 (left), $A_V$\,=\,1.2--1.7 (center), and $A_V$\,$\geq$\,$1.7$ (right).}
\label{fig:radec_control_field}
\end{figure*}

Results of this analysis for the control field are illustrated in Fig.\,\ref{fig:radec_control_field}. The ``upper sequence locus'', forced on the photometric diagrams for the control field in analogy to what found for the population of the science field, contains {388 sources, mostly comprised in the lowest A$_V$ group}. The spatial distribution of field stars in the various A$_V$ groups do not exhibit any remarkable feature, and no distinct behavior is observed between the ``foreground population'' {(middle panel of the uppermost row in Fig.\,\ref{fig:radec_control_field})} and the ``background population'' {(first panel of the uppermost row, and middle row panels in Fig.\,\ref{fig:radec_control_field})}. By extrapolating the trend depicted in the density curve in Fig.\,\ref{fig:radec_density_curve} to our control field, we would expect to find there no more than $\sim${50} candidate cluster members. {The 388 sources, in the control field, selected photometrically as if they were part of the clustered population are likely dominated by foreground stars. Indeed, as mentioned earlier, the control field is more densely populated than the science field, and an estimation of foreground contaminants similar to that performed in Sect.\,\ref{sec:contaminants} indicates that at least 25\% of the 388 sources in the control field are MS dwarfs located between 0 and 350~pc. This estimate may actually be a lower limit to the number of MS dwarfs in the area, as suggested by the Besan\c{c}on model of stellar population synthesis of the Galaxy \citep{robin2003}, which predicts a population of around 300 M-type dwarfs distributed between 0 and 350~pc in the solid angle subtended by our control field, with apparent magnitudes comprised within the photometric ranges probed in our catalog, and assuming a diffuse extinction of 0.7~mag/kpc. In addition, a comparison in dereddened $J$-band photometry between the complete source catalog of the control field and the sample of 388 stars selected above the ``upper sequence'' photometric locus shows that the latter are on average brighter, by about 3~$\sigma$, than the typical $J$-mag level measured across the entire M-type population of the field. This suggest that giants may also contribute significantly to the sample of 388 sources above the main field locus on the ($r-i$, $r$) diagram.} Therefore, we conclude that the application of our investigation method to the control field did not reveal any sign of clustered populations, consistently with expectations.

\section{Conclusions} \label{sec:conclusions}

We reported on a blind photometric study of a large (4~deg$^2$) area around the young cluster NGC~2264. This study follows on from the new conceptual approach presented in \citet{damiani2018} to investigate the structure and stellar content of a given field by using deep multi-wavelength photometry of its M-type population. We gathered an extensive $r,i,J$ catalog of the region, making use of existing large-scale surveys, notably Pan-STARRS1 in the optical and UKIDSS in the near-infrared. We then mapped the stellar color locus on the ($i-J$, $r-i$) diagram to identify and extract M-type stars in the field, for which a direct and empirical estimate of individual reddening (hence distance) can be obtained from $r,i,J$ data. We used the color-color diagram to measure the A$_V$ of individual M-type stars and built an averaged reddening map of the region; we then combined the ($r-i$, $r$)  and (R.A., Dec) diagrams to inspect the photometric and spatial properties of stars in our sample as a function of their A$_V$, in order to reconstruct the structure of the NGC~2264 region and explore the nature of its population.

Assuming no previous knowledge on the NGC~2264 region, we were able to identify two distinct populations within the field: a diffuse field population, and a clustered population. The latter appears to be spatially associated with a concentration of obscuring material, responsible for a gap between the photometric loci of foreground stars and low-reddening background stars on the color-magnitude diagram. This peculiar feature allowed us to derive an estimate of the distance (800--900~pc) and age {($\sim$0.5--5~Myr)} of the clustered population, by comparison with reference color-magnitude sequences and PMS isochrones. Such estimates are globally consistent with the literature parameters for the NGC~2264 cluster (760~pc and $\sim$3--5~Myr). Our clustered population exhibits a hierarchical structure, with two main clumps along its vertical extension and several smaller-scale substructures. An a posteriori comparison with the NGC~2264 subregions listed in \citet{sung08, sung09} revealed an excellent agreement between the observed spatial features of the clustered population and the locations of the known NGC~2264 subclusters. {Our data selection enables us to detect $\sim$1~Myr--old stars at the distance of the NGC~2264 cluster down to masses of 0.1~$M_\odot$ for A$_V$\,$\leq$\,2~mag (A$_V$ up to 4~mag for M$_\star$$\sim$\,0.2~$M_\odot$, and up to 6~mag for M$_\star$$\sim$\,0.35~$M_\odot$). Although very embedded, Class~0/I sources in the region are not accessible with our approach, we could recover the spatial components of NGC~2264 down to the Spokes cluster, previously identified as the star-forming site of most recent formation within the region from mid-IR surveys.} 

About 84\% of the sources in our clustered population are known members or candidate members of the NGC~2264 cluster, or are included in point source catalogs of the NGC~2264 region, albeit without membership information. Only about 4\% of objects in our clustered population were previously reported to be field stars instead; this corroborates the strength of our observational approach to statistically identify a young clustered population amidst a diffuse population of field stars. Thanks to the wide, deep and homogeneous field coverage of our sample, we were also able to identify a new population of about a hundred M-type stars, with indications of statistical membership to the cluster. This group of objects are located at a distance of about 10--15~pc from the cluster cores, that is, beyond the region that is usually targeted in focused studies of NGC~2264. This finding supports previous inferences that NGC~2264 members are spread over a much larger radius than the 2--2.5~pc around the active star-forming sites which constitute the bulk of the cluster (e.g., \citealp{venuti2014}).   

The method of investigation examined in \citet{damiani2018} and successfully tested here to rederive the NGC~2264 cluster has an enormous potential for studies of young star clusters in the era of large-scale, homogeneous photometric surveys. Specifically, it can be readily implemented to statistically identify and map young distant ($\geq$ 1--2~kpc) clusters, that will become accessible for the first time with campaigns such as Pan-STARRS and LSST, and for which extensive spectroscopic surveys and X-ray surveys would not be feasible due to prohibitive integration times. In particular, based on the expected survey performance, a single LSST frame will be able to capture young (1--10~Myr), late M-type ($\geq$M3) stars with low-to-moderate reddening (A$_V$~{\small $\lesssim$}~1) at distances from 200--300~pc to 2--5~kpc, and their earlier-type counterparts at distances from 0.5--1~kpc to $\sim$10--20~kpc and beyond. Synergies with proper motion surveys, such as {\it Gaia} (see, for instance, the work by \citealp{prisinzano2018} on the Vela Molecular Ridge) and LSST itself, will enable detailed characterization of the photometrically selected populations, to probe the dynamics of star cluster formation and evolution. Moreover, the specific focus on M-type stars, often affected by selection biases even in well-known young clusters, will enable a complete stellar census of many star-forming regions over the entire mass spectrum. This is of direct interest to a broad range of topics, from reconstructing the initial mass function, to compiling catalogs of suitable targets to investigate the earliest stages of planetary systems.

\begin{acknowledgements}
We wish to thank the anonymous referee for their careful manuscript reading. This work makes use of data products from the Pan-STARRS1, IPHAS, SDSS, UKIDSS, and 2MASS surveys. This work is also based on data from MegaPrime/MegaCam at the Canada-France-Hawaii Telescope (CFHT). L.V. acknowledges support from the Institutional Strategy of the University of T\"ubingen (Deutsche Forschungsgemeinschaft, ZUK 63) while this manuscript was being prepared.
\end{acknowledgements}

\bibliographystyle{aa}
\bibliography{references}

\end{document}